\documentclass[10pt,a4paper,reqno]{amsart}

\makeatletter
\g@addto@macro{\endabstract}{\@setabstract}
\newcommand{\authorfootnotes}{\renewcommand\thefootnote{\@fnsymbol\c@footnote}}%
\makeatother

\usepackage{amssymb}
\usepackage{mathtools}
\usepackage[utf8]{inputenc}
\usepackage{bm}
\usepackage{bbm}
\usepackage{amsmath}
\usepackage{amsfonts}
\makeatletter
\def\amsbb{\use@mathgroup \M@U \symAMSb}
\makeatother

\usepackage{comment}
\usepackage{graphicx}
\usepackage{caption}
\usepackage{subcaption}

\usepackage{bbold}

\newcommand{\bga}{\begin{aligned}}
	\newcommand{\ena}{\end{aligned}}

\newcommand{\bge}{\begin{enumerate}}
	\newcommand{\ene}{\end{enumerate}}

\usepackage{dutchcal}
\usepackage{fancyhdr}
\usepackage{subcaption}

\usepackage{pgfplots}
\pgfplotsset{compat=1.15}

\usepackage{booktabs}
\usepackage{xcolor}

\definecolor{webgreen}{rgb}{0,.5,0}
\definecolor{webbrown}{rgb}{.6,0,0}
\definecolor{RoyalBlue}{cmyk}{1, 0.50, 0, 0}

\newcommand{\red}[1]{{\color{red} #1}}
\newcommand{\blue}[1]{{\color{blue} #1}}

\newcommand{\hide}[1]{}

\usepackage{tikz}
\usepackage{pict2e}
\usepackage{todonotes}
\usetikzlibrary{decorations.markings}

\DeclareSymbolFont{bbold}{U}{bbold}{m}{n}
\DeclareSymbolFontAlphabet{\mathbbold}{bbold}

\newcommand{\aaa}{\textrm{a}}

\newcommand{\E}{{\mathbb E}}

\newcommand{\C}{{\mathbb C}}

\newcommand{\Z}{{\mathbb Z}}

\newcommand{\N}{{\mathbb N}}

\newcommand{\T}{{\mathbb T}}

\newcommand{\al}{\alpha}
\newcommand{\be}{\beta}

\newcommand{\ga}{\gamma}
\newcommand{\Ga}{\Gamma}
\newcommand{\La}{\Lambda}

\newcommand{\de}{\delta}

\newcommand{\De}{\Delta}
\newcommand{\om}{\omega}
\newcommand{\Om}{\Omega}
\newcommand{\ze}{\zeta}

\newcommand{\di}{\displaystyle}

\newcommand{\ic}{\textrm{i}}
\newcommand{\dd}{\textrm{d}}

\newcommand{\qasq}{\quad \text{as} \quad}
\newcommand{\qandq}{\quad \text{and} \quad}

\usepackage{etoolbox}

\makeatletter
\pretocmd{\section}{\addtocontents{toc}{\protect\addvspace{1\p@}}}{}{}
\pretocmd{\subsection}{\addtocontents{toc}{\protect\addvspace{1\p@}}}{}{}
\pretocmd{\subsubsection}{\addtocontents{toc}{\protect\addvspace{1\p@}}}{}{}
\makeatother

\usepackage{comment}

\newtheorem{theorem}{Theorem}

\newtheorem{remark}[theorem]{Remark}
\newtheorem{lemma}[theorem]{Lemma}

\newtheorem{corollary}{Corollary}[theorem]

\numberwithin{equation}{section}
\numberwithin{theorem}{section}
\numberwithin{notation}{section}

\usepackage[T1]{fontenc}

\usepackage{mathtools}

\makeatletter
\DeclareRobustCommand\widecheck[1]{{\mathpalette\@widecheck{#1}}}
\def\@widecheck#1#2{%
	\setbox\z@\hbox{\m@th$#1#2$}%
	\setbox\tw@\hbox{\m@th$#1%
		\widehat{%
			\vrule\@width\z@\@height\ht\z@
			\vrule\@height\z@\@width\wd\z@}$}%
	\dp\tw@-\ht\z@
	\@tempdima\ht\z@ \advance\@tempdima2\ht\tw@ \divide\@tempdima\thr@@
	\setbox\tw@\hbox{%
		\raise\@tempdima\hbox{\scalebox{1}[-1]{\lower\@tempdima\box
				\tw@}}}%
	{\ooalign{\box\tw@ \cr \box\z@}}}
\makeatother

\usepackage[mathscr]{euscript}
\usepackage{amsmath}

\ExplSyntaxOn

\NewDocumentCommand{\pFq}{O{}mmmmm}
{
	\group_begin:
	\keys_set:nn { hypergeometric } { #1 }
	\hypergeometric_print:nnnnn { #2 } { #3 } { #4 } { #5 } { #6 }
	\group_end:
}
\NewDocumentCommand{\hypergeometricsetup}{m}
{
	\keys_set:nn { hypergeometric } { #1 }
}

\tl_new:N \l_hypergeometric_divider_tl
\tl_new:N \l_hypergeometric_left_tl
\tl_new:N \l_hypergeometric_right_tl

\keys_define:nn { hypergeometric }
{
	symbol .tl_set:N = \l_hypergeometric_symbol_tl,
	symbol .initial:n = F,
	separator .tl_set:N = \l_hypergeometric_separator_tl,
	separator .initial:n = {},
	skip .tl_set:N = \l_hypergeometric_skip_tl,
	skip .initial:n = 8,
	divider .choice:,
	divider/semicolon .code:n = \tl_set:Nn \l_hypergeometric_divider_tl { \;; },
	divider/bar .code:n = \tl_set:Nn \l_hypergeometric_divider_tl { \;\middle|\; },
	divider .initial:n = semicolon,
	fences .choice:,
	fences/brack .code:n = 
	\tl_set:Nn \l_hypergeometric_left_tl {[}
	\tl_set:Nn \l_hypergeometric_right_tl {]},
	fences/parens .code:n = 
	\tl_set:Nn \l_hypergeometric_left_tl {(}
	\tl_set:Nn \l_hypergeometric_right_tl {)},
	fences .initial:n = brack,
}

\cs_new_protected:Nn \hypergeometric_print:nnnnn
{
	{} \sb {#1} \l_hypergeometric_symbol_tl \sb { #2 }
	\left\l_hypergeometric_left_tl
	\genfrac .. % no delimiters
	{0pt} % no line
	{} % default style
	{ \__hypergeometric_process:n { #3 } } % numerator
	{ \__hypergeometric_process:n { #4 } } % denominator
	\l_hypergeometric_divider_tl
	#5
	\right\l_hypergeometric_right_tl
}

\cs_new_protected:Nn \__hypergeometric_process:n
{
	\clist_use:nn { #1 }
	{
		{\l_hypergeometric_separator_tl}
		\mspace { \l_hypergeometric_skip_tl mu }
	}
}

\ExplSyntaxOff

\usepackage{yfonts}

\usepackage{xcolor}
\definecolor{webgreen}{rgb}{0,.5,0}
\definecolor{webbrown}{rgb}{.6,0,0}
\definecolor{RoyalBlue}{cmyk}{1, 0.50, 0, 0}
\usepackage[colorlinks=true, breaklinks=true, urlcolor=webbrown, linkcolor=RoyalBlue, citecolor=webgreen,backref=page]{hyperref}

\usepackage{amssymb,amsthm,amsxtra}
\usepackage{epic,eepic}
\usepackage{epsfig}
\newcommand\psymmU{%
	\begin{picture}(1,1)(0,0)%
		\allinethickness{0.5pt}%
		\path(0,0)(0,1)(1,1)(1,0)(0,0)%
\end{picture}}
\newcommand\psymmUU{%
	\begin{picture}(1,1)(0,0)%
		\allinethickness{0.5pt}%
		\path(0,0)(0,1)(1,1)(1,0)(0,0)%
		\put(0.5,0.5){\makebox(0,0){$\cdot$}}%
\end{picture}}
\newcommand\psymmO{%
	\begin{picture}(1,1)(0,0)%
		\allinethickness{0.5pt}%
		\path(0,0)(0,1)(1,1)(1,0)(0,0)%
		\path(0,0)(1,1)%
\end{picture}}
\newcommand\psymmS{%
	\begin{picture}(1,1)(0,0)%
		\allinethickness{0.5pt}%
		\path(0,0)(0,1)(1,1)(1,0)(0,0)%
		\path(1,0)(0,1)%
\end{picture}}
\newcommand\psymmu{%
	\begin{picture}(1,1)(0,0)%
		\allinethickness{0.5pt}%
		\path(0,0)(0,1)(1,1)(1,0)(0,0)%
		\path(0,0)(1,1)%
		\path(0,1)(1,0)%
\end{picture}}

\newbox\tsymmUbox
\newbox\tsymmUUbox
\newbox\tsymmObox
\newbox\tsymmSbox
\newbox\tsymmubox
\setbox\tsymmUbox =\hbox{\kern0.75pt\setlength{\unitlength}{6pt}\psymmU \kern0.75pt}
\setbox\tsymmUUbox=\hbox{\kern0.75pt\setlength{\unitlength}{6pt}\psymmUU\kern0.75pt}
\setbox\tsymmObox =\hbox{\kern0.75pt\setlength{\unitlength}{6pt}\psymmO \kern0.75pt}
\setbox\tsymmSbox =\hbox{\kern0.75pt\setlength{\unitlength}{6pt}\psymmS \kern0.75pt}
\setbox\tsymmubox =\hbox{\kern0.75pt\setlength{\unitlength}{6pt}\psymmu \kern0.75pt}

\newbox\symmUbox
\newbox\symmUUbox
\newbox\symmObox
\newbox\symmSbox
\newbox\symmubox
\setbox\symmUbox =\hbox{\kern0.75pt\setlength{\unitlength}{4.5pt}\psymmU \kern0.75pt}
\setbox\symmUUbox=\hbox{\kern0.75pt\setlength{\unitlength}{4.5pt}\psymmUU\kern0.75pt}
\setbox\symmObox =\hbox{\kern0.75pt\setlength{\unitlength}{4.5pt}\psymmO \kern0.75pt}
\setbox\symmSbox =\hbox{\kern0.75pt\setlength{\unitlength}{4.5pt}\psymmS \kern0.75pt}
\setbox\symmubox =\hbox{\kern0.75pt\setlength{\unitlength}{4.5pt}\psymmu \kern0.75pt}

\begin{document}
	
	\tikzset{->-/.style={decoration={
				markings,
				mark=at position #1 with {\arrow{latex}}},postaction={decorate}}}
	
	\tikzset{-<-/.style={decoration={
				markings,
				mark=at position #1 with {\arrowreversed{latex}}},postaction={decorate}}}
	
	\title{A Riemann-Hilbert approach to asymptotic analysis of Toeplitz+Hankel determinants II}
	
	\maketitle
	
	\begin{center}
		\authorfootnotes
		Roozbeh Gharakhloo\footnote{Mathematics Department,	University of California Santa Cruz, Santa Cruz, CA, USA.	e-mail: roozbeh@ucsc.edu},
		Alexander Its\footnote{Department of Mathematical Sciences, Indiana University Indianapolis, Indianapolis, IN, USA. e-mail: aits@iu.edu}  \par \bigskip
	\end{center}
	
	\begin{abstract}
		In this article, we continue the development of the Riemann–Hilbert formalism for studying the asymptotics of Toeplitz+Hankel determinants with non-identical symbols, which we initiated in \cite{GI}. In \cite{GI}, we showed that the Riemann–Hilbert problem we formulated admits the Deift–Zhou nonlinear steepest descent analysis, but with a special restriction on the winding numbers of the associated symbols. In particular, the most natural case, namely zero winding numbers, is not allowed. A principal goal of this paper is to develop a framework that extends the asymptotic analysis of Toeplitz+Hankel determinants to a broader range of winding-number configurations. As an application, we consider the case in which the winding numbers of the Szegő-type Toeplitz and Hankel symbols are zero and one, respectively, and compute the asymptotics of the norms of the corresponding system of orthogonal polynomials.
	\end{abstract}
	
	\begin{itemize}
		\item[] \footnotesize \textbf{2020 Mathematics Subject Classification:} 15B05, 30E15, 30E25, 41A60, 42C05, 82B20.
		
		\vspace{.2cm}
		
		\item[] \textbf{Keywords:}  Riemann–Hilbert problems · Toeplitz+Hankel determinants · asymptotic analysis \footnotesize 
	\end{itemize}
	
	\tableofcontents
	
	\section{Introduction}
	
Structured matrices and their determinants, such as Toeplitz, Hankel, and their various extensions, play a central role in diverse areas of mathematics and physics, especially in random matrix theory and statistical mechanics. These include classical Toeplitz and Hankel matrices, combinations like Toeplitz+Hankel, bordered and framed variants, and other generalizations such as slant Toeplitz matrices. There is a vast literature on the study of these structured determinants; see, e.g.,~\cite{BE,BE1,BE2,BE3,BE4,BEGIL,Bottcher-Silbermann,Bottcher-Silbermann1,Charlier, DIK,G24,GL24,GW,Krasovsky, Krasovsky1} and references therein.

	The $n \times n$ Toeplitz and Hankel  matrices associated respectively to the symbols $\phi$ and $w$, supported on the unit circle $\mathbb{T}$ are respectively defined as  \begin{equation}\label{Toeplitz}
		T_{n}[\phi;r]:=\{\phi_{j-k+r}\}, \qquad  j,k=0,\cdots, n-1, \qquad  \phi_k=\int_{\T} z^{-k}\phi(z) \frac{dz}{2\pi i z},
	\end{equation} and 
	\begin{equation}\label{Hankel}	    	H_{n}[w;s]:=\{w_{j+k+s}\}, \qquad  j,k=0,\cdots, n-1, \qquad  w_k=\int_{\T} z^{-k}w(z) \frac{dz}{2\pi i z},
	\end{equation}
	for fixed \textit{offset} values $r,s \in \Z$. If the Hankel symbol $w$ is supported on a subset $I$ of the real line, then $w_k$ in \eqref{Hankel} are instead given by
	\begin{equation}\label{moments of w}
		w_k=\int_{I} x^{k}w(x) dx.
	\end{equation}
	
		The Toeplitz+Hankel determinant associated with symbols $\phi$ and $w$, respectively generating Toeplitz and Hankel components with the offset pair $(r,s) \in \Z \times \Z$, is of the form $$D_n[\phi,w;r,s] := \det \left\{ \phi_{j-k+r} + w_{j+k+s} \right\}^{n-1}_{j,k=0},$$ i.e.
	\begin{equation}\label{Det}
		D_n[\phi,w;r,s] =		\det \begin{pmatrix}
			\phi_r+w_s & \phi_{r-1}+w_{s+1} & \cdots & \phi_{r-n+1}+w_{s+n-1} \\
			\phi_{r+1}+w_{s+1} & \phi_{r}+w_{s+2} & \cdots  & \phi_{r-n+2}+w_{s+n}  \\
			\vdots & \vdots & \ddots & \vdots\\
			\phi_{r+n-1}+w_{s+n-1} & \phi_{r+n-2}+w_{s+n} & \cdots  & \phi_{r}+w_{s+2n-2} 
		\end{pmatrix}.
	\end{equation} 
	
				Our ultimate goal is to obtain the large-$n$ asymptotics of $D_n[\phi,w;r,s]$ for arbitrary offset pairs and to determine how the analytical properties of the symbols $\phi$ and $w$ affect this asymptotics.

			This paper is a sequel to our first paper \cite{GI} where we studied the offset pair $(r,s)=(1,1)$ for the Szeg{\H o}-type symbols — that is, nonvanishing smooth symbols on the unit circle with zero winding number which admit an analytic continuation to a neighborhood of the circle. In \cite{GI} we computed the asymptotics of $D_n[\phi,w;1,1]$ up to the constant.  At the end of \cite{GI} we formulated a number of open questions. Among those open questions was the extension of analysis  to other offset pairs $(r,s)$. In this paper we provide a Riemann-Hilbert framework for this extension for any offset pair. We explicitly work out the details for the offset pairs $(r,s)\in \{(0,0), (0,1), (0,2)\}$ which are of more interest due to their appearance in applications. Additionally, for the offset pair $(0,1)$, we asymptotically solve the associated Riemann-Hilbert problem and obtain the asymptotics for the norms of the associated orthogonal polynomials.

Below, we outline the main motivations for considering such offset extensions. 
We first review key applications in which principal objects are characterized 
by Toeplitz--Hankel determinants with various offset pairs, treating separately 
the cases $w = \phi$ and $w \ne \phi$. 
Although the main results of this work pertain to the latter case, 
we also review the case $w = \phi$ and its applications 
for completeness and to place the discussion in a broader context.

\subsection{Identical Toeplitz and Hankel symbols}
When the Toeplitz and Hankel symbols are identical, that is when $\phi=w$, the applications and asymptotic properties of Toeplitz+Hankel determinants have been studied extensively by several authors. In particular, E.~Basor and T.~Ehrhardt have made significant contributions to understanding various aspects of these determinants in a series of papers~\cite{BE1,BE2,BE3,BE4} using operator-theoretic methods. In most of these works, they focus on the offset pair $(0,1)$, except in~\cite{BE3}, where they establish Szeg{\H o}-type limit theorems for $D_n[\phi,\phi;0,s]$ with $s \geq 1$. 

We now briefly outline several key applications of Toeplitz+Hankel determinants $D_n[\phi,\phi;0,s]$. For a compact group $G$, define
\[
\mathbb{E}_{U \in G} f(U)
\]
as the integral of $f(U)$ with respect to the normalized Haar measure on $G$\footnote{When $G$ is the orthogonal group, 
	$\mathbb{E}_{U\in O^\pm(l)} f(U)$ denotes the integral of $f$ over the coset of $O(l)$ with determinant $\pm 1$.}.
Matrix integrals over the classical groups $G(N)$ are connected to a wide range of fields, including combinatorics~\cite{BaikRains}, quantum field theory~\cite{GarciaTierzT+H}, number theory~\cite{KeatingSnaith}, and integrable systems~\cite{AdlerVanMoerbeke2001}.

In their seminal work~\cite{BaikRains} on increasing subsequences of permutations under certain symmetry constraints, Baik and Rains established precise connections between Toeplitz+Hankel determinants $D_n[\phi,\phi;0,s]$ and integrals over classical groups:
\begin{theorem}\label{thm:intstodets}\cite{BaikRains}
	Let $g(z)$ be any function on the unit circle such that the integrals
	\[
	\iota_j = \frac{1}{2\pi} \int_{[0,2\pi]} g(e^{i\theta})g(e^{-i\theta})
	\, e^{ij\theta} \, d\theta
	\]
	are well defined. Then
	{
		\allowdisplaybreaks
		\begin{align}
			\mathbb{E}_{U\in O^+(2l)} \det(g(U))
			&= \frac{1}{2}
			\det(\iota_{j-k}+\iota_{j+k})_{0\le j,k<l}, \\
			\mathbb{E}_{U\in O^-(2l)} \det(g(U))
			&= g(1)g(-1)
			\det(\iota_{j-k}-\iota_{j+k+2})_{0\le j,k<l-1}, \\
			\mathbb{E}_{U\in O^+(2l+1)} \det(g(U))
			&= g(1)
			\det(\iota_{j-k}-\iota_{j+k+1})_{0\le j,k<l}, \\
			\mathbb{E}_{U\in O^-(2l+1)} \det(g(U))
			&= g(-1)
			\det(\iota_{j-k}+\iota_{j+k+1})_{0\le j,k<l}, \\
			\mathbb{E}_{U\in Sp(2l)} \det(g(U))
			&= \det(\iota_{j-k}-\iota_{j+k+2})_{0\le j,k<l},
		\end{align}
	}
	except that $\mathbb{E}_{U\in O^+(0)} \det(g(U)) = 1$.
\end{theorem}

We would like to highlight that different choices of the offset pair $(0,s)$ correspond to averages over different classical groups. Through Theorems~1.2 and~2.5 of~\cite{BaikRains}, these integrals are related to the probability that the longest increasing subsequences of involutions with specific symmetry properties have length at most $2l$ or $2l+1$. 

Averages over classical groups, and thus Toeplitz+Hankel determinants, also arise in the description of the ground state density matrix $\rho_{N+1}(x,y)$ of the impenetrable Bose gas~\cite{LogGases, FF}. In particular, $\rho_{N+1}(x,y)$ can be represented as
\begin{itemize}
	\item   a $U(N)$ average (and thus a pure Toeplitz determinant) for periodic boundary conditions,
	\item  a $Sp(N)$ average for Dirichlet boundary conditions,
	\item an $O^+(2N)$ average for Neumann boundary conditions, and
	\item an $O^+(2N+1)$ average for mixed Dirichlet–Neumann boundary conditions.
\end{itemize}
Forrester and Frankel combined these characterizations with the Baik–Rains theorem~\cite{BaikRains} and the asymptotic results of Basor and Ehrhardt for Toeplitz+Hankel determinants with Fisher–Hartwig singularities~\cite{BE2} to obtain large-$N$ asymptotic formulas for $\rho_{N+1}(x,y)$ under Dirichlet, Neumann, and mixed Dirichlet–Neumann boundary conditions.

Toeplitz+Hankel determinants $D_n[\phi,\phi;0,s]$ for various values of the offset parameter $s$ appear in other contexts as well. For instance, the characterizations of Theorem~\ref{thm:intstodets} were applied in~\cite{ABKmoments} to compute \textit{moments of moments} of characteristic polynomials of orthogonal and symplectic groups, with connections to analytic number theory. In another application,~\cite{GarciaTierzT+H} evaluated the corresponding Toeplitz+Hankel determinants for symplectic and orthogonal matrix integrals, obtaining explicit expressions for the partition functions, Wilson loops, and Hopf links of Chern--Simons theory on $S^3$. Moreover, in~\cite[Theorem 7.1]{Stembridge}, certain $k \times k$ Toeplitz+Hankel determinants were shown to characterize generating functions for enumerating column‑strict tableaux with at most $2k$ and $2k+1$ rows, corresponding respectively to offset pairs $(0,-1)$ and $(0,0)$.

Finally, we note that for the large-size asymptotics of Toeplitz+Hankel determinants with identical symbols, in addition to the operator-theoretic approaches developed by E.~Basor and T.~Ehrhardt \cite{BE1, BE2, BE3}, a Riemann--Hilbert approach has also been developed and successfully applied in~\cite{DIK}. In particular, \cite{DIK} derives precise large-$n$ asymptotic formulae for $D_n[\phi,\phi;r,s]$ with a Fisher--Hartwig symbol~$\phi$ and offset pairs $(r,s) \in \{(0,0),(0,1),(0,2)\}$, using a $2\times 2$ Riemann--Hilbert framework.

		\subsection{Distinct Toeplitz and Hankel symbols}
			
			Besides the inherent and natural motivation to obtain the asymptotics of $D_n[\phi,w;r,s]$ in the absence of the condition $\phi=w$, such asymptotic results for Toeplitz+Hankel determinants with distinct symbols are further justified due to their appearance in important applications. Below, we highlight two such applcations that have significantly motivated this project. Regarding the asymptotics of $D_n[\phi,w;r,s]$ with $w \neq \phi$, there are two works \cite{BE} and \cite{GI}, respectively studying determinants \[D_n[\phi, d \phi; 0, 1 ], \qandq D_n[\phi, d \phi; 1, 1 ] \] 
			with Szeg{\H o}-type functions $\phi$ and $d$, where  the following extra condition is satisfied by $d$:
			\begin{equation}\label{condition on d}
				d(z) d(z^{-1}) \equiv 1, \qquad z \in \T.
			\end{equation}
			The application from Statistical Mechanics highlighted below in Section \ref{Sec Ising Zig Zag} involves symbols $\phi$ and $d$ which enjoy these properties, where the offset pairs of interest are $(r,s) \in \{(0,1),(0,2)\}$.

		\subsubsection{ Ising Model on the  Zig-Zag Layered Half-Plane}\label{Sec Ising Zig Zag}
			
	 In \cite{Chelkak}, Chelkak, Hongler, and Mahfouf considered the Ising model on the so-called \textit{zig-zag layered half-plane} $H_+^{\diamondsuit}$, which is the left half-plane on the $45^{\circ}$-rotated square grid (see Figure 3. of \cite{Chelkak}) with $+$ boundary conditions along the rightmost column and at infinity. The \textit{one-point function} 
			
			\begin{equation}
				M_n = \E^{+}_{H^{\diamondsuit}} [\sigma_{(-2n-\frac{1}{2},0)}] 
			\end{equation} 
			is the \textit{magnetization} in the $(2n)$-th column. Let the parameters $\theta$ and $q<1$ be defined by 
			
			\noindent\begin{minipage}{.5\linewidth}
				\begin{alignat*}{2}
					& \tan \frac{\theta}{2}  && = \exp \left[ -\frac{2J}{KT} \right],   
				\end{alignat*}	
			\end{minipage}
			\begin{minipage}{.5\linewidth}
				\begin{alignat*}{2}
					&q  &&= \tan \theta, 
				\end{alignat*}	
			\end{minipage}
			where $J$ is the nearest neighbor coupling constant, $K$ is the Boltzmann constant, and $T$ is the temperature.  Define the functions $\mathcal{r}$, $\phi$ and $d$ as 
			
			\noindent\begin{minipage}{.5\linewidth}
				\begin{alignat*}{2}
					& \mathcal{r}(\theta) &&:= 1 - \frac{\cos^2 \theta_1}{\cos^2 \theta} \in (-q^2,1),  
				\end{alignat*}	
			\end{minipage}
			\begin{minipage}{.5\linewidth}
				\begin{alignat*}{2}
					&\phi(z;q)  &&= |1-q^2 z|, 
				\end{alignat*}	
			\end{minipage}
			and
			\begin{equation}
				d(z;q) := -\frac{\left(\mathcal{r}z-q^2\right)\left(q^2z-1\right)}{\left(z-q^2\right)\left(q^2z-\mathcal{r}\right)},
			\end{equation}
			where in the last expression we think of $r$ being a function of the independent variable $q$, and $\theta_1$ is interpreted as the \textit{boundary magnetic field} \cite{Chelkak} and is considered as a fixed parameter. Recalling the condition \eqref{condition on d}, it can be checked that $d(z;q) d(z^{-1};q) \equiv 1$. Define the symbol $w$ to be 
			\begin{equation}
				w(z;q) := d(z;q) \phi(z;q),
			\end{equation}
			(see \cite{GI,BE}). If $\mathcal{r}\neq 0$, we define 
			\begin{equation}\label{aaa}
				\aaa := \frac{q^2}{\mathcal{r}}.
			\end{equation}
			The critical value of external field $h$ is specified by the condition that $\aaa=1$:
			\[ h=h_{\mbox{\footnotesize cr}}(q) \iff \aaa=1, \]
			see Remark 4.6 of \cite{Chelkak}. Let
			\begin{equation}
				\gamma(z;q) := \begin{cases}
					\di \frac{c(q)}{1-\aaa z}, & \aaa<1, \\[10pt]
					0, & \aaa \geq 1,
				\end{cases} 
			\end{equation}
			where
			\begin{equation}
				c(q) = (\mathcal{r}^2-q^4)\mathcal{r}^{-3/2}(\mathcal{r}-q^4)^{-1/2}.
			\end{equation}
			 Also, let us define
			\begin{equation}
				v(z;q) := w(z;q) + (1-\mathcal{r})^{3/2} \gamma(z;q). 
			\end{equation}
			Now we can recall the relevant result from \cite{Chelkak}.

			\begin{theorem}\cite{Chelkak} It holds that the magnetization in the $(2m)$-th column has the follwing Toeplitz+Hankel determinant representation
				\begin{equation}
					M_n = \left(1-\mathcal{r}\right)^{-3/2} \det \left[ \phi_{k-j} + w_{k+j} + (1-r)^{3/2} \gamma_{k+j} \right]^{n-1}_{k,j=0},
				\end{equation}
				where 
				\begin{equation}
					f_m = \int_{\T} f(z)z^{-m}\frac{\mathrm{d} z}{2 \pi \ic z}, \qquad f \in \{\phi,w,\gamma\},
				\end{equation}
				is the $m$-th Fourier coefficient of the symbol $f$.
				
			\end{theorem}

			Notice that when $\aaa \geq 1$ we have the simpler Toeplitz+Hankel determinant
			\begin{equation}
				M_n =  \left(1-\mathcal{r}\right)^{-3/2} \det \left[ \phi_{k-j} + w_{k+j} \right]^{n-1}_{k,j=0} = \left(1-r\right)^{-3/2}  D_n[\phi,w;0,0],
			\end{equation}
			while when $\aaa<1$ we have 
			\begin{equation}
				M_n =  \left(1-\mathcal{r}\right)^{-3/2} \det \left[ \phi_{k-j} + v_{k+j} \right]^{n-1}_{k,j=0}= \left(1-r\right)^{-3/2} D_n[\phi,v;0,0].
			\end{equation}
			For $z\in \T$, let us write $\phi$ as 
			\[ \phi(z;q) = (1-q^2z)^{1/2} \overline{(1-q^2z)^{1/2}} = (1-q^2z)^{1/2}(1-q^2z^{-1})^{1/2} = -\ic \sqrt{\frac{(z-q^{-2})(z-q^2)}{z}}, \]
			where, for $\al \in \{0,q^2,q^{-2}\}$, the branch cuts of $\sqrt{z-\al}$ are chosen to be $[\al,+\infty)$, and the branches are fixed by $0<\arg(z-\al)<2\pi$. Also, notice that since $0<q<1$, $\phi$ has no widing number.
			
			First, let us focus on the case $a\geq 1$ and the determinant $D_n[\phi,w;0,0]$. Consider two cases: $\mathcal{r}\neq 0$, and  $\mathcal{r}=0$.
			
			\begin{enumerate}
				\item  \textbf{The case $\mathcal{r} \neq 0$.} If $\mathcal{r}\neq 0$, that is if $\mathcal{r} \in (-q^2,0) \cup (0,1)$, we define $\aaa$ by \eqref{aaa} and $d(z)$ can be written as
				\begin{equation}
					d(z;q) = 
					-\frac{\left(z-\aaa \right)\left(q^2z-1\right)}{\left(z-q^2\right)\left(\aaa z-1\right)}.
				\end{equation}
				Let $q_{\mbox{\footnotesize cr}}$ be such that $\aaa(q_{\mbox{\footnotesize cr}})=1$.	In this case $d(z;q)$ reduces to
				\begin{equation}
					d(z;q_{\mbox{\footnotesize cr}}) = 
					-\frac{q_{\mbox{\footnotesize cr}}^2z-1}{z-q_{\mbox{\footnotesize cr}}^2}.
				\end{equation}
				and thus 
				\begin{equation}
					\textrm{wind} \ d(z;q)  = \begin{cases}
						-1, & \aaa=1, \\
						-2, & \aaa>1.
					\end{cases}
				\end{equation}
				\begin{itemize}
					\item[(1A)] For the case $\mathcal{r}\neq0$ and $\aaa>1$, consider the function \begin{equation}
						\mathcal{d}(z;q) := z^2  d(z;q)
					\end{equation}
					which has zero winding number and also satisfies $\mathcal{d}(z;q) \mathcal{d}(1/z;q) \equiv 1$. So the Toeplitz+Hankel determinant to study is
					
					\begin{equation}
						M_n = \det \left( T_n[\phi] + H_n[z^{-2} \mathcal{d} \phi] \right) = D_n(\phi,\mathcal{d}\phi;0,2),
					\end{equation}
					since $(z^{-2} \mathcal{d} \phi)_j = ( \mathcal{d} \phi)_{j+2}$. Therefore, when $\mathcal{r}\neq 0$ and  $\aaa>1$,  the resulting Toeplitz+Hankel determinant has symbol pair $(\phi, \mathcal{d}\phi)$, with offsets $r=0$ and $s=2$, where both $\phi$ and $\mathcal{d}$ are of Szeg{\H o}-type.
					\item[(1B)] For the case $\mathcal{r}\neq0$ and $\aaa=1$, consider  the function \begin{equation}
						\mathbb{d}(z) := z d(z;q_{\mbox{\footnotesize cr}}).
					\end{equation}
					which has zero winding number and also satisfies $\mathbb{d}(z) \mathbb{d}(1/z) \equiv 1$. So the Toeplitz+Hankel determinant to study is
					\begin{equation}
						M_n = \det \left( T_n[\phi] + H_n[z^{-1} \mathbb{d} \phi] \right)= D_n(\phi,\mathbb{d}\phi;0,1).
					\end{equation}
					So when $\mathcal{r}\neq 0$ and  $\aaa=1$, the associated Toeplitz+Hankel determinant has symbol pair $(\phi, \mathbb{d}\phi)$ with offsets $r=0$ and $s=1$, where both $\phi$ and $\mathbb{d}$ are of  Szeg{\H o}-type.
					
				\end{itemize}

				\item \textbf{The case $\mathcal{r} = 0$.} Let $q_0$ be such that $\mathcal{r}(q_0)=0$. In this case we have 
				\begin{equation}
					\dd(z;q_0) := \frac{q_0^2z-1}{z\left(z-q_0^2\right)},
				\end{equation}
				in which case, the winding number is still $-2$. Similar to the case (1A), consider the function \begin{equation}
					\mathfrak{d}(z) := z^2 \dd(z;q_0)
				\end{equation}
				which has zero winding number and also satisfies $\mathfrak{d}(z) \mathfrak{d}(1/z) \equiv 1$. So in this case one needs to study
				\begin{equation}
					M_n = \det \left( T_n[\phi] + H_n[z^{-2} \mathfrak{d} \phi] \right)= D_n(\phi,\mathfrak{d}\phi;0,2),
				\end{equation}
				which is a Toeplitz+Hankel determinant with symbol pair $(\phi, \mathfrak{d}\phi)$ and offsets $r=0$ and $s=2$, where both $\phi$ and $\mathfrak{d}$ are of Szeg{\H o}-type.
			\end{enumerate}

			\subsubsection{Spectral Analysis of (pure) Hankel Matrices and Offset Pairs $(0,s), s \in \N_0$} The asymptotics of eigenvalues of a class of pure Toeplitz determinants were studied in \cite{DIK1}, relying on the fact that the characteristic polynomial
			\[
			\det \!\left[\lambda I -T_n[\phi;r]\right]
			\]
			retains a Toeplitz structure. A natural problem, then, is to investigate the large-size asymptotic behavior of the characteristic polynomial
			\[
			 \det \!\left[\lambda I -H_n[w;s]\right]
			\]
			of the Hankel matrix \(H_n[w;s]\). Unlike in the Toeplitz case, this characteristic polynomial no longer inherits the structure of \(H_n\); rather, it is represented as a Toeplitz+Hankel determinant:
			\[
			 \det \!\left[\lambda I -H_n[w;s]\right] \equiv D_n(\lambda, w; 0, s),
			\]
			with the Toeplitz symbol being the \emph{constant} function \(\lambda\).
			
			In fact, this observation was the original motivation behind the Toeplitz+Hankel program initiated in \cite{GI}, continued in the present work, and to be developed further in upcoming papers. This line of inquiry led us to study Toeplitz+Hankel determinants with non-coinciding symbols \(\phi \neq w\). The main challenge in this direction is not related to offset pairs, but to the factorization of the model Riemann--Hilbert problem introduced in \cite{GI}, where the symbols do \emph{not} satisfy the relation \(w = d\phi\), unlike the Ising model application discussed earlier. Should such a factorization be achieved, the issue of offset pairs can then be successfully addressed using the methods developed in this paper.
			
			Two cases are particularly of interest:
			
			\begin{itemize}
				\item \textbf{Fourier case:} If the matrix \(H_n[w;s]\) is generated by Fourier coefficients of \(w \in L^1(\mathbb{T})\), then each offset pair satisfies \((0,s) \neq (1,1)\), calling for further analysis beyond what was carried out in \cite[Section 2]{GI}.
				\item \textbf{Moment case:} If the matrix \(H_n[w;s]\) is generated by the moments of \(w \in L^1(I)\) for some subset \(I \subset \mathbb{R}\), then each offset pair satisfies \((0,s) \neq (1,s)\), and hence requires further investigation compared to the treatment in \cite[Section 3]{GI}.
			\end{itemize}
			
			We will return to the first challenge outlined above in a forthcoming publication. In the present work, our focus is on developing ideas to address the various offset arrangements.

\subsection{Backround}\label{Sec Background} 
\begin{comment}
	content...
	Let $\phi$ be a function defined on the unit circle $\T$. For  $z\in \T$ write $\phi(z)=|\phi(z)|\exp{[2\pi i b(z)}]$ for some choice of $b$, then the increment of $b$ as the result of a counter-clockwise circuit around $\T$ is an integer solely dependent on $\phi$ which is normally referred to as the \textit{winding number} of the symbol $\phi$ and plays an important role in the analysis of the generated determinants as will be discusses in \S ??.
	
	Let $\phi$ be a symbol defined on the unit circle with winding number $r \in \Z$. We can write $\phi(z) \equiv z^r \psi(z)$, where $\psi$ has no winding number. The associated Toeplitz determinant is 
	\begin{equation}
		D_n[\phi]:= \underset{n\times n}{\det} \{\phi_{j-k}\} \equiv \det \begin{pmatrix}
		\psi_{-r}  & \psi_{-r-1} & \cdots & \psi_{-n+1-r} \cr
		\psi_{1-r} & \psi_{-r} & \cdots & \psi_{-n+2-r}  \cr
		\vdots & \ddots & \ddots  & \vdots   \cr
		\psi_{n-1-r} & \psi_{n-2-r} & \cdots & \psi_{-r}   \cr
		\end{pmatrix},
	\end{equation}
	where
	\begin{equation}\label{FC}
		\phi_j = \int_{\T} z^{-j}\phi(z) \frac{\dd z}{2 \pi \ic z}.
	\end{equation}
	  Like this case, when $\psi$ has no winding number, the integer $-r$ is also referred to as the \textit{offset} of the Toeplitz determinant $\underset{n\times n}{\det} \{\psi_{j-k-r}\}$, in other words, the winding number of the symbol $\phi$ is, up to a minus sign, the offset of the determinant generated by the associated symbol $\psi$ with zero winding number. 
\end{comment}
	Suppose that $\phi$ has zero winding number with Fourier coefficients $\phi_j$ given by \eqref{Toeplitz}, and $w_j$'s be either: 
	\begin{itemize}
		\item the Fourier coefficients of a symbol $w$ with zero winding number, given by \eqref{Hankel}, or
		\item  the moments \begin{equation}
				w_j = \int_{I} x^jw(x)\dd x,
		\end{equation}
	\end{itemize}
of a weight $w$ supported on some subset $I$ of the real line. Similar to the cases of pure Toeplitz and pure Hankel determinants,
in addition to operator-theoretic techniques 
(see \cite{BE, BE1, BE2, BE3} and references therein), 
an alternative approach to analyzing the large\mbox{-}$n$ asymptotics 
of $D_n[\phi,w;r,s]$---or of the associated system of (bi)orthogonal polynomials---
is provided by the Riemann--Hilbert formulation. In what follows we shall describe this formulation in details. 

It was observed in \cite{GI} that the determinant (\ref{Det}) is related to the system  $\{\mathcal{P}_n(z;r,s)\}_{n \in \Z_{\geq 0}}$, $\deg \mathcal{P}_n(z;r,s)=n$, of monic polynomials determined by the orthogonality relations\footnote{{\bf Notation.} Throughout the paper we will frequently use the notation $\tilde{f}(z)$. to denote $f(z^{-1})$.}
  \begin{equation}\label{T+H OP}
	\int_{\T} \mathcal{P}_n(z;r,s)z^{-k-r}\phi(z) \frac{\dd z}{2\pi \ic z} +	\int_{\T} \mathcal{P}_n(z;r,s)z^{k+s} \tilde{w}(z) \frac{\dd z}{2\pi \ic z}  = h^{(r,s)}_n \delta_{n,k},   \ \ \ \ k=0,1,\cdots,n. \end{equation}
These polynomials exist and are unique if the Toeplitz+Hankel determinants \eqref{Det} are non-zero. Indeed, if $D_n[\phi,w;r,s] \neq 0$, the polynomials $\mathcal{P}_{n}$ can be explicitly written as \begin{equation}\label{T+H OP Det rep}
	\mathcal{P}_n(z;r,s) := \frac{1}{D_n[\phi,w;r,s]} \det \begin{pmatrix}
		\phi_r+w_s & \phi_{r-1}+w_{s+1} & \cdots &  \phi_{r-n}+w_{s+n} \\
		\phi_{r+1}+w_{s+1} & \phi_{r}+w_{s+2} & \cdots  &  \phi_{r-n+1}+w_{s+n+1} \\
		\vdots & \vdots & \ddots & \vdots\\
		\phi_{r+n-1}+w_{s+n-1} & \phi_{r+n-2}+w_{s+n} & \cdots  &  \phi_{r-1}+w_{s+2n-1}  \\
		1 & z & \cdots  & z^n
	\end{pmatrix},
\end{equation}
while the uniqueness follows from the fact that the linear system which determines the vector of coefficients $ \boldsymbol{a}:= \left(a_0, \cdots, a_{n-1}\right)^T$ of the polynomials $	\mathcal{P}_n(z;r,s) = z^n + \sum_{k=0}^{n-1} a_k z^k$ is of the form $\left\{ \phi_{j-k+r} + w_{j+k+s} \right\}^{n-1}_{j,k=0} \boldsymbol{a} = \boldsymbol{b}$, and thus can be inverted if $D_n[\phi,w;r,s] \neq 0$.  Using \eqref{T+H OP} and \eqref{T+H OP Det rep} we find
	\begin{equation}\label{T+H h_n and Det}
	h^{(r,s)}_n=\frac{D_{n+1}(\phi,w;r,s)}{D_{n}(\phi,w;r,s)}.
\end{equation}
\begin{comment}
	content...

\red{Maybe Not necessary:}\blue{Let us here remind the Riemann-Hilbert problems for the pure-Toeplitz and Toeplitz+Hankel determinants. The Riemann-Hilbert problem for the pure Toeplitz determinants \cite{BDJ} with offset $r \in \Z$ is the problem of finding a $2\times 2$ matrix-valued function $Y$ satisfying
	
	\begin{itemize}
		\item  \textbf{RH-Y1} \qquad $Y(\cdot;n,r):\C\setminus \T \to \C^{2\times2}$ is analytic,
		\item \textbf{RH-Y2} \qquad  For $z\in  \T$, we have $Y_+(z;n,r)=Y_-(z;n,r)J_{Y}(z;r)$, where
		\begin{equation*}
		J_{Y}(z;r)=\begin{pmatrix}
		1 & z^{-n-r}\phi(z) \\
		0 & 1
		\end{pmatrix},
		\end{equation*}
		
		\item \textbf{RH-Y3} \qquad  As $z \to \infty$
		
		\begin{equation*}
		Y(z;n,r)=\big( I + O(z^{-1}) \big)\begin{pmatrix}
			z^n &  0\\
			0 & z^{-n}
		\end{pmatrix}
		\end{equation*}
	\end{itemize}}
\end{comment}
Define now the $2 \times 2$ matrix valued function, 
\begin{equation}\label{OP Rep of Solution}
	\mathcal{Y}(z;n,r,s)=\begin{pmatrix}
		\mathcal{P}_n(z;r,s) & \di  \int_{\T} \frac{\xi^{s}  \tilde{w}(\xi) 	 \mathcal{P}_n(\xi;r,s) + \xi^{r}\tilde{\phi}(\xi)\tilde{\mathcal{P}}_n(\xi;r,s)}{\xi-z}\frac{\dd\xi }{2\pi \ic \xi} \\
		-\di \frac{1}{h^{(r,s)}_{n-1}}\mathcal{P}_{n-1}(z;r,s) & -\di \frac{1}{h^{(r,s)}_{n-1}} \int_{\T} \frac{\xi^{s}  \tilde{w}(\xi) 	 \mathcal{P}_{n-1}(\xi;r,s) + \xi^{r}\tilde{\phi}(\xi)\tilde{\mathcal{P}}_{n-1}(\xi;r,s)}{\xi-z}\frac{\dd\xi }{2\pi \ic \xi}
	\end{pmatrix}.
\end{equation}
In \cite{GI}, it was shown that $\mathcal{Y}$ satisfies the  following Riemann-Hilbert type analytical problem:
\begin{itemize}
	\item \textbf{RH-$\mathcal{Y}$1} \quad  $\mathcal{Y}$ is holomorphic in $\C \setminus \T. $
	\item \textbf{RH-$\mathcal{Y}$2} \quad  For a given function $f$, and an oriented contour $\Ga$, we write $f_{+}(z)$ (resp. $f_-(z)$) to denote the limiting value of $f(\ze)$, as $\ze$ approaches $z \in \Ga$ from the left (resp. right) hand side of the oriented contour $\Ga$ with respect to its orientation. For $z \in \T$ we have  \begin{equation}\label{022}
		\mathcal{Y}_+^{(1)}(z;n,r,s)=\mathcal{Y}^{(1)}_-(z;n,r,s),
	\end{equation} 
	and \begin{equation}\label{0023}
		\mathcal{Y}_+^{(2)}(z;n,r,s)=\mathcal{Y}^{(2)}_-(z;n,r,s) +  z^{-1+s}  \tilde{w}(z) \mathcal{Y}^{(1)}_-(z;n,r,s) + z^{-1+r}\tilde{\phi}(z)\mathcal{Y}^{(1)}_-(z^{-1};n,r,s),
	\end{equation} 
	where $\T$ is positively oriented in the counter-clockwise direction. 
	\item \textbf{RH-$\mathcal{Y}$3} \quad As $z \to \infty$, $\mathcal{Y}$ satisfies  
	\begin{equation}\label{Yinfty}
		\mathcal{Y}(z;n,r,s)=\left( I + O(z^{-1}) \right) z^{n \sigma_3} = \begin{pmatrix}
			z^n+ O(z^{n-1}) &  O(z^{-n-1}) \\
			O(z^{n-1}) & z^{-n} +  O(z^{-n-1})
		\end{pmatrix},
	\end{equation}
	
\end{itemize}
where $\mathcal{Y}^{(1)}$ and $\mathcal{Y}^{(2)}$ are the first and second columns of $\mathcal{Y}$ respectively.  The relation of this problem
to the Toeplitz + Hankel determinants with offset $r,s \in \Z$ is given by the following theorem.

	\begin{theorem}\label{THM RHP formulation circle}\cite[Theorem 2.1]{GI}
	The following statements are true.
	\begin{enumerate}
		\item Suppose that $D_n[\phi,w;r,s],D_{n-1}[\phi,w;r,s] \neq 0$. Then, the Riemann-Hilbert problem \textbf{RH-$\mathcal{Y}$1} through \textbf{RH-$\mathcal{Y}$3} is uniquely solvable and its solution $\mathcal{Y}$ is  defined by \eqref{OP Rep of Solution}.
		Moreover,
		\begin{equation}\label{T+H h_n00}
			h^{(r,s)}_{n-1} = - \lim_{z \to \infty} z^{n-1}/\mathcal{Y}_{21}(z; n,r,s).    
		\end{equation}
		\item Suppose that  the Riemann-Hilbert problem \textbf{RH-$\mathcal{Y}$1} through \textbf{RH-$\mathcal{Y}$3}
		has a unique solution. Then $D_n[\phi,w;r,s] \neq 0 $, $\mbox{rank}\, (T_{n-1}[\phi;r] + H_{n-1}[w;s]) \geq n-2$, and $\mathcal{P}_n(z;r,s)
		= \mathcal{Y}_{11}(z;n,r,s)$.
		\item Suppose that the Riemann-Hilbert problem \textbf{RH-$\mathcal{Y}$1} through \textbf{RH-$\mathcal{Y}$3}
		has a unique solution. Suppose also that 
		\begin{equation*}\label{condition on y21}
			\lim_{z \to \infty}\mathcal{Y}_{21}(z;n,r,s)z^{-n+1}  \neq 0.
		\end{equation*}
		Then, as before, $D_n[\phi,w;r,s] \neq 0$,  $\mathcal{P}_n(z;r,s)
		= \mathcal{Y}_{11}(z;n,r,s)$, and, in addition,
		\begin{equation*}
			D_{n-1}[\phi,w;r,s] \neq 0, \quad h^{(r,s)}_{n-1} = -\lim_{z\to \infty} \mathcal{Y}^{-1}_{21}(z;n,r,s) z^{n-1},\quad
			\mathcal{P}_{n-1}(z;r,s) = -h^{(r,s)}_{n-1}\mathcal{Y}_{21}(z;n,r,s).
		\end{equation*}
	\end{enumerate}
\end{theorem}

	\begin{corollary}\label{Corollary 2.1.1}\cite[Corollary 2.2]{GI} Suppose that the $\mathcal{Y}$-RH problem has a unique solution for $n$ and $n-1$.
	Then 
	\begin{equation*}
		D_n[\phi,w;r,s] \neq 0, \quad D_{n-1}[\phi,w;r,s] \neq 0, \quad \mbox{and} \quad h^{(r,s)}_{n-1} \neq 0,
	\end{equation*}
	where $h^{(r,s)}_{n-1}$ can be reconstructed form the RHP data as
	\begin{equation}\label{T+H h_n}
		h^{(r,s)}_{n-1} = - \lim_{z \to \infty} z^{n-1}/\mathcal{Y}_{21}(z;n,r,s).  
	\end{equation}
\end{corollary}

The above ``Riemann-Hilbert  type'' problem can be transformed to a genuine Riemann-Hilbert problem, i.e. to the problem whose jump conditions can be written in the usual matrix-multiplication form. To this end, in \cite{GI}, the following \textit{enlargement} of the $\mathcal{Y}$-RHP was considered  
 \begin{equation}\label{X naught to Y}
	\widehat{X}(z;n,r,s) := \begin{pmatrix}
		\mathcal{Y}^{(1)}(z;n,r,s), \widetilde{\mathcal{Y}}^{(1)}(z;n,r,s), \mathcal{Y}^{(2)}(z;n,r,s), \widetilde{\mathcal{Y}}^{(2)}(z;n,r,s)
	\end{pmatrix}.
\end{equation}
From (\ref{022}), (\ref{0023}) and (\ref{Yinfty}) we obtain the following  $2\times4$ Riemann-Hilbert problem for $\widehat{X}$:

\begin{itemize}
	\item  \textbf{RH-}$\widehat{X}$\textbf{1} \ \ \  $\widehat{X}$ is holomorphic in $\C \setminus \left(\T \cup\{0\}\right)$.
	\item  \textbf{RH-}$\widehat{X}$\textbf{2} \quad For $z \in \T$, $\widehat{X}$ satisfies 
	\begin{equation}\label{JXo}
		\widehat{X}_+(z;n,r,s)=\widehat{X}_-(z;n,r,s) \begin{pmatrix}
			1 & 0 & z^{s-1}\tilde{w}(z) & -z^{-r+1}\phi(z) \\
			0 & 1 & z^{r-1}\tilde{\phi}(z) & -z^{-s+1}w(z) \\
			0 & 0 & 1 & 0 \\
			0 & 0 & 0 & 1
		\end{pmatrix}.
	\end{equation}
	\color{black}
	\item  \textbf{RH-}$\widehat{X}$\textbf{3} \quad As $z \to \infty$ we have  \begin{equation}\label{n07}
		\widehat{X}(z;n,r,s) = \begin{pmatrix}
			1 + O(z^{-1}) & C_1(n,r,s)+O(z^{-1}) & O(z^{-1}) & C_3(n,r,s) + O(z^{-1}) \\
			O(z^{-1}) & C_2(n,r,s)+O(z^{-1}) & 1 + O(z^{-1}) & C_4(n,r,s) + O(z^{-1})
		\end{pmatrix}\begin{pmatrix}
			z^n & 0 & 0 & 0\\
			0 & 1 & 0 & 0 \\
			0 & 0 & z^{-n} & 0 \\
			0 & 0 & 0 & 1
		\end{pmatrix}.
	\end{equation}
	
	\item  \textbf{RH-}$\widehat{X}$\textbf{4} \quad As $z \to 0$ we have  \begin{equation}\label{n088}
		\widehat{X}(z;n,r,s) = \begin{pmatrix}
			C_1(n,r,s) + O(z) & 1 + O(z) & C_3(n,r,s)+O(z) & O(z)  \\
			C_2(n,r,s) + O(z) & O(z) & C_4(n,r,s)+O(z) & 1 + O(z) 
		\end{pmatrix}\begin{pmatrix}
			1 & 0 & 0 & 0\\
			0 & z^{-n} & 0 & 0 \\
			0 & 0 & 1 & 0 \\
			0 & 0 & 0 & z^n
		\end{pmatrix},
	\end{equation}
\end{itemize}
where \begin{equation*}
	C_1(n,r,s)= \mathcal{Y}_{11}(0;n,r,s), \ \ \ C_3(n,r,s)=  \mathcal{Y}_{12}(0;n,r,s), \ \ \ C_2(n,r,s)= \mathcal{Y}_{21}(0;n,r,s), \ \ \ C_4(n,r,s)=  \mathcal{Y}_{22}(0;n,r,s).
\end{equation*} 

It is a natural next  step to associate with the above formulated $2 \times 4$ Riemann-Hilbert problem the following, canonically normalized
 at $z = \infty$,  $4 \times 4$ square Riemann-Hilbert problem:

\begin{itemize}
	\item \textbf{RH-X1} \quad  $X(\cdot;n,r,s): \C\setminus \left(\T \cup \{0\} \right) \to \C^{4\times4}$ is analytic.
	
	\item \textbf{RH-X2} \quad For $z\in  \T$, we have $X_+(z;n,r,s)=X_-(z;n,r,s)J_{X}(z;r,s)$, where \begin{equation}\label{hatW-jumpab}
	J_{X}(z;r,s)= \begin{pmatrix}
	1 & 0 & z^{s-1} \tilde{w}(z) & - z^{1-r} \phi(z) \\
	0 & 1 & z^{r-1} \tilde{\phi}(z) & - z^{1-s} w(z) \\
	0 & 0 & 1 & 0 \\
	0 & 0 & 0 & 1
	\end{pmatrix}, 
	\end{equation} 
	
	\item \textbf{RH-X3}  \quad As $z \to \infty$  \begin{equation*}\label{hatXinfinityab}
	X(z;n,r,s)=\left(\di I+ O(z^{-1})\right)\begin{pmatrix}
	z^n & 0 & 0 & 0\\
	0 & 1 & 0 & 0 \\
	0 & 0 & z^{-n} & 0 \\
	0 & 0 & 0 & 1
	\end{pmatrix},  
	\end{equation*} 
	
	\item \textbf{RH-X4} \quad As $z \to 0$ 
	\begin{equation*}\label{hatXzeroab}
	X(z;n,r,s)=P(n,r,s)\left(I+O(z)\right)\begin{pmatrix}
	1 & 0 & 0 & 0\\
	0 & z^{-n} & 0 & 0 \\
	0 & 0 & 1 & 0 \\
	0 & 0 & 0 & z^n
	\end{pmatrix}.
	\end{equation*}
\end{itemize}
Note that the matrix $P(n,r,s)$ is not a priori prescribed, and thus by the standard Liouville theorem-based arguments one can show that: 

\begin{lemma}\label{Lemma X unique}
	The solution of the Riemann-Hilbert problem \textbf{RH-X1} through \textbf{RH-X4} is unique, if it exists.
\end{lemma} 
\begin{remark}\normalfont It should be noted that the  points $z=\infty$ and $z=0$ are isolated singular points of $X(z;n,r,s)$;
therefore, the  above asymptotic series  at $z=\infty$ and $z=0$, are the convergent power series of the forms,
$$
\left(\di I+ O(z^{-1})\right) = \left( \di I+\frac{ \overset{\infty}{X}_1}{z}+\frac{\overset{\infty}{X}_2}{z^2} + O(z^{-3})\right),
$$and
$$
\left(\di I+ O(z)\right)=\left(I+\overset{\circ}{X}_1z+\overset{\circ}{X}_2z^2+O(z^3)\right).
$$
\end{remark}
The connection between $\widehat{X}$ and $X$ - Riemann Hilbert problems is given by the following  lemma
whose proof is given in Section 2.2 of  \cite{GI}.
\begin{lemma}\label{12}\
%cite[Lemma 2.7]{GI}
	The solution to the $\widehat{X}$-RHP can be reconstructed from the solution of the $X$-RHP using 
		\begin{equation}\label{24-to-44}
		 \widehat{X}(z;n,r,s)=\mathfrak{R}(z;n,r,s) X(z;n,r,s),
	\end{equation} 
where	\begin{equation}\label{mathfrakR}
		\mathfrak{R}(z;n,r,s) = \begin{pmatrix}
			1  & C_1(n,r,s) & 0 & C_3(n,r,s) \\
			0 & C_2(n,r,s) & 1 & C_4(n,r,s)
		\end{pmatrix}.
	\end{equation}
Moreover, the following linear system for solving $C_j(n,r,s)$ in terms of $P_{jk}(n,r,s)$
	\begin{equation}\label{C's-to-P}
		\begin{pmatrix}
			1  & C_1(n,r,s) & 0 & C_3(n,r,s) \\
			0 & C_2(n,r,s) & 1 & C_4(n,r,s)
		\end{pmatrix} = \begin{pmatrix}
			C_1(n,r,s) & 1 & C_3(n,r,s) & 0  \\
			C_2(n,r,s)  & 0 & C_4(n,r,s) & 1 
		\end{pmatrix} P^{-1}(n,r,s),
	\end{equation} is well defined and is uniquely solvable if at least one of the following inequalities is true:
	\begin{align}
	P_{22}(n,r,s)P_{44}(n,r,s) - P_{42}(n,r,s)P_{24}(n,r,s) & \neq 0,\label{Csolvcond01} \\
	(1-P_{21}(n,r,s))P_{42}(n,r,s) + P_{22}(n,r,s)P_{41}(n,r,s) &\neq 0,\label{gencond} \\
	(1-P_{43}(n,r,s))P_{22}(n,r,s) + P_{23}(n,r,s)P_{42}(n,r,s) &\neq 0,\label{gencond1}\\
	(1-P_{21}(n,r,s))P_{44}(n,r,s) +P_{41}(n,r,s)P_{24}(n,r,s) &\neq 0,\label{Csolvcond1}\\
	(1-P_{21}(n,r,s))(P_{43}(n,r,s) - 1) +P_{41}(n,r,s)P_{23}(n,r,s) &\neq 0,\label{Csolvcond2}\\
	(1-P_{43}(n,r,s))P_{24}(n,r,s) + P_{23}(n,r,s)P_{44}(n,r,s) &\neq 0.\label{gencond2}
\end{align}
\end{lemma}

	\begin{lemma}\label{Lemma Unique reconstruction of Y from X}\cite[Lemma 2.9]{GI}
	Suppose that the solution of the  $X$-RHP exists. Then, if at least one of the conditions \eqref{Csolvcond01} through \eqref{gencond2} holds, one can uniquely reconstruct the solution of the $\mathcal{Y}$-RHP.
\end{lemma}
\begin{remark} \normalfont
The reconstruction goes through equations (\ref{24-to-44}) and (\ref{X naught to Y}). The crucial issue is to 
prove the uniqueness of the solution of the   $\widehat{X}$ - RHP ($C_j$ are not a priory prescribed).
\end{remark}
\begin{corollary}\label{corollary2.8.1}\cite[Corollary 2.10]{GI}
	Suppose that the solution of the  $X$-RHP exists for $n$ and $n-1$, then if at least one of the conditions \eqref{Csolvcond01} through \eqref{gencond2} holds also for  $n$ and $n-1$, then we have
	\begin{equation*}
		D_n[\phi,w;r,s] \neq 0, \qquad D_{n-1}[\phi,w;r,s]  \neq 0, \qquad \mbox{and} \qquad h^{(r,s)}_{n-1}\neq0.
	\end{equation*}
	Moreover,
	\begin{equation}\label{T+H h_n000}
		h^{(r,s)}_{n-1} = - \lim_{z \to \infty} z^{n-1}/\mathcal{Y}_{21}(z;n,r,s).    
	\end{equation}	
	where in  the present context,
	$$
	\mathcal{Y}_{21}(z;n,r,s) = C_2(n,r,s)X_{21}(z;n,r,s) +  X_{31}(z;n,r,s)  +  	C_4(n,r,s)X_{41}(z;n,r,s).
	$$
\end{corollary}
	
In a similar way, as detailed in \cite{GI},	the analogue of the Riemann-Hilbert problem \textbf{RH-X1} through \textbf{RH-X4}  for the Toeplitz+Hankel determinants with offset $r,s \in \Z$, when $\phi_j$ is still given by \eqref{Toeplitz} but $w_j$'s are instead  of \eqref{Hankel} given by
 \begin{equation}
	w_j = \int_{a}^{b} x^jw(x)\dd x, \qquad 0<a<b<1,
\end{equation} is the problem of finding the $4 \times 4$ matrix-valued function $Z(z;n,r,s)$ satisfying 

\begin{itemize}
	\item \textbf{RH-Z1} \quad  $Z(\cdot;n,r,s): \C \setminus \left( \T \cup [a,b] \cup [b^{-1},a^{-1}] \cup \{0\} \right) \to \C^{4\times4}$ is analytic,
	
	\item \textbf{RH-Z2} \quad For $z\in \Sigma := \T \cup [a,b] \cup [b^{-1},a^{-1}]$, we have $Z_+(z;n,r,s)=Z_-(z;n,r,s)J_{Z}(z;r,s)$, where \begin{equation*}\label{2by4jumps Hankel on I}
	J_{Z}(z;r,s) = \begin{cases}
	\begin{pmatrix}
	1 & 0 & 0 & -z^{1-r}\phi(z) \\
	0 & 1 & z^{r-1}\tilde{\phi}(z) & 0 \\
	0 & 0 & 1 & 0 \\
	0 & 0 & 0 & 1
	\end{pmatrix}, & z \in \T, \\
	\begin{pmatrix}
	1 & 0 & 2\pi ix^sw(x) & 0 \\
	0 & 1 & 0 & 0 \\
	0 & 0 & 1 & 0 \\
	0 & 0 & 0 & 1
	\end{pmatrix}, & z\equiv x \in (a,b), \\
	\begin{pmatrix}
	1 & 0 & 0 & 0 \\
	0 & 1 & 0 & -2 \pi i x^{-s} \tilde{w}(x) \\
	0 & 0 & 1 & 0 \\
	0 & 0 & 0 & 1
	\end{pmatrix}, & z\equiv  x \in (b^{-1},a^{-1}). 
	\end{cases}
	\end{equation*} 
	
	\item \textbf{RH-Z3}  \quad As $z \to \infty$  \begin{equation*}\label{hatXinfinityabc}
	Z(z;n,r,s)=\left(\di I+O(z^{-1})\right)\begin{pmatrix}
	z^n & 0 & 0 & 0\\
	0 & 1 & 0 & 0 \\
	0 & 0 & z^{-n} & 0 \\
	0 & 0 & 0 & 1
	\end{pmatrix},  
	\end{equation*} 
	
	\item \textbf{RH-Z4} \quad As $z \to 0$ 
	\begin{equation*}\label{hatXzeroabc}
	Z(z;n,r,s)=Q(n,r,s)\left(I+O(z)\right)\begin{pmatrix}
	1 & 0 & 0 & 0\\
	0 & z^{-n} & 0 & 0 \\
	0 & 0 & 1 & 0 \\
	0 & 0 & 0 & z^n
	\end{pmatrix}.
	\end{equation*}
\end{itemize}
In this case the associated system of orthogonal polynomials are characterized by
	\begin{equation*}
	\int_{\T} Q_n(z;r,s)z^{-k-r}\phi(z) \frac{dz}{2\pi i z} + \int^{b}_{a} Q_n(x;r,s)x^{k+s} w(x) dx  = h_n \delta_{n,k},   \qquad  k=0,1,\cdots, n.
\end{equation*}

The analogues of Theorem \ref{THM RHP formulation circle}, Corollary \ref{Corollary 2.1.1},  Lemma \ref{Lemma Unique reconstruction of Y from X}, and Corollary \ref{corollary2.8.1}, for \textbf{RH-Z1} through \textbf{RH-Z4} are detailed in \cite{GI} in Theorem 3.1, Corollary 3.2, Lemma 3.5, Lemma 3.6, and Corollary 3.7. The analog of  Lemma \ref{12} is proven in Section 3.1.

Lemma \ref{Lemma Unique reconstruction of Y from X} describes the (asymptotic) solvability of \textbf{RH-X1} through \textbf{RH-X4} as the sufficient condition for the existence 
%Corollary \ref{corollary2.8.1}, 
and uniqueness of the orthogonal polynomials $\{\mathcal{P}_n(z;r,s)\}$. For particular choices of the offset values, the Riemann-Hilbert problems for $\textbf{X}$ and $\textbf{Z}$ are amenable to the Deift-Zhou non-linear steepest descent analysis. Such requirements on the offset values are important in the Deift-Zhou method, when one does the so-called \textit{lens-opening} transformation to construct the \textit{global parametrix}. To that end,  the Deift-Zhou nonlinear steepest descent analysis of \textbf{RH-X1} through \textbf{RH-X4} and that of \textbf{RH-Z1} through \textbf{RH-Z4} were worked out in \cite{GI} respectively when $r=s=1$ and $(r,s) \in \{1\}\times \Z$. 

	\section{Main Results}

Let us particularly denote the functions $X(z;n,1,1)$ and $Z(z;n,1,s)$, which are amenable for  the Deift-Zhou nonlinear steepest descent analysis \cite{GI} respectively by $\mathscr{X}(z;n)$ and $\mathscr{Z}(z;n,s)$, and thus, these functions respectively satisfy the following Riemann-Hilbert problems

\begin{itemize}
	\item \textbf{RH-$\mathscr{X}$1} \quad  $\mathscr{X}$ is holomorphic in the complement of $\T \cup \{0\}$.
	
	\item \textbf{RH-$\mathscr{X}$2} \quad   For $z \in \T$, $\mathscr{X}$ satisfies \begin{equation*}\label{X-jump}
	\mathscr{X}_+(z;n)=\mathscr{X}_-(z;n) \begin{pmatrix}
	1 & 0 & \tilde{w}(z) & -\phi(z) \\
	0 & 1 & \tilde{\phi}(z) & -w(z) \\
	0 & 0 & 1 & 0 \\
	0 & 0 & 0 & 1
	\end{pmatrix}, 
	\end{equation*} 
	
	\item \textbf{RH-$\mathscr{X}$3}  \quad As $z \to \infty$, we have \begin{equation*}\label{Xinfinity}
	\mathscr{X}(z;n)=\left(\di I+\frac{ \overset{\infty}{  \mathscr{X}}_1}{z}+\frac{\overset{\infty}{  \mathscr{X}}_2}{z^2} + O(z^{-3})\right)\begin{pmatrix}
	z^n & 0 & 0 & 0\\
	0 & 1 & 0 & 0 \\
	0 & 0 & z^{-n} & 0 \\
	0 & 0 & 0 & 1
	\end{pmatrix},  
	\end{equation*} 
	
	\item \textbf{RH-$\mathscr{X}$4} \quad As $z \to 0$, we have
	\begin{equation*}\label{Xzero}
	\mathscr{X}(z;n)=P(n)\left(I+\overset{\circ}{  \mathscr{X}}_1z+\overset{\circ}{  \mathscr{X}}_2z^2+O(z^3)\right)\begin{pmatrix}
	1 & 0 & 0 & 0\\
	0 & z^{-n} & 0 & 0 \\
	0 & 0 & 1 & 0 \\
	0 & 0 & 0 & z^n
	\end{pmatrix},
	\end{equation*}
\end{itemize}
and 
\begin{itemize}
	\item \textbf{RH-$\mathscr{Z}$1} \quad  $\mathscr{Z}(\cdot;n,s): \C \setminus \left( \T \cup [a,b] \cup [b^{-1},a^{-1}] \cup \{0\} \right) \to \C^{4\times4}$ is analytic,
	
	\item \textbf{RH-$\mathscr{Z}$2} \quad For $z\in \Sigma := \T \cup [a,b] \cup [b^{-1},a^{-1}]$, we have $\mathscr{Z}_+(z;n,s)=\mathscr{Z}_-(z;n,s)J_{\mathscr{Z}}(z,s)$, where \begin{equation*}\label{2by4jumps Hankel on I}
		J_{\mathscr{Z}}(z;s) = \begin{cases}
			\begin{pmatrix}
				1 & 0 & 0 & -\phi(z) \\
				0 & 1 & \tilde{\phi}(z) & 0 \\
				0 & 0 & 1 & 0 \\
				0 & 0 & 0 & 1
			\end{pmatrix}, & z \in \T, \\
			\begin{pmatrix}
				1 & 0 & 2\pi ix^sw(x) & 0 \\
				0 & 1 & 0 & 0 \\
				0 & 0 & 1 & 0 \\
				0 & 0 & 0 & 1
			\end{pmatrix}, & z\equiv x \in (a,b), \\
			\begin{pmatrix}
				1 & 0 & 0 & 0 \\
				0 & 1 & 0 & -2 \pi i x^{-s} \tilde{w}(x) \\
				0 & 0 & 1 & 0 \\
				0 & 0 & 0 & 1
			\end{pmatrix}, & z\equiv  x \in (b^{-1},a^{-1}). 
		\end{cases}
	\end{equation*} 
	
	\item \textbf{RH-$\mathscr{Z}$3}  \quad As $z \to \infty$  \begin{equation*}\label{hatXinfinityabc}
		\mathscr{Z}(z;n,s)=\left(\di I+O(z^{-1})\right)\begin{pmatrix}
			z^n & 0 & 0 & 0\\
			0 & 1 & 0 & 0 \\
			0 & 0 & z^{-n} & 0 \\
			0 & 0 & 0 & 1
		\end{pmatrix},  
	\end{equation*} 
	
	\item \textbf{RH-$\mathscr{Z}$4} \quad As $z \to 0$ 
	\begin{equation*}\label{hatXzeroabc}
		\mathscr{Z}(z;n,s)=Q(n,1,s)\left(I+O(z)\right)\begin{pmatrix}
			1 & 0 & 0 & 0\\
			0 & z^{-n} & 0 & 0 \\
			0 & 0 & 1 & 0 \\
			0 & 0 & 0 & z^n
		\end{pmatrix}.
	\end{equation*}
\end{itemize}

Our principal methodological idea which we propose  to use for the asymptotic analysis
of the solutions of $X$ and $Z$ - RHPs  with arbitrary $r,s$ is to algebraically connect these RHPs to $\mathscr{X}$ and 
$\mathscr{Z}$ - RHPs, respectively. In more details, in the case of the $X$-RHP, we suggest to consider the following transformations,
\begin{equation}\label{W}
	\mathscr{W}(z;n,r,s) := X(z;n,r,s)  \begin{pmatrix}
	1 & 0 & 0 & 0\\
	0 & z^{r-s} & 0 & 0 \\
	0 & 0 & z^{1-s} & 0 \\
	0 & 0 & 0 & z^{r-1}
	\end{pmatrix},
\end{equation}
and
\begin{equation}\label{V}
\mathscr{V}(z;n,r,s) := X(z;n,r,s)  \begin{pmatrix}
z^{s-r} & 0 & 0 & 0\\
0 & 1 & 0 & 0 \\
0 & 0 & z^{1-r} & 0 \\
0 & 0 & 0 & z^{s-1}
\end{pmatrix},
\end{equation}
which  coincide when $r=s$.  The key observation is that these functions, both, satisfy the same jump condition on the unit circle as $\mathscr{X}$, indeed
\begin{equation}
\begin{split}
	\mathscr{W}^{-1}_-(z;n,r,s) \mathscr{W}_+(z;n,r,s) & = \begin{pmatrix}
	1 & 0 & 0 & 0\\
	0 & z^{s-r} & 0 & 0 \\
	0 & 0 & z^{s-1} & 0 \\
	0 & 0 & 0 & z^{1-r}
	\end{pmatrix} \begin{pmatrix}
	1 & 0 & z^{s-1} \tilde{w}(z) & - z^{1-r} \phi(z) \\
	0 & 1 & z^{r-1} \tilde{\phi}(z) & - z^{1-s} w(z) \\
	0 & 0 & 1 & 0 \\
	0 & 0 & 0 & 1
	\end{pmatrix} \\ & \times \begin{pmatrix}
	1 & 0 & 0 & 0\\
	0 & z^{r-s} & 0 & 0 \\
	0 & 0 & z^{1-s} & 0 \\
	0 & 0 & 0 & z^{r-1}
	\end{pmatrix} = \begin{pmatrix}
	1 & 0 & \tilde{w}(z) & -\phi(z) \\
	0 & 1 & \tilde{\phi}(z) & -w(z) \\
	0 & 0 & 1 & 0 \\
	0 & 0 & 0 & 1
	\end{pmatrix},
\end{split}	
\end{equation}
and
\begin{equation}
\begin{split}
\mathscr{V}^{-1}_-(z;n,r,s) \mathscr{V}_+(z;n,r,s) & = \begin{pmatrix}
z^{r-s} & 0 & 0 & 0\\
0 & 1 & 0 & 0 \\
0 & 0 & z^{r-1} & 0 \\
0 & 0 & 0 & z^{1-s}
\end{pmatrix} \begin{pmatrix}
1 & 0 & z^{s-1} \tilde{w}(z) & - z^{1-r} \phi(z) \\
0 & 1 & z^{r-1} \tilde{\phi}(z) & - z^{1-s} w(z) \\
0 & 0 & 1 & 0 \\
0 & 0 & 0 & 1
\end{pmatrix} \\ & \times \begin{pmatrix}
z^{s-r} & 0 & 0 & 0\\
0 & 1 & 0 & 0 \\
0 & 0 & z^{1-r} & 0 \\
0 & 0 & 0 & z^{s-1}
\end{pmatrix}  = \begin{pmatrix}
1 & 0 & \tilde{w}(z) & -\phi(z) \\
0 & 1 & \tilde{\phi}(z) & -w(z) \\
0 & 0 & 1 & 0 \\
0 & 0 & 0 & 1
\end{pmatrix}. 
\end{split}	
\end{equation}
This observation means that the functions

\noindent\begin{minipage}{.5\linewidth}
\begin{equation}
	\mathscr{R}(z;n,r,s):=\mathscr{W}(z;n,r,s) \mathscr{X}^{-1}(z;n)
\end{equation}
\end{minipage}	
\begin{minipage}{.5\linewidth}
\begin{equation}
R(z;n,r,s):=\mathscr{V}(z;n,r,s) \mathscr{X}^{-1}(z;n)
\end{equation}	
\end{minipage}

\medskip

\noindent are both rational functions in $z$, with singular behavior only at zero and infinity. If one of the rational functions $\mathscr R$ or $R$ is completely determined using the data from the $\mathscr X$-RHP, then accordingly, either

\begin{equation}\label{X R X 1}
X(z;n,r,s)  =	\mathscr{R}(z;n,r,s) \mathscr{X}(z;n)  \begin{pmatrix}
1 & 0 & 0 & 0\\
0 & z^{s-r} & 0 & 0 \\
0 & 0 & z^{s-1} & 0 \\
0 & 0 & 0 & z^{1-r}
\end{pmatrix},
\end{equation}
or
\begin{equation}\label{X R X}
X(z;n,r,s) =	R(z;n,r,s) \mathscr{X}(z;n)  \begin{pmatrix}
z^{r-s} & 0 & 0 & 0\\
0 & 1 & 0 & 0 \\
0 & 0 & z^{r-1} & 0 \\
0 & 0 & 0 & z^{1-s}
\end{pmatrix}
\end{equation}	
directly relates the solution of the $X$-RHP to the solution of the $\mathscr X$-RHP which admits a successful nonlinear steepest descent analysis \cite{GI}, and thus paves the path for an asymptotic analysis of the $X$-RHP. 

Similarly, we could also consider
\begin{equation}\label{K}
	\mathscr{K}(z;n,r,s) := Z(z;n,r,s)  \begin{pmatrix}
		1 & 0 & 0 & 0\\
		0 & z^{r-1} & 0 & 0 \\
		0 & 0 & 1 & 0 \\
		0 & 0 & 0 & z^{r-1}
	\end{pmatrix},
\end{equation}
and
\begin{equation}\label{N}
	\mathscr{N}(z;n,r,s) := Z(z;n,r,s)   \begin{pmatrix}
		z^{1-r} & 0 & 0 & 0\\
		0 & 1 & 0 & 0 \\
		0 & 0 & z^{1-r} & 0 \\
		0 & 0 & 0 & 1
	\end{pmatrix}.
\end{equation}

It can be simply checked that both $\mathscr{K}$ and $\mathscr{N}$ satisfy the same jump conditions as $\mathscr{Z}$ does (see \textbf{RH-$\mathscr{Z}$2}), and therefore the functions

\noindent\begin{minipage}{.5\linewidth}
	\begin{equation}
		\mathscr{S}(z;n,r,s):=\mathscr{K}(z;n,r,s) \mathscr{Z}^{-1}(z;n,s)
	\end{equation}
\end{minipage}	
\begin{minipage}{.5\linewidth}
	\begin{equation}
		S(z;n,r,s):=\mathscr{N}(z;n,r,s) \mathscr{Z}^{-1}(z;n,s)
	\end{equation}	
\end{minipage}

\medskip

\noindent are both rational functions in $z$, with singular behavior only at zero and infinity. Provided that either $\mathscr{S}$ or $S$ are given explicitly by the data from the solution of the Riemann-Hilbert problem \textbf{RH-$\mathscr{Z}$1} through \textbf{RH-$\mathscr{Z}$4}, then one of 
\begin{equation}\label{Z S Z 1}
Z(z;n,r,s)  =	\mathscr{S}(z;n,r,s) \mathscr{Z}(z;n,s)  \begin{pmatrix}
1 & 0 & 0 & 0\\
0 & z^{1-r} & 0 & 0 \\
0 & 0 & 1 & 0 \\
0 & 0 & 0 & z^{1-r}
\end{pmatrix},
\end{equation}or
	\begin{equation}\label{Z S Z}
Z(z;n,r,s) =	S(z;n,r,s) \mathscr{Z}(z;n,s)  \begin{pmatrix}
z^{r-1} & 0 & 0 & 0\\
0 & 1 & 0 & 0 \\
0 & 0 & z^{r-1} & 0 \\
0 & 0 & 0 & 1
\end{pmatrix}
\end{equation}	
expresses the solution of the $Z$-RHP in terms of the solution of the $\mathscr Z$-RHP which admits a successful nonlinear steepest descent analysis \cite{GI}, and thus provides a pathway for asymptotic analysis of the $Z$-RHP. 

In this paper we show how to effectively use this idea by considering a few examples. In our exposition we focus on determining one of $\mathscr{R}$ or $R$ (depending on the particular choice of $(r,s) \in \Z^2$) explicitly in terms of the data from the solution of the $\mathscr{X}$-RHP. It will be evident that the determination of one of $\mathscr{S}$ or $S$ (depending on the choice of $r \in \Z$) can be achieved in an identical way.

Our first choice of the offsets will be $r = 0$ and $s = 1$. This choice is of special interest, 
both because it is used in the operator\mbox{-}theoretic approach of \cite{BE} 
and because it appears in the statistical\mbox{-}mechanics application 
discussed in Section~\ref{Sec Ising Zig Zag}.  We obtain a connection between the solution of the $X$-RHP with $r=0$ and $s=1$, which we denote by $\mathscr {U}(z;n)$, and the solution of the $\mathscr X$-RHP which is amenable for  the Deift-Zhou nonlinear steepest descent analysis. This will pave the way for the asymptotic analysis of $\mathscr{U}$ and eventually $h^{(0,1)}_n$, details of which are presented in Section \ref{Sec asym norms r=0 s=1}. 
By its definition, the matrix valued function $\mathscr {U}(z;n)$ is the solution of the following RHP

\begin{itemize}
	\item \textbf{RH-$\mathscr{U}$1} \quad  $\mathscr{U}$ is holomorphic in the complement of $\T \cup \{0\}$.
	
	\item \textbf{RH-$\mathscr{U}$2} \quad   For $z \in \T$, $\mathscr{U}$ satisfies \begin{equation*}\label{Z-jump}
	\mathscr{U}_+(z;n)=\mathscr{U}_-(z;n) \begin{pmatrix}
	1 & 0 &  \tilde{w}(z) & - z \phi(z) \\
	0 & 1 & z^{-1} \tilde{\phi}(z) & -  w(z) \\
	0 & 0 & 1 & 0 \\
	0 & 0 & 0 & 1
	\end{pmatrix}, 
	\end{equation*} 
	
	\item \textbf{RH-$\mathscr{U}$3}  \quad As $z \to \infty$ we have \begin{equation*}\label{Zinfinity}
	\mathscr{U}(z;n)=\left(\di I+\frac{ \overset{\infty}{  \mathscr{U}}_1}{z}+\frac{\overset{\infty}{  \mathscr{U}}_2}{z^2} + O(z^{-3})\right)\begin{pmatrix}
	z^n & 0 & 0 & 0\\
	0 & 1 & 0 & 0 \\
	0 & 0 & z^{-n} & 0 \\
	0 & 0 & 0 & 1
	\end{pmatrix},  
	\end{equation*} 
	
	\item \textbf{RH-$\mathscr{U}$4} \quad As $z \to 0$ we have
	\begin{equation*}\label{Zzero}
	\mathscr{U}(z;n)=\widehat{\mathscr{U}}\left(I+\overset{\circ}{  \mathscr{U}}_1z+\overset{\circ}{  \mathscr{U}}_2z^2+O(z^3)\right)\begin{pmatrix}
	1 & 0 & 0 & 0\\
	0 & z^{-n} & 0 & 0 \\
	0 & 0 & 1 & 0 \\
	0 & 0 & 0 & z^n
	\end{pmatrix}.
	\end{equation*}
\end{itemize}

In \S \ref{sec thm r=0 s=1} we prove the following Theorem which represents the solution of the $\mathscr{U}$-RHP in terms of the $\mathscr{X}$-RHP.

\begin{theorem}\label{thm r=0 s=1}
	The solution $\mathscr{U}$ to the Riemann-Hilbert problem  \textbf{RH-$\mathscr{U}$1} through \textbf{RH-$\mathscr{U}$4} can be expressed in terms of the data extracted from the solution $\mathscr{X}$ of the Riemann-Hilbert problem \textbf{RH-$\mathscr{X}$1} through \textbf{RH-$\mathscr{X}$4} as 
	
	\begin{equation}
	\mathscr{U}(z)= R(z)\mathscr{X}(z) \begin{pmatrix}
	z^{-1} & 0 & 0 & 0\\
	0 & 1 & 0 & 0 \\
	0 & 0 & z^{-1} & 0 \\
	0 & 0 & 0 & 1
	\end{pmatrix},
	\end{equation}
	where
	\begin{equation}
	R(z) =  \begin{pmatrix}
	\overset{\infty}{\mathscr{U}}_{1,11} - \overset{\infty}{\mathscr{X}}_{1,11} & -\overset{\infty}{\mathscr{X}}_{1,12} & 	\overset{\infty}{\mathscr{U}}_{1,13} - \overset{\infty}{\mathscr{X}}_{1,13} & - \overset{\infty}{\mathscr{X}}_{1,14}\\
	\overset{\infty}{\mathscr{U}}_{1,21} & 1 & \overset{\infty}{\mathscr{U}}_{1,23} & 0 \\
	\overset{\infty}{\mathscr{U}}_{1,31} - \overset{\infty}{\mathscr{X}}_{1,31} & -\overset{\infty}{\mathscr{X}}_{1,32} & 	\overset{\infty}{\mathscr{U}}_{1,33} - \overset{\infty}{\mathscr{X}}_{1,33} & - \overset{\infty}{\mathscr{X}}_{1,34}\\
	\overset{\infty}{\mathscr{U}}_{1,41} & 0 & \overset{\infty}{\mathscr{U}}_{1,43} & 1
	\end{pmatrix} + z \begin{pmatrix}
	1 & 0 & 0 & 0\\
	0 & 0 & 0  & 0 \\
	0 & 0 & 1  & 0 \\
	0 & 0 & 0  & 0
	\end{pmatrix},
	\end{equation}
	and  $\{ \overset{\infty}{\mathscr{U}}_{1,j1}, \overset{\infty}{\mathscr{U}}_{1,j3} \}^4_{j=1}$ are explicitly given in terms of the following data from the $\mathscr X$-RHP
	
	\noindent\begin{minipage}{.5\linewidth}
		\begin{alignat*}{2}
		&\overset{\infty}{\mathscr{U}}_{1,11} &&= \overset{\infty}{\mathscr{X}}_{1,11} + \frac{P_{33} \underset{\footnotesize{j \in\{2,4\}}}{\sum} \overset{\infty}{\mathscr{X}}_{1,1j} P_{j1} -P_{31}\underset{\footnotesize{j \in\{2,4\}}}{\sum} \overset{\infty}{\mathscr{X}}_{1,1j} P_{j3}}{P_{11}P_{33}  -P_{13}P_{31}},  \\
		& \overset{\infty}{\mathscr{U}}_{1,13}&&= \overset{\infty}{\mathscr{X}}_{1,13}+ \frac{P_{11}\underset{\footnotesize{j \in\{2,4\}}}{\sum} \overset{\infty}{\mathscr{X}}_{1,1j} P_{j3}-P_{13}\underset{\footnotesize{j \in\{2,4\}}}{\sum} \overset{\infty}{\mathscr{X}}_{1,1j} P_{j1}}{P_{11}P_{33}  -P_{13}P_{31}}, 
		\end{alignat*}	
	\end{minipage}	
	\begin{minipage}{.5\linewidth}
		\begin{alignat*}{2}
		&\overset{\infty}{\mathscr{U}}_{1,21} &&= \frac{P_{31}P_{23}  -P_{33}P_{21}}{P_{11}P_{33}  -P_{13}P_{31}}, \\
		& \overset{\infty}{\mathscr{U}}_{1,23} &&= \frac{P_{13}P_{21}  -P_{11}P_{23}}{P_{11}P_{33}  -P_{13}P_{31}}, 
		\end{alignat*}	
	\end{minipage}
	
	\noindent\begin{minipage}{.5\linewidth}
		\begin{alignat*}{2}
		&\overset{\infty}{\mathscr{U}}_{1,31} &&= \overset{\infty}{\mathscr{X}}_{1,31} + \frac{P_{33}\underset{\footnotesize{j \in\{2,4\}}}{\sum} \overset{\infty}{\mathscr{X}}_{1,3j} P_{j1} -P_{31}\underset{\footnotesize{j \in\{2,4\}}}{\sum} \overset{\infty}{\mathscr{X}}_{1,3j} P_{j3}}{P_{11}P_{33}  -P_{13}P_{31}},  \\
		& \overset{\infty}{\mathscr{U}}_{1,33} &&= \overset{\infty}{\mathscr{X}}_{1,33} + \frac{P_{11}\underset{\footnotesize{j \in\{2,4\}}}{\sum} \overset{\infty}{\mathscr{X}}_{1,3j} P_{j3}-P_{13}\underset{\footnotesize{j \in\{2,4\}}}{\sum} \overset{\infty}{\mathscr{X}}_{1,3j} P_{j1}}{P_{11}P_{33}  -P_{13}P_{31}}, 
		\end{alignat*}	
	\end{minipage}	
	\begin{minipage}{.5\linewidth}
		\begin{alignat*}{2}
		&\overset{\infty}{\mathscr{U}}_{1,41} &&= \frac{P_{43}P_{31}  -P_{33}P_{41}}{P_{11}P_{33}  -P_{13}P_{31}},  \\
		& \overset{\infty}{\mathscr{U}}_{1,43} &&= \frac{P_{13}P_{41}  -P_{11}P_{43}}{P_{11}P_{33}  -P_{13}P_{31}},
		\end{alignat*}	
	\end{minipage}\\

\noindent where we have assumed a generic condition,
$$
P_{11}P_{33}  -P_{13}P_{31}\neq 0,
$$
and in all objects we have suppressed the dependence on $n$.
\end{theorem}

Our next example is the $X$-RHP with  $r=s=0$, and  whose solution we shall denote by $\mathscr {Y}(z;n)$. By its definition, 
 $\mathscr Y$ satisfies the following RHP,

\begin{itemize}
	\item \textbf{RH-$\mathscr{Y}$1} \quad  $\mathscr{Y}$ is holomorphic in the complement of $\T \cup \{0\}$.
	
	\item \textbf{RH-$\mathscr{Y}$2} \quad   For $z \in \T$, $\mathscr{Y}$ satisfies \begin{equation*}\label{Y-jump}
	\mathscr{Y}_+(z;n)=\mathscr{Y}_-(z;n) \begin{pmatrix}
	1 & 0 & z^{-1} \tilde{w}(z) & - z \phi(z) \\
	0 & 1 & z^{-1} \tilde{\phi}(z) & - z w(z) \\
	0 & 0 & 1 & 0 \\
	0 & 0 & 0 & 1
	\end{pmatrix}, 
	\end{equation*} 
	
	\item \textbf{RH-$\mathscr{Y}$3}  \quad As $z \to \infty$ we have \begin{equation*}\label{Yinfinity}
	\mathscr{Y}(z;n)=\left(\di I+\frac{ \overset{\infty}{  \mathscr{Y}}_1}{z}+\frac{\overset{\infty}{  \mathscr{Y}}_2}{z^2} + O(z^{-3})\right)\begin{pmatrix}
	z^n & 0 & 0 & 0\\
	0 & 1 & 0 & 0 \\
	0 & 0 & z^{-n} & 0 \\
	0 & 0 & 0 & 1
	\end{pmatrix},  
	\end{equation*} 
	
	\item \textbf{RH-$\mathscr{Y}$4} \quad As $z \to 0$ we have
	\begin{equation*}\label{Yzero}
	\mathscr{Y}(z;n)=\widehat{\mathscr{Y}}\left(I+\overset{\circ}{  \mathscr{Y}}_1z+\overset{\circ}{  \mathscr{Y}}_2z^2+O(z^3)\right)\begin{pmatrix}
	1 & 0 & 0 & 0\\
	0 & z^{-n} & 0 & 0 \\
	0 & 0 & 1 & 0 \\
	0 & 0 & 0 & z^n
	\end{pmatrix}.
	\end{equation*}
\end{itemize}

In \S \ref{sec thm r=s=0} we prove:
\begin{theorem}\label{thm r=s=0}
	The solution $\mathscr{Y}$ to the Riemann-Hilbert problem  \textbf{RH-$\mathscr{Y}$1} through \textbf{RH-$\mathscr{Y}$4} can be expressed in terms of the data extracted from the solution $\mathscr{X}$ of the Riemann-Hilbert problem \textbf{RH-$\mathscr{X}$1} through \textbf{RH-$\mathscr{X}$4} as 
	
	\begin{equation}
	\mathscr{Y}(z)= \mathscr{R}(z)\mathscr{X}(z) \begin{pmatrix}
	1 & 0 & 0 & 0\\
	0 & 1 & 0 & 0 \\
	0 & 0 & z^{-1} & 0 \\
	0 & 0 & 0 & z
	\end{pmatrix},
	\end{equation}
	where
	\begin{equation} \begin{split}
	\mathscr{R}(z) & =  \frac{1}{z} \begin{pmatrix}
	\widehat{\mathscr{Y}}_{14}  P_{32} & \widehat{\mathscr{Y}}_{14}  P_{31} & \widehat{\mathscr{Y}}_{14}  P_{34} & \widehat{\mathscr{Y}}_{14}  P_{33} \\
	\widehat{\mathscr{Y}}_{24}  P_{32} & \widehat{\mathscr{Y}}_{24}  P_{31} & \widehat{\mathscr{Y}}_{24}  P_{34} & \widehat{\mathscr{Y}}_{24}  P_{33} \\
	\widehat{\mathscr{Y}}_{34}  P_{32} & \widehat{\mathscr{Y}}_{34}  P_{31} & \widehat{\mathscr{Y}}_{34}  P_{34} & \widehat{\mathscr{Y}}_{34}  P_{33} \\
	\widehat{\mathscr{Y}}_{44}  P_{32} & \widehat{\mathscr{Y}}_{44}  P_{31} & \widehat{\mathscr{Y}}_{44}  P_{34} & \widehat{\mathscr{Y}}_{44}  P_{33} \\
	\end{pmatrix} + \begin{pmatrix}
	1 & 0 & \overset{\infty}{\mathscr{Y}}_{1,13} & 0\\
	0 & 1 & \overset{\infty}{\mathscr{Y}}_{1,23} & 0 \\
	-\overset{\infty}{\mathscr{X}}_{1,31} & -\overset{\infty}{\mathscr{X}}_{1,32} &  \overset{\infty}{\mathscr{Y}}_{1,33} - \overset{\infty}{\mathscr{X}}_{1,33} & - \overset{\infty}{\mathscr{X}}_{1,34} \\
	0 & 0 & \overset{\infty}{\mathscr{Y}}_{1,43} & 0
	\end{pmatrix} \\ & + z \begin{pmatrix}
	0 & 0 & 0 & 0\\
	0 & 0 & 0  & 0 \\
	0 & 0 & 1  & 0 \\
	0 & 0 & 0  & 0
	\end{pmatrix}, 
\end{split}
	\end{equation}
	and $\left\{ \widehat{\mathscr{Y}}_{j4} , \overset{\infty}{\mathscr{Y}}_{1,j3} \right\}^4_{j=1}$ are explicitly given in terms of the following data from the $\mathscr X$-RHP
	
	\noindent\begin{minipage}{.5\linewidth}
		\begin{alignat*}{2}
		&\widehat{\mathscr{Y}}_{14} &&= \frac{P_{13}\overset{\infty}{\mathscr{X}}_{1,34}-P_{33}\overset{\infty}{\mathscr{X}}_{1,14}}{P^2_{33}  -\overset{\infty}{\mathscr{X}}_{1,34}\overset{\circ}{\mathscr{X}}_{1,43}},  \\
			&\widehat{\mathscr{Y}}_{24} &&= \frac{P_{23}\overset{\infty}{\mathscr{X}}_{1,34}-P_{33}\overset{\infty}{\mathscr{X}}_{1,24}}{P^2_{33}  -\overset{\infty}{\mathscr{X}}_{1,34}\overset{\circ}{\mathscr{X}}_{1,43}}, 
		\end{alignat*}	
	\end{minipage}	
	\begin{minipage}{.5\linewidth}
		\begin{alignat*}{2}
		&\overset{\infty}{\mathscr{Y}}_{1,13} &&= \frac{\overset{\circ}{\mathscr{X}}_{1,43} \overset{\infty}{\mathscr{X}}_{1,14}-P_{33}P_{13}}{P^2_{33}  -\overset{\infty}{\mathscr{X}}_{1,34}\overset{\circ}{\mathscr{X}}_{1,43}},  \\
		& \overset{\infty}{\mathscr{Y}}_{1,23} &&= \frac{\overset{\circ}{\mathscr{X}}_{1,43} \overset{\infty}{\mathscr{X}}_{1,24}-P_{33}P_{23}}{P^2_{33}  -\overset{\infty}{\mathscr{X}}_{1,34}\overset{\circ}{\mathscr{X}}_{1,43}}, 
		\end{alignat*}	
	\end{minipage}
	
	\noindent\begin{minipage}{.5\linewidth}
		\begin{alignat*}{2}
		&\widehat{\mathscr{Y}}_{44} &&= \frac{P_{33}}{P^2_{33}  -\overset{\infty}{\mathscr{X}}_{1,34}\overset{\circ}{\mathscr{X}}_{1,43}},  \\
		& \widehat{\mathscr{Y}}_{34} &&= \frac{\De}{P^2_{33}  -\overset{\infty}{\mathscr{X}}_{1,34}\overset{\circ}{\mathscr{X}}_{1,43}}, 
		\end{alignat*}	
	\end{minipage}	
	\begin{minipage}{.5\linewidth}
		\begin{alignat*}{2}
		&\overset{\infty}{\mathscr{Y}}_{1,43} &&= \frac{-\overset{\circ}{\mathscr{X}}_{1,43}}{P^2_{33}  -\overset{\infty}{\mathscr{X}}_{1,34}\overset{\circ}{\mathscr{X}}_{1,43}},  \\
		& \overset{\infty}{\mathscr{Y}}_{1,33} &&= \frac{\La}{P^2_{33}  -\overset{\infty}{\mathscr{X}}_{1,34}\overset{\circ}{\mathscr{X}}_{1,43}},  
		\end{alignat*}	
	\end{minipage}
	with 
	\begin{equation*}
	\De = P_{33} \left( -\overset{\infty}{\mathscr{X}}_{2,34} + \overset{\infty}{\mathscr{X}}_{1,31} \overset{\infty}{\mathscr{X}}_{1,14} + \overset{\infty}{\mathscr{X}}_{1,32} \overset{\infty}{\mathscr{X}}_{1,24}  +  \overset{\infty}{\mathscr{X}}_{1,34}\overset{\infty}{\mathscr{X}}_{1,44} \right)   - \overset{\infty}{\mathscr{X}}_{1,34} \left( \overset{\infty}{\mathscr{X}}_{1,31}P_{13} + \overset{\infty}{\mathscr{X}}_{1,32}P_{23} + \overset{\infty}{\mathscr{X}}_{1,34}P_{43} \right),
	\end{equation*}
	and
	\begin{equation*}\label{La}
	\begin{split}
	\La & = -\overset{\circ}{  \mathscr{X}}_{1,43} \left( - \overset{\infty}{\mathscr{X}}_{2,34} + \overset{\infty}{\mathscr{X}}_{1,31} \overset{\infty}{\mathscr{X}}_{1,14} + \overset{\infty}{\mathscr{X}}_{1,32} \overset{\infty}{\mathscr{X}}_{1,24}  + \overset{\infty}{\mathscr{X}}_{1,34}\overset{\infty}{\mathscr{X}}_{1,33} + \overset{\infty}{\mathscr{X}}_{1,34}\overset{\infty}{\mathscr{X}}_{1,44}  \right) \\ & +P_{33} \left( \overset{\infty}{\mathscr{X}}_{1,31}P_{13} +\overset{\infty}{\mathscr{X}}_{1,32}P_{23}+\overset{\infty}{\mathscr{X}}_{1,33}P_{33}+\overset{\infty}{\mathscr{X}}_{1,34}P_{43}  \right).
	\end{split}
	\end{equation*}
\noindent Here,  we have assumed a generic condition,
$$
P^2_{33}  -\overset{\infty}{\mathscr{X}}_{1,34}\overset{\circ}{\mathscr{X}}_{1,43} \neq 0
$$
and in all objects the dependence on $n$ is suppressed. 
\end{theorem}

Our third choice will be $r=0$ and $s=2$ . As we have already explained  in the introduction, these offsets  appear in the theory of Ising model on the zig-zag layered half-plane. Denoting   $ X(z;n,0,2)$ as $\mathscr{T}(z;n)$ we shall arrive this time to the following  RHP.
%We elaborate this interesting appearance further in \S \ref{Sec Chelkak}. The function $\mathscr{T}(z;n)  := X(z;n,0,2)$ satisfies
\begin{itemize}
	\item \textbf{RH-$\mathscr{T}$1} \quad  $\mathscr{T}$ is holomorphic in the complement of $\T \cup \{0\}$.
	
	\item \textbf{RH-$\mathscr{T}$2} \quad   For $z \in \T$, $\mathscr{T}$ satisfies \begin{equation*}\label{hatW-jumpa}
		\mathscr{T}_+(z;n)=\mathscr{T}_-(z;n) \begin{pmatrix}
			1 & 0 & z \tilde{w}(z) & - z \phi(z) \\
			0 & 1 & z^{-1} \tilde{\phi}(z) & - z^{-1} w(z) \\
			0 & 0 & 1 & 0 \\
			0 & 0 & 0 & 1
		\end{pmatrix}, 
	\end{equation*} 
	
	\item \textbf{RH-$\mathscr{T}$3}  \quad As $z \to \infty$ we have \begin{equation*}\label{hatXinfinityaa}
		\mathscr{T}(z;n)=\left(\di I+\frac{\overset{\infty}{\mathscr{T}}_{1}}{z} +\frac{\overset{\infty}{\mathscr{T}}_{2}}{z^2} + O(z^{-3})\right)\begin{pmatrix}
			z^n & 0 & 0 & 0\\
			0 & 1 & 0 & 0 \\
			0 & 0 & z^{-n} & 0 \\
			0 & 0 & 0 & 1
		\end{pmatrix},  
	\end{equation*} 
	
	\item \textbf{RH-$\mathscr{T}$4} \quad As $z \to 0$ we have
	\begin{equation*}\label{hatXzeroaa}
		\mathscr{T}(z;n)=\widehat{\mathscr{T}}\left(I+\overset{\circ}{\mathscr{T}}_{1} z +\overset{\circ}{\mathscr{T}}_{2} z^2 + O(z^{3})\right)\begin{pmatrix}
			1 & 0 & 0 & 0\\
			0 & z^{-n} & 0 & 0 \\
			0 & 0 & 1 & 0 \\
			0 & 0 & 0 & z^n
		\end{pmatrix}.
	\end{equation*}
\end{itemize}

In \S\ref{sec thm r=0 s=2} we prove:
\begin{theorem}\label{thm r=0 s=2}
	The solution $\mathscr{T}$ to the Riemann-Hilbert problem  \textbf{RH-$\mathscr{T}$1} through \textbf{RH-$\mathscr{T}$4} can be expressed in terms of the data extracted from the solution $\mathscr{X}$ of the Riemann-Hilbert problem \textbf{RH-$\mathscr{X}$1} through \textbf{RH-$\mathscr{X}$4} as
	\begin{equation}\label{T R X}
		\mathscr{T}(z;n)= R(z)\mathscr{X}(z;n) \begin{pmatrix}
			z^{-2} & 0 & 0 & 0\\
			0 & 1 & 0 & 0 \\
			0 & 0 & z^{-1} & 0 \\
			0 & 0 & 0 & z^{-1}
		\end{pmatrix},
	\end{equation} with
\begin{equation}
	R(z)=z^2 \begin{pmatrix}
	1 & 0 & 0 & 0\\
	0 & 0 & 0 & 0 \\
	0 & 0 & 0 & 0 \\
	0 & 0 & 0 & 0
	\end{pmatrix} + z E + B, \label{RT}
\end{equation}
where
\begin{equation}
	E = \begin{pmatrix}
	\overset{\infty}{\mathscr{T}}_{1,11}- \overset{\infty}{\mathscr{X}}_{1,11} & -\overset{\infty}{\mathscr{X}}_{1,12} & -\overset{\infty}{\mathscr{X}}_{1,13} & -\overset{\infty}{\mathscr{X}}_{1,14}\\
	\overset{\infty}{\mathscr{T}}_{1,21} & 0 & 0 & 0 \\
	\overset{\infty}{\mathscr{T}}_{1,31} & 0 & 1 & 0 \\
	\overset{\infty}{\mathscr{T}}_{1,41} & 0 & 0 & 1
	\end{pmatrix},
\end{equation}
and
\begin{equation} \begin{split}
B & =  \begin{pmatrix}
\overset{\infty}{\mathscr{T}}_{2,11} + \left[ \overset{\infty}{\mathscr{X}}_1^2 \right]_{11} - \overset{\infty}{  \mathscr{X}}_{2,11} &  \left[ \overset{\infty}{\mathscr{X}}_1^2 \right]_{12} - \overset{\infty}{  \mathscr{X}}_{2,12} &  \overset{\infty}{\mathscr{T}}_{1,13} + \left[ \overset{\infty}{\mathscr{X}}_1^2 \right]_{13} - \overset{\infty}{  \mathscr{X}}_{2,13} & \overset{\infty}{\mathscr{T}}_{1,14} + \left[ \overset{\infty}{\mathscr{X}}_1^2 \right]_{14} - \overset{\infty}{  \mathscr{X}}_{2,14} \\
\overset{\infty}{\mathscr{T}}_{2,21} & 1 & \overset{\infty}{\mathscr{T}}_{1,23} & \overset{\infty}{\mathscr{T}}_{1,24} \\
\overset{\infty}{\mathscr{T}}_{2,31}- \overset{\infty}{\mathscr{X}}_{1,31} & - \overset{\infty}{\mathscr{X}}_{1,32} & \overset{\infty}{\mathscr{T}}_{1,33}- \overset{\infty}{\mathscr{X}}_{1,33} & \overset{\infty}{\mathscr{T}}_{1,34} - \overset{\infty}{\mathscr{X}}_{1,34} \\
\overset{\infty}{\mathscr{T}}_{2,41}- \overset{\infty}{\mathscr{X}}_{1,41} & - \overset{\infty}{\mathscr{X}}_{1,42} & \overset{\infty}{\mathscr{T}}_{1,43} - \overset{\infty}{\mathscr{X}}_{1,43} & \overset{\infty}{\mathscr{T}}_{1,44} - \overset{\infty}{\mathscr{X}}_{1,44}
\end{pmatrix}  \\ & 	- \begin{pmatrix}
\overset{\infty}{\mathscr{T}}_{1,11}\overset{\infty}{\mathscr{X}}_{1,11} &  \overset{\infty}{\mathscr{T}}_{1,11}\overset{\infty}{\mathscr{X}}_{1,12} &  \overset{\infty}{\mathscr{T}}_{1,11}\overset{\infty}{\mathscr{X}}_{1,13} &  \overset{\infty}{\mathscr{T}}_{1,11}\overset{\infty}{\mathscr{X}}_{1,14}\\
\overset{\infty}{\mathscr{T}}_{1,21}\overset{\infty}{\mathscr{X}}_{1,11} &  \overset{\infty}{\mathscr{T}}_{1,21}\overset{\infty}{\mathscr{X}}_{1,12} &  \overset{\infty}{\mathscr{T}}_{1,21}\overset{\infty}{\mathscr{X}}_{1,13} &  \overset{\infty}{\mathscr{T}}_{1,21}\overset{\infty}{\mathscr{X}}_{1,14}\\
\overset{\infty}{\mathscr{T}}_{1,31}\overset{\infty}{\mathscr{X}}_{1,11} &  \overset{\infty}{\mathscr{T}}_{1,31}\overset{\infty}{\mathscr{X}}_{1,12} &  \overset{\infty}{\mathscr{T}}_{1,31}\overset{\infty}{\mathscr{X}}_{1,13} &  \overset{\infty}{\mathscr{T}}_{1,31}\overset{\infty}{\mathscr{X}}_{1,14}\\
\overset{\infty}{\mathscr{T}}_{1,41}\overset{\infty}{\mathscr{X}}_{1,11} &  \overset{\infty}{\mathscr{T}}_{1,41}\overset{\infty}{\mathscr{X}}_{1,12} &  \overset{\infty}{\mathscr{T}}_{1,41}\overset{\infty}{\mathscr{X}}_{1,13} &  \overset{\infty}{\mathscr{T}}_{1,41}\overset{\infty}{\mathscr{X}}_{1,14}
\end{pmatrix} . 
\end{split}
\end{equation}
In the above formula, the sixteen unknowns $\left\{ \overset{\infty}{\mathscr{T}}_{2,j1}, \overset{\infty}{\mathscr{T}}_{1,j1}, \overset{\infty}{\mathscr{T}}_{1,j3}, \overset{\infty}{\mathscr{T}}_{1,j4} \right\}^4_{j=1}$ are explicitly given in terms of the $\mathscr{X}$-RHP data as described below. Consider
\begin{equation}
\mathscr A := P\overset{\circ}{\mathscr{X}}_1, \qquad \mathscr B := \overset{\infty}{\mathscr{X}}_1 P \qquad \mathscr C := \overset{\infty}{  \mathscr{X}}_2 - \overset{\infty}{  \mathscr{X}}_1^2 , \qquad \mathscr{D} :=  P -\overset{\infty}{\mathscr{X}}_{1}\mathscr A,
\end{equation}
and the objects
\noindent\begin{minipage}{.3\linewidth}
	\begin{alignat*}{2}
		& \al  && := \left(\frac{\mathscr{A}_{11} \mathscr{B}_{11}}{P_{11}} + \mathscr{D}_{11}\right)^{-1},  \\
		& \om_{jk} &&:=\frac{P_{1j} P_{k1}}{P_{11}}, 
	\end{alignat*}	
\end{minipage}
\noindent\begin{minipage}{.3\linewidth}
	\begin{alignat*}{2}
			&	\theta &&:=\frac{\mathscr{A}_{11}}{P_{11}}, \\
	&\eta_j && :=\frac{P_{1j}}{P_{11}},
	\end{alignat*}	
\end{minipage}
\noindent\begin{minipage}{.3\linewidth}
	\begin{alignat*}{2}
		& \rho_j &&:= \mathscr{A}_{j1} - \frac{\mathscr{A}_{11} P_{j1}}{P_{11}},   \\
		&\nu_j && := \frac{\mathscr{B}_{11} P_{1j}}{P_{11}} - \mathscr{B}_{1j} .
	\end{alignat*}	
\end{minipage}

assuming that they are well defined.
\noindent Using these objects define

\noindent\begin{minipage}{.5\linewidth}
	\begin{alignat*}{2}
	& \mathscr{M}_{jk}  && := -\al\rho_k \nu_j-\om_{jk}+P_{kj}, 
	\end{alignat*}	
\end{minipage}
\begin{minipage}{.5\linewidth}
	\begin{alignat*}{2}
	&f_j(x,y,z)  &&:= \al\nu_j
	\left(z-\theta x\right)+\eta_j x-y, 
	\end{alignat*}	
\end{minipage}
and the following four functions
\begin{align}
\mathscr{F}_1(x,y,w,z) & =  \frac{x}{P_{11}} + \frac{\al \mathscr{B}_{11}}{P_{11}}  \left(z-\theta x\right)  + \left( \left( \frac{P_{41}+\al \mathscr{B}_{11}\rho_4}{P_{11}\Delta} \right) \mathscr{M}_{43}   -\left(   \frac{P_{31}+\al \mathscr{B}_{11}\rho_3}{P_{11}\Delta}   \right)  \mathscr{M}_{44}   \right) f_3(x,y,z) \\ &  + \left(    \left(   \frac{P_{31}+\al \mathscr{B}_{11}\rho_3}{P_{11}\Delta}   \right)   \mathscr{M}_{34}   - \left( \frac{P_{41}+\al \mathscr{B}_{11}\rho_4}{P_{11}\Delta} \right) \mathscr{M}_{33}  \right) f_4(x,w,z), \nonumber \\
\mathscr{F}_2(x,y,w,z) & = \al \left( z - \theta x + \left(\frac{\rho_4 \mathscr{M}_{43} - \rho_3\mathscr{M}_{44}}{\Delta} \right) f_3(x,y,z) + \left(  \frac{\rho_3\mathscr{M}_{34}   -\rho_4 \mathscr{M}_{33}}{\Delta}   \right) f_4(x,w,z)   \right), \\
\mathscr{F}_3(x,y,w,z) &=\frac{ 1}{\Delta}  \left( \frac{1}{\mathscr{M}_{43}} f_3(x,y,z)   - \mathscr{M}_{34} f_4(x,w,z)    \right),  \\
\mathscr{F}_4(x,y,w,z) & = \frac{\mathscr{M}_{33} f_4(x,w,z) - f_3(x,y,z)}{\Delta},
\end{align}
where
\begin{equation}
	\Delta := \mathscr{M}_{34}\mathscr{M}_{43}  -\mathscr{M}_{33}\mathscr{M}_{44}.
\end{equation}
and it is assumed that $\Delta\neq 0$.
Finally, the $\mathscr{T}$-RHP data $\left\{ \overset{\infty}{\mathscr{T}}_{2,j1}, \overset{\infty}{\mathscr{T}}_{1,j1}, \overset{\infty}{\mathscr{T}}_{1,j3}, \overset{\infty}{\mathscr{T}}_{1,j4} \right\}^4_{j=1}$ describing the constant matrices $B$ and $E$ in \eqref{RT} are determined by the $\mathscr{X}$-RHP data as

	\noindent\begin{minipage}{.5\linewidth}
	\begin{alignat*}{2}
	&\overset{\infty}{\mathscr{T}}_{2,11} &&= \mathscr{F}_1\left(\left[\mathscr{C}P\right]_{11} ,
	\left[\mathscr{C}P\right]_{13} ,
	\left[\mathscr{C}P\right]_{14} ,
	\left[\mathscr{C}\mathscr{A} + \mathscr{B} \right]_{11}\right),  \\
	&\overset{\infty}{\mathscr{T}}_{1,11} &&= \mathscr{F}_2\left(\left[\mathscr{C}P\right]_{11} ,
	\left[\mathscr{C}P\right]_{13} ,
	\left[\mathscr{C}P\right]_{14} ,
	\left[\mathscr{C}\mathscr{A} + \mathscr{B} \right]_{11}\right), 
	\end{alignat*}	
\end{minipage}
	\begin{minipage}{.5\linewidth}
	\begin{alignat*}{2}
	&\overset{\infty}{\mathscr{T}}_{1,13} &&= \mathscr{F}_3\left(\left[\mathscr{C}P\right]_{11} ,
	\left[\mathscr{C}P\right]_{13} ,
	\left[\mathscr{C}P\right]_{14} ,
	\left[\mathscr{C}\mathscr{A} + \mathscr{B} \right]_{11}\right),  \\
	& \overset{\infty}{\mathscr{T}}_{1,14} &&= \mathscr{F}_4\left(\left[\mathscr{C}P\right]_{11} ,
	\left[\mathscr{C}P\right]_{13} ,
	\left[\mathscr{C}P\right]_{14} ,
	\left[\mathscr{C}\mathscr{A} + \mathscr{B} \right]_{11}\right),  
	\end{alignat*}	
\end{minipage}

	\noindent\begin{minipage}{.5\linewidth}
	\begin{alignat*}{2}
	&\overset{\infty}{\mathscr{T}}_{2,21} &&= \mathscr{F}_1\left(-P_{21} ,
	-P_{23} ,
	-P_{24} ,
	-\mathscr{A}_{21}\right),  \\
	&\overset{\infty}{\mathscr{T}}_{1,21} &&= \mathscr{F}_2\left(-P_{21} ,
	-P_{23} ,
	-P_{24} ,
	-\mathscr{A}_{21}\right), 
	\end{alignat*}	
\end{minipage}
\begin{minipage}{.5\linewidth}
	\begin{alignat*}{2}
	&\overset{\infty}{\mathscr{T}}_{1,23} &&= \mathscr{F}_3\left(-P_{21} ,
	-P_{23} ,
	-P_{24} ,
	-\mathscr{A}_{21}\right),  \\
	& \overset{\infty}{\mathscr{T}}_{1,24} &&= \mathscr{F}_4\left(-P_{21} ,
	-P_{23} ,
	-P_{24} ,
	-\mathscr{A}_{21}\right), 
	\end{alignat*}	
\end{minipage}

	\noindent\begin{minipage}{.5\linewidth}
	\begin{alignat*}{2}
	&\overset{\infty}{\mathscr{T}}_{2,31} &&= \mathscr{F}_1\left(\mathscr{B}_{31} ,
	\mathscr{B}_{33} ,
	\mathscr{B}_{34} ,
	-\mathscr{D}_{31}\right),  \\
	&\overset{\infty}{\mathscr{T}}_{1,31} &&= \mathscr{F}_2\left(\mathscr{B}_{31} ,
	\mathscr{B}_{33} ,
	\mathscr{B}_{34} ,
	-\mathscr{D}_{31}\right), 
	\end{alignat*}	
\end{minipage}
\begin{minipage}{.5\linewidth}
	\begin{alignat*}{2}
	&\overset{\infty}{\mathscr{T}}_{1,33} &&= \mathscr{F}_3\left(\mathscr{B}_{31} ,
	\mathscr{B}_{33} ,
	\mathscr{B}_{34} ,
	-\mathscr{D}_{31}\right),  \\
	& \overset{\infty}{\mathscr{T}}_{1,34} &&= \mathscr{F}_4\left(\mathscr{B}_{31} ,
	\mathscr{B}_{33} ,
	\mathscr{B}_{34} ,
	-\mathscr{D}_{31}\right),  
	\end{alignat*}	
\end{minipage}

	\noindent\begin{minipage}{.5\linewidth}
	\begin{alignat*}{2}
	&\overset{\infty}{\mathscr{T}}_{2,41} &&= \mathscr{F}_1\left(\mathscr{B}_{41} ,
	\mathscr{B}_{43} ,
	\mathscr{B}_{44} ,
	-\mathscr{D}_{41}\right),  \\
	&\overset{\infty}{\mathscr{T}}_{1,41} &&= \mathscr{F}_2\left(\mathscr{B}_{41} ,
	\mathscr{B}_{43} ,
	\mathscr{B}_{44} ,
	-\mathscr{D}_{41}\right),  
	\end{alignat*}	
\end{minipage}
\begin{minipage}{.5\linewidth}
	\begin{alignat*}{2}
	&\overset{\infty}{\mathscr{T}}_{1,43} &&= \mathscr{F}_3\left(\mathscr{B}_{41} ,
	\mathscr{B}_{43} ,
	\mathscr{B}_{44} ,
	-\mathscr{D}_{41}\right),  \\
	& \overset{\infty}{\mathscr{T}}_{1,44} &&= \mathscr{F}_4\left(\mathscr{B}_{41} ,
	\mathscr{B}_{43} ,
	\mathscr{B}_{44} ,
	-\mathscr{D}_{41}\right). 
	\end{alignat*}
	\end{minipage}
	Again, the dependence of all objects on $n$ 	is suppressed. 
\end{theorem}

In Section \S \ref{Sec asym norms r=0 s=1} we will  illustrate a general procedure of how Theorems \ref{thm r=0 s=1}, \ref{thm r=s=0},  and \ref{thm r=0 s=2} can be used  to find the asymptotics of the norms $h^{(r,s)}_{n}$ for  $(r,s) \in \{(0,0),(0,1),(0,2)\}$. In addition to these theorems, we will need two more ingredients.

First, we  will need the following general fact concerning the $X$-RHP.
\footnote{Part (b) of this Theorem was used in the case of $(r,s)=(1,1)$ in \cite{GI}.}:

\begin{theorem}\label{Thm WW}
	For any choice of $(r,s)\in \Z\times\Z$, it holds that 
	\begin{itemize}
		\item[(a)] $J^{-1}_X(z;r,s)=W J_X(z^{-1};r,s) W$,
		\item[(b)] $P^{-1}(n,r,s)=W P(n;r,s) W$, and
		\item[(c)] $\overset{\circ}{X}_1(n,r,s)=W \overset{\infty}{X}_1(n,r,s) W$,		where $W$ is the following permutation matrix:
		\begin{equation}\label{WWW}
			W=\begin{pmatrix}
				0 & 1 & 0 & 0 \\
				1& 0 & 0 & 0 \\
				0 & 0 & 0 & 1 \\
				0 & 0 & 1 & 0\\
			\end{pmatrix}.
		\end{equation}
	\end{itemize}
\end{theorem}
We give a proof of this general Theorem in \S\ref{proof Thm{WW}}.

The second ingredient which we will use in  \S \ref{Sec asym norms r=0 s=1} is the asymptotic results concerning the 
$X$-RHP with $(r,s)=(1,1)$, i.e. the $ \mathscr{X}$-RHP, obtained in \cite{GI}. Here are the details.

Given the Szeg{\H o}-type symbols $\phi(z)$ and $w(z) = d(z)\phi(z)$, let
 	\begin{equation}\label{annulus}
	U_0 := \left\{z : r_{i} < |z| < r_{o}: \quad 0 < r_{i} < 1 < r_{o}\right\},
\end{equation}    
be the neighborhood of the unit circle where both functions, $\phi(z)$ and $d(z)$ are analytic and denote 
\begin{equation}\label{r0}
	r_{0} := \max\{r_{i}, r^{-1}_{o}\}.
\end{equation} Define
\begin{equation}\label{al be}
	\al(z)=\exp \left[ \frac{1}{2 \pi i } \int_{\T} \frac{\ln(\phi(\tau))}{\tau-z}d\tau \right], \qquad 
	\be(z)=\exp \left[ \frac{1}{2 \pi i } \int_{\T} \frac{\ln(d(\tau))}{\tau-z}d\tau \right],
\end{equation}
\begin{equation}\label{C rho}
	C_{\rho}(z) = -\frac{1}{2\pi i } \int_{\T} \frac{1}{\be_-(\tau) \be_+(\tau) \tilde{\al}_-(\tau) \al_+(\tau)(\tau-z)}d\tau,
\end{equation}
\begin{equation}
	g_{23}(z) = - \frac{\al(0) \tilde{d}(z) \be(z) }{\tilde{\al}(z)}, \qquad g_{43}(z) = - \al^2(0) \be(z) \left( \frac{\al(z) }{\tilde{\phi}(z)} + \frac{\tilde{d}(z) C_{\rho}(z)}{\tilde{\al}(z)} \right),
\end{equation}
\begin{equation}\label{T+H integrals in the norm}
	R_{1,23}(z;n) = \frac{1}{2\pi i}\int_{\Gamma'_i} \frac{\mu^ng_{23}(\mu)}{\mu-z}d\mu, \qquad         R_{1,43}(z;n) = \frac{1}{2\pi i}\int_{\Gamma'_i} \frac{\mu^ng_{43}(\mu)}{\mu-z}d\mu,
\end{equation}    
and finally
\begin{equation}\label{En0001}
	\mathcal{E}(n)=  \frac{2}{\al(0)}R_{1,43}(0;n)-C_{\rho}(0)R_{1,23}(0;n).
\end{equation}
In \eqref{T+H integrals in the norm}, the contour $\Gamma'_i$ is a circle, oriented counter-clockwise, with radius $r' \in (r_0,1)$. 
\begin{theorem}\label{T+H main thm}\cite[Theorem 1.1]{GI}
	Suppose that $\phi(e^{i\theta})$ is smooth and nonzero on the unit circle with zero winding number, which admits an analytic continuation in a neighborhood of the unit circle.  Let $w=d\phi$, where $d$ satisfies all the properties of $\phi$ in addition to $d(e^{i\theta})d(e^{-i\theta})=1$, for all $\theta \in [0,2\pi)$. Suppose that
	there exists $C>0$ such that for sufficiently large $n$, 
	\begin{equation}\label{Enneq0}
		|\mathcal{E}(n)| \geq C\mathfrak{r}^{n}, \quad \mbox{for some}\,\, \mathfrak{r}:  \quad r_{0}\leq \mathfrak{r} < 1, 
	\end{equation}
	where $ \mathcal{E}(n)$ is the functional of the weights $\phi$ and $w$ defined in  (\ref{En0001}). Then, for sufficiently large $n$
	the determinant $D_{n}(\phi,w;1,1) \neq0$ and the asymptotics of 
	\begin{equation*}
		h^{(1,1)}_{n-1} \equiv \frac{D_{n}(\phi,w;1,1)}{D_{n-1}(\phi,w;1,1)},
	\end{equation*} 
	is given by
	\begin{equation}\label{T+H main result}
		h^{(1,1)}_{n-1}= -\al(0) \frac{\mathcal{E}(n)}{\mathcal{E}(n-1)}(1+ O{(e^{-c_1n})}), \qquad n \to \infty,
	\end{equation}
	where $c_1 = -\log\left(\frac{r^{2}_1}{\mathfrak{r}}\right)  >0$, and $r_1$ is any number satisfying the conditions: $ \mathfrak{r} < r_1 < 1$ and $ r^2_1 < \mathfrak{r}$\footnote{See section 4.2 of \cite{GI} for the requirements on $\mathfrak{r}$ and $r_1$.}.
\end{theorem}

We conclude this section by presenting the asymptotics of $h^{(0,1)}$ which will be obtained in  \S\ref{Sec asym norms r=0 s=1} based on
Theorem \ref{thm r=0 s=1}. We treat this as a case study: the same procedure can be used in view of Theorems \ref{thm r=s=0} and \ref{thm r=0 s=2} to find asymptotic expressions for $h^{(0,0)}_n$ and $h^{(0,2)}_n$.

\begin{theorem}\label{main thm}
		Suppose that $\phi(e^{i\theta})$ is smooth and nonzero on the unit circle with zero winding number, which admits an analytic continuation in a neighborhood of the unit circle.  Let $w=d\phi$, where $d$ satisfies all the properties of $\phi$ in addition to $d(e^{i\theta})d(e^{-i\theta})=1$, for all $\theta \in [0,2\pi)$. Let also $\al(z)$, $\be(z)$ and $C_{\rho}(z)$ be defined via \eqref{al be}-\eqref{C rho}, and fix $r_*$ to be be any number satisfying the condition 
$r_0 < r_* < 1$, where $r_0$ is defined in \eqref{r0}.		Define
		\begin{equation}\label{T+H entries of R_1}
			\begin{split}
				& \mathcal{R}_{1,jk}(z;n) := \frac{1}{2\pi i}\int_{\Gamma'_i} \frac{\mu^ng_{jk}(\mu)}{\mu-z}d\mu, \qquad \ \  jk=12,14, \\
				&  \mathcal{R}_{1,jk}(z;n) := \frac{1}{2\pi i}\int_{\Gamma'_o} \frac{\mu^{-n}g_{jk}(\mu)}{\mu-z}d\mu, \qquad jk=32,34,
			\end{split}
		\end{equation} where the contours $\Gamma'_i$ and $\Gamma'_o$ are circles, oriented counter-clockwise, with radii $r_* \in (r_0,1)$ and $1/r_*$ respectively, and
		\begin{equation*}\label{T+H g_ij's}
			\begin{split}
				& g_{12}(z) = - \frac{\al(z)}{\phi(z)\be(z)} - \frac{\tilde{w}(z) C_{\rho}(z)}{\phi(z)\be(z)\tilde{\al}(z)}, \qquad  g_{14}(z) =  \frac{\tilde{w}(z)}{\phi(z)\be(z)\tilde{\al}(z) \al(0) }, \\
				&  g_{32}(z) = - \frac{1}{\al(0)\tilde{\phi}(z)} \left( \frac{\tilde{\al}(z)}{\be(z)} - w(z)\tilde{\al}^2(z)\be(z)\al(z)C_{\rho}(z) \right), \qquad g_{34}(z) =  \frac{w(z) \tilde{\al}^2(z)\be(z)\al(z) }{\tilde{\phi}(z)\al^2(0)}. \\
			\end{split}
		\end{equation*} Assume further that $\phi$ and $d$ be symbols for which there exist $\widehat{n} \in \N$, constants $r_1,r_2 \in   (r^3_*,r^2_*)$, $r_3,r_4 \in  (r^2_*,r_*)$, and  $C_j>0$, $j=1,\ldots,4$,  such that
		\begin{eqnarray}
			|\mathcal{R}_{1,32}(0;n)\mathcal{R}_{1,14}(0;n)| & \geq C_1r^{n}_1, \\
			|\mathcal{R}_{1,32}(0;n)\mathcal{R}_{1,14}(0;n) - \mathcal{R}_{1,12}(0;n)\mathcal{R}_{1,34}(0;n)| & \geq C_2r^{n}_2, \label{key1}\\
			\left|\frac{\mathcal{R}_{1,12}(0;n)}{\al(0)} - \mathcal{R}_{1,32}(0;n-1)\right| &\geq C_3r^{n}_3, \label{key2}\\
			\left|-C_{\rho}(0)\al(0)\mathcal{R}_{1,34}(0;n)-\mathcal{R}_{1,32}(0;n) + \mathcal{R}_{1,32}(0;n-1)\right| &\geq C_4r^{n}_4,
		\end{eqnarray} for all $n > \widehat{n}$. Let 		$\mathfrak{c} := \min\{c_1,c_2,c_3,c_4\}$ where
		
\noindent\begin{minipage}{.23\linewidth}
	\begin{alignat*}{2}
		&	c_{1} &&:= - \log\left(\frac{r^{3}_*}{r_1}\right)>0,  
	\end{alignat*}	
\end{minipage}
\noindent\begin{minipage}{.23\linewidth}
	\begin{alignat*}{2}
		&	c_{2} &&:=- \log\left(\frac{r^{3}_*}{r_2}\right)>0,  
	\end{alignat*}	
\end{minipage}
\noindent\begin{minipage}{.23\linewidth}
	\begin{alignat*}{2}
		&	c_{3} &&:= - \log\left(\frac{r^{2}_*}{r_3}\right)>0 ,  
	\end{alignat*}	
\end{minipage}
\noindent\begin{minipage}{.23\linewidth}
	\begin{alignat*}{2}
		&	c_{4} &&:= - \log\left(\frac{r^{2}_*}{r_4}\right)>0.  
	\end{alignat*}	
\end{minipage}		

Then, for sufficiently large $n$ the determinant $D_{n}(\phi,w;0,1) \neq0$ and the asymptotics of 
\begin{equation*}
	h^{(0,1)}_{n-1} \equiv \frac{D_{n}(\phi,w;0,1)}{D_{n-1}(\phi,w;0,1)},
\end{equation*} 
is given by
				\begin{equation}\label{h asymp 2intro}
			h^{(0,1)}_{n-1} =\frac{\left(   \al(0) \mathcal{R}_{1,32}(0;n-1) - \mathcal{R}_{1,12}(0;n) \right)\left( \mathcal{R}_{1,32}(0;n)\mathcal{R}_{1,14}(0;n) - \mathcal{R}_{1,12}(0;n)\mathcal{R}_{1,34}(0;n) \right)}{\mathcal{R}_{1,32}(0;n)\mathcal{R}_{1,14}(0;n)\left( C_{\rho}(0)\al(0)\mathcal{R}_{1,34}(0;n)+\mathcal{R}_{1,32}(0;n) - \mathcal{R}_{1,32}(0;n-1) \right)} \left( 1 + O(e^{-\mathfrak{c}n}) \right),
		\end{equation}  as $n \to \infty$.
\end{theorem}
\begin{remark}\normalfont
	As detailed in Section \ref{sec asymp h_n 01}, the conditions \eqref{key1} and \eqref{key2} ensure that the condition \eqref{gencond} and thus the statement of Corollary \ref{corollary2.8.1} is in place. This justifies the statement in Theorem \ref{main thm} that for sufficiently large $n$ the determinant $D_{n}(\phi,w;0,1) \neq0$.
\end{remark}

\begin{remark} \normalfont
Being applied to the Ising Model in the half plane studied in \cite{Chelkak} and described in \S\ref{Sec Ising Zig Zag},
equation (\ref{h asymp 2intro}) yields the estimate
$$
h^{(0,1)}_{n-1} = 1 + O(e^{-\mathfrak{c}n}). 
$$
This in turn would mean that the corresponding magnetization $M_n$ is approaching a constant as $n\rightarrow \infty$,
\begin{equation}\label{last1}
M_n = D_n(\phi,\mathbb{d}\phi;0,1) \sim \mbox{constant}, \quad n \rightarrow \infty.
\end{equation}
We shall present the details in the forthcoming publication where we also hope to produce the exact value of the constant in
(\ref{last1}).
\end{remark}
\section{Proof of Theorem \ref{thm r=0 s=1}}\label{sec thm r=0 s=1}

Let us recall the Riemann-Hilbert problem satisfied by $\mathscr U$:
\begin{itemize}
	\item \textbf{RH-$\mathscr{U}$1} \quad  $\mathscr{U}$ is holomorphic in the complement of $\T \cup \{0\}$.
	\item \textbf{RH-$\mathscr{U}$2} \quad   For $z \in \T$, $\mathscr{U}$ satisfies \begin{equation*}\label{UZ-jump}
	\mathscr{U}_+(z;n)=\mathscr{U}_-(z;n) \begin{pmatrix}
	1 & 0 &  \tilde{w}(z) & - z \phi(z) \\
	0 & 1 & z^{-1} \tilde{\phi}(z) & -  w(z) \\
	0 & 0 & 1 & 0 \\
	0 & 0 & 0 & 1
	\end{pmatrix}, 
	\end{equation*} 
	\item \textbf{RH-$\mathscr{U}$3}  \quad As $z \to \infty$ we have \begin{equation*}\label{UZinfinity}
	\mathscr{U}(z;n)=\left(\di I+\frac{ \overset{\infty}{  \mathscr{U}}_1}{z}+\frac{\overset{\infty}{  \mathscr{U}}_2}{z^2} + O(z^{-3})\right)\begin{pmatrix}
	z^n & 0 & 0 & 0\\
	0 & 1 & 0 & 0 \\
	0 & 0 & z^{-n} & 0 \\
	0 & 0 & 0 & 1
	\end{pmatrix},  
	\end{equation*} 
	\item \textbf{RH-$\mathscr{U}$4} \quad As $z \to 0$ we have
	\begin{equation*}\label{UZzero}
	\mathscr{U}(z;n)=\widehat{\mathscr{U}}\left(I+\overset{\circ}{  \mathscr{U}}_1z+\overset{\circ}{  \mathscr{U}}_2z^2+O(z^3)\right)\begin{pmatrix}
	1 & 0 & 0 & 0\\
	0 & z^{-n} & 0 & 0 \\
	0 & 0 & 1 & 0 \\
	0 & 0 & 0 & z^n
	\end{pmatrix}.
	\end{equation*}
\end{itemize} Recall also the function
\begin{equation}\label{VU}
\mathscr{V}(z;n) := \mathscr{U}(z;n)  \begin{pmatrix}
z & 0 & 0 & 0\\
0 & 1 & 0 & 0 \\
0 & 0 & z & 0 \\
0 & 0 & 0 & 1
\end{pmatrix}.
\end{equation} This function satisfies the following Riemann-Hilbert problem
\begin{itemize}
	\item \textbf{RH-$\mathscr{V}$1} \quad  $\mathscr{V}$ is holomorphic in the complement of $\T \cup \{0\}$.
	\item \textbf{RH-$\mathscr{V}$2} \quad   For $z \in \T$, $\mathscr{V}$ satisfies \begin{equation*}\label{VUZ-jump}
	\mathscr{V}_+(z;n)=\mathscr{V}_-(z;n) \begin{pmatrix}
	1 & 0 &  \tilde{w}(z) & -  \phi(z) \\
	0 & 1 &  \tilde{\phi}(z) & -  w(z) \\
	0 & 0 & 1 & 0 \\
	0 & 0 & 0 & 1
	\end{pmatrix}, 
	\end{equation*} 
	\item \textbf{RH-$\mathscr{V}$3}  \quad As $z \to \infty$ we have \begin{equation*}\label{VUZinfinity}
	\mathscr{V}(z;n)=\left(\di I+\frac{ \overset{\infty}{  \mathscr{U}}_1}{z}+\frac{\overset{\infty}{  \mathscr{U}}_2}{z^2} + O(z^{-3})\right)\begin{pmatrix}
	z^n & 0 & 0 & 0\\
	0 & 1 & 0 & 0 \\
	0 & 0 & z^{-n} & 0 \\
	0 & 0 & 0 & 1
	\end{pmatrix}\begin{pmatrix}
	z & 0 & 0 & 0\\
	0 & 1 & 0 & 0 \\
	0 & 0 & z & 0 \\
	0 & 0 & 0 & 1
	\end{pmatrix},  
	\end{equation*} 
	\item \textbf{RH-$\mathscr{V}$4} \quad As $z \to 0$ we have
	\begin{equation*}\label{VUZzero}
	\mathscr{V}(z;n)=\widehat{\mathscr{U}}\left(I+\overset{\circ}{  \mathscr{U}}_1z+\overset{\circ}{  \mathscr{U}}_2z^2+O(z^3)\right)\begin{pmatrix}
	1 & 0 & 0 & 0\\
	0 & z^{-n} & 0 & 0 \\
	0 & 0 & 1 & 0 \\
	0 & 0 & 0 & z^n
	\end{pmatrix}\begin{pmatrix}
	z & 0 & 0 & 0\\
	0 & 1 & 0 & 0 \\
	0 & 0 & z & 0 \\
	0 & 0 & 0 & 1
	\end{pmatrix}.
	\end{equation*}
\end{itemize}
Since $\mathscr{V}$ and $\mathscr{X}$ have the same jump matrices on the unit circle, their ratio \begin{equation}\label{R S X V}
R(z;n) = \mathscr{V}(z;n) \mathscr{X}^{-1}(z;n)
\end{equation} must be a rational function with singular behavior only at zero and $\infty$. From \textbf{RH-$\mathscr{X}$4} and \textbf{RH-$\mathscr{V}$4} we readily observe that $R(z;n)$ is holomorphic at zero and thus is an entire function. Let us consider its behavior at $\infty$ using \textbf{RH-$\mathscr{X}$3} and \textbf{RH-$\mathscr{V}$3}; we indeed have
\begin{equation}
\begin{split}
	R(z;n) & = \left(\di I+\frac{ \overset{\infty}{  \mathscr{U}}_1}{z}+\frac{\overset{\infty}{  \mathscr{U}}_2}{z^2} + O(z^{-3})\right)\begin{pmatrix}
	z & 0 & 0 & 0\\
	0 & 1 & 0 & 0 \\
	0 & 0 & z & 0 \\
	0 & 0 & 0 & 1
	\end{pmatrix}\left(\di I - \frac{ \overset{\infty}{  \mathscr{X}}_1}{z}	+ \frac{ \overset{\infty}{  \mathscr{X}}_1^2 - \overset{\infty}{  \mathscr{X}}_2}{z^2} 	+ O(z^{-3})\right) \\ & = z\begin{pmatrix}
	1 & 0 & 0 & 0\\
	0 & 0 & 0 & 0 \\
	0 & 0 & 1 & 0 \\
	0 & 0 & 0 & 0
	\end{pmatrix} + \begin{pmatrix}
	0 & 0 & 0 & 0\\
	0 & 1 & 0 & 0 \\
	0 & 0 & 0 & 0 \\
	0 & 0 & 0 & 1
	\end{pmatrix}  +  \overset{\infty}{\mathscr{U}}_1 \begin{pmatrix}
	1 & 0 & 0 & 0\\
	0 & 0 & 0 & 0 \\
	0 & 0 & 1 & 0 \\
	0 & 0 & 0 & 0
	\end{pmatrix} - \begin{pmatrix}
	1 & 0 & 0 & 0\\
	0 & 0 & 0 & 0 \\
	0 & 0 & 1 & 0 \\
	0 & 0 & 0 & 0
	\end{pmatrix} \overset{\infty}{  \mathscr{X}}_1 + O(z^{-1}), 
	\end{split} 
\end{equation}
and thus 
\begin{equation}\label{RUX}
	R(z;n) = z\begin{pmatrix}
	1 & 0 & 0 & 0\\
	0 & 0 & 0 & 0 \\
	0 & 0 & 1 & 0 \\
	0 & 0 & 0 & 0
	\end{pmatrix} + \begin{pmatrix}
	0 & 0 & 0 & 0\\
	0 & 1 & 0 & 0 \\
	0 & 0 & 0 & 0 \\
	0 & 0 & 0 & 1
	\end{pmatrix}  +  \overset{\infty}{\mathscr{U}}_1 \begin{pmatrix}
	1 & 0 & 0 & 0\\
	0 & 0 & 0 & 0 \\
	0 & 0 & 1 & 0 \\
	0 & 0 & 0 & 0
	\end{pmatrix} - \begin{pmatrix}
	1 & 0 & 0 & 0\\
	0 & 0 & 0 & 0 \\
	0 & 0 & 1 & 0 \\
	0 & 0 & 0 & 0
	\end{pmatrix} \overset{\infty}{  \mathscr{X}}_1.
\end{equation}
It follows that to detemine $R(z;n)$, and, as a consequence,  $\mathscr{U}(z;n)$, in terms of the $\mathscr{X}$-RHP data, we only need to find the eight unknowns in the first and the third columns of $ \overset{\infty}{\mathscr{U}}_1$ in terms of the $\mathscr{X}$-RHP data. To that end, we use the above expression for $R(z;n)$ in 	\begin{equation}\label{X R X U}
\mathscr{U}(z;n) =	R(z;n) \mathscr{X}(z;n)  \begin{pmatrix}
z^{-1} & 0 & 0 & 0\\
0 & 1 & 0 & 0 \\
0 & 0 & z^{-1} & 0 \\
0 & 0 & 0 & 1
\end{pmatrix},
\end{equation}
which is a combination of \eqref{VU} and \eqref{R S X V}, and try to match the behavior with \textbf{RH-$\mathscr{U}$3} and \textbf{RH-$\mathscr{U}$4}.
As far as the behavior \textbf{RH-$\mathscr{U}$3} at infinity is concerned, it holds automatically - because of the indicated in (\ref{RUX})  structure of matrix function $R(z;n)$. Hence we have to hope that the matching the behavior with \textbf{RH-$\mathscr{U}$4} will give us  all  the eight unknowns in the first and third columns of $ \overset{\infty}{\mathscr{U}}_1$. It turns out to be indeed the case. To see this, let us rewrite \eqref{RUX} as
\begin{equation}
R(z;n) = z\begin{pmatrix}
1 & 0 & 0 & 0\\
0 & 0 & 0 & 0 \\
0 & 0 & 1 & 0 \\
0 & 0 & 0 & 0
\end{pmatrix} + A,
\end{equation}
where
\begin{equation}
	A := \begin{pmatrix}
	 \overset{\infty}{\mathscr{U}}_{1,11} - \overset{\infty}{  \mathscr{X}}_{1,11}& - \overset{\infty}{  \mathscr{X}}_{1,12} &  \overset{\infty}{\mathscr{U}}_{1,13} - \overset{\infty}{  \mathscr{X}}_{1,13} & - \overset{\infty}{  \mathscr{X}}_{1,14}\\
	 \overset{\infty}{\mathscr{U}}_{1,21} & 1 &  \overset{\infty}{\mathscr{U}}_{1,23} & 0 \\
	 \overset{\infty}{\mathscr{U}}_{1,31} - \overset{\infty}{  \mathscr{X}}_{1,31} & - \overset{\infty}{  \mathscr{X}}_{1,32} &  \overset{\infty}{\mathscr{U}}_{1,33} - \overset{\infty}{  \mathscr{X}}_{1,33} & - \overset{\infty}{  \mathscr{X}}_{1,34} \\
	 \overset{\infty}{\mathscr{U}}_{1,41} & 0 &  \overset{\infty}{\mathscr{U}}_{1,43} & 1
	\end{pmatrix}.
\end{equation}
Using \textbf{RH-$\mathscr{X}$4}, the behavior of \eqref{X R X U} as $z \to 0$ reads
\begin{equation}\label{X R X U inf}
\begin{split}
\mathscr{U}(z;n) & =	\left[ z\begin{pmatrix}
1 & 0 & 0 & 0\\
0 & 0 & 0 & 0 \\
0 & 0 & 1 & 0 \\
0 & 0 & 0 & 0
\end{pmatrix} + A\right] P(n)\left(I+\overset{\circ}{  \mathscr{X}}_1z+\overset{\circ}{  \mathscr{X}}_2z^2+O(z^3)\right)\begin{pmatrix}
1 & 0 & 0 & 0\\
0 & z^{-n} & 0 & 0 \\
0 & 0 & 1 & 0 \\
0 & 0 & 0 & z^n
\end{pmatrix}  \begin{pmatrix}
z^{-1} & 0 & 0 & 0\\
0 & 1 & 0 & 0 \\
0 & 0 & z^{-1} & 0 \\
0 & 0 & 0 & 1
\end{pmatrix} \\ & = \left[ z\begin{pmatrix}
1 & 0 & 0 & 0\\
0 & 0 & 0 & 0 \\
0 & 0 & 1 & 0 \\
0 & 0 & 0 & 0
\end{pmatrix} + A\right] P(n)\left(I+\overset{\circ}{  \mathscr{X}}_1z+\overset{\circ}{  \mathscr{X}}_2z^2+O(z^3)\right)\begin{pmatrix}
z^{-1} & 0 & 0 & 0\\
0 & 1 & 0 & 0 \\
0 & 0 & z^{-1} & 0 \\
0 & 0 & 0 & 1
\end{pmatrix}\begin{pmatrix}
1 & 0 & 0 & 0\\
0 & z^{-n} & 0 & 0 \\
0 & 0 & 1 & 0 \\
0 & 0 & 0 & z^n
\end{pmatrix}.  
\end{split}
\end{equation}
Comparing this with \textbf{RH-$\mathscr{U}$4} suggests that 
\begin{equation}
	\begin{split}
		\left[ z\begin{pmatrix}
			1 & 0 & 0 & 0\\
			0 & 0 & 0 & 0 \\
			0 & 0 & 1 & 0 \\
			0 & 0 & 0 & 0
		\end{pmatrix} + A\right] & P(n)\left(I+\overset{\circ}{  \mathscr{X}}_1z+\overset{\circ}{  \mathscr{X}}_2z^2+O(z^3)\right)\begin{pmatrix}
			z^{-1} & 0 & 0 & 0\\
			0 & 1 & 0 & 0 \\
			0 & 0 & z^{-1} & 0 \\
			0 & 0 & 0 & 1
		\end{pmatrix} \\ & = \widehat{\mathscr{U}}\left(I+\overset{\circ}{  \mathscr{U}}_1z+\overset{\circ}{  \mathscr{U}}_2z^2+O(z^3)\right).
	\end{split}
\end{equation}
Since the right hand side has no $\frac{1}{z}$ behavior, the left hand side does not as well. This condition gives the needed eight equations to determine the eight unknowns in $A$. Indeed, we need to have
\begin{equation}\label{conditions in u problem}
	A P(n) \begin{pmatrix}
		1 & 0 & 0 & 0\\
		0 & 0 & 0 & 0 \\
		0 & 0 & 1 & 0 \\
		0 & 0 & 0 & 0
	\end{pmatrix} = 0,
\end{equation}
which means that all eight entries in the first and third columns of $A P(n)$ must vanish. Thus we have 
\begin{align}
	\left( A P(n) \right)_{11} & = \sum_{j=1}^{4} A_{1j} P_{j1}(n) = \left( \overset{\infty}{\mathscr{U}}_{1,11} - \overset{\infty}{  \mathscr{X}}_{1,11}  \right) P_{11}(n) - \overset{\infty}{  \mathscr{X}}_{1,12} P_{21}(n) + \left(  \overset{\infty}{\mathscr{U}}_{1,13} - \overset{\infty}{  \mathscr{X}}_{1,13}  \right) P_{31}(n) \nonumber \\ & - \overset{\infty}{  \mathscr{X}}_{1,14} P_{41}(n) = 0, \label{AP11} \\
	\left( A P(n) \right)_{13} & = \sum_{j=1}^{4} A_{1j} P_{j3}(n) = \left( \overset{\infty}{\mathscr{U}}_{1,11} - \overset{\infty}{  \mathscr{X}}_{1,11}  \right) P_{13}(n) - \overset{\infty}{  \mathscr{X}}_{1,12} P_{23}(n) + \left(  \overset{\infty}{\mathscr{U}}_{1,13} - \overset{\infty}{  \mathscr{X}}_{1,13}  \right) P_{33}(n) \nonumber \\ & - \overset{\infty}{  \mathscr{X}}_{1,14} P_{43}(n) = 0, \label{AP13}
\end{align}
\begin{align}
		\left( A P(n) \right)_{21} & = \sum_{j=1}^{4} A_{2j} P_{j1}(n) =  \overset{\infty}{\mathscr{U}}_{1,21}  P_{11}(n) +  P_{21}(n) +   \overset{\infty}{\mathscr{U}}_{1,23}  P_{31}(n)  = 0, \label{AP21} \\
	\left( A P(n) \right)_{23} & = \sum_{j=1}^{4} A_{2j} P_{j3}(n) =  \overset{\infty}{\mathscr{U}}_{1,21}  P_{13}(n) +  P_{23}(n) +   \overset{\infty}{\mathscr{U}}_{1,23}  P_{33}(n)  = 0, \label{AP23} \\
		\left( A P(n) \right)_{31} & = \sum_{j=1}^{4} A_{3j} P_{j1}(n) = \left( \overset{\infty}{\mathscr{U}}_{1,31} - \overset{\infty}{  \mathscr{X}}_{1,31}  \right) P_{11}(n) - \overset{\infty}{  \mathscr{X}}_{1,32} P_{21}(n) + \left(  \overset{\infty}{\mathscr{U}}_{1,33} - \overset{\infty}{  \mathscr{X}}_{1,33}  \right) P_{31}(n) \nonumber \\ &- \overset{\infty}{  \mathscr{X}}_{1,34} P_{41}(n) = 0, \label{AP31} \\
	\left( A P(n) \right)_{33} & = \sum_{j=1}^{4} A_{3j} P_{j3}(n) = \left( \overset{\infty}{\mathscr{U}}_{1,31} - \overset{\infty}{  \mathscr{X}}_{1,31}  \right) P_{13}(n) - \overset{\infty}{  \mathscr{X}}_{1,32} P_{23}(n) + \left(  \overset{\infty}{\mathscr{U}}_{1,33} - \overset{\infty}{  \mathscr{X}}_{1,33}  \right) P_{33}(n) \nonumber \\ & - \overset{\infty}{  \mathscr{X}}_{1,34} P_{43}(n) = 0, \label{AP33} \\
	\left( A P(n) \right)_{41} & = \sum_{j=1}^{4} A_{4j} P_{j1}(n) =  \overset{\infty}{\mathscr{U}}_{1,41}  P_{11}(n)  +   \overset{\infty}{\mathscr{U}}_{1,43}  P_{31}(n) +  P_{41}(n) = 0, \label{AP41} \\
\left( A P(n) \right)_{43} & = \sum_{j=1}^{4} A_{4j} P_{j3}(n) =  \overset{\infty}{\mathscr{U}}_{1,41}  P_{13}(n)  +   \overset{\infty}{\mathscr{U}}_{1,43}  P_{33}(n) +  P_{43}(n) = 0. \label{AP43}
\end{align}
We view these eight equations as four mutually decoupled systems for determining the unknowns, for example $\overset{\infty}{\mathscr{U}}_{1,11}$ and $\overset{\infty}{\mathscr{U}}_{1,13}$ can be found by solving the system  \eqref{AP11}-\eqref{AP13}, $\overset{\infty}{\mathscr{U}}_{1,21}$ and $\overset{\infty}{\mathscr{U}}_{1,23}$ can be found by solving the system \eqref{AP21}-\eqref{AP23} , and so on.  We will skip these quite routine calculations whose result
is the formulae for $\overset{\infty}{\mathscr{U}}_{1,jk}$ given in Theorem \ref{thm r=0 s=1}. We have thus proven the theorem.
\begin{comment}
	content...

Let $\Om_4$ denote the space of $4\times 4$ matrix-valued functions of a complex variable. Consider the operators $T_{\uparrow}, T_{\downarrow} : \Om_4 \to \Om_4$ defined by

\begin{equation}
	T_{\uparrow} : f(z) \mapsto f(z) \begin{pmatrix}
	1 & 0 & 0 & 0\\
	0 & 1 & 0 & 0 \\
	0 & 0 & z^{-1} & 0 \\
	0 & 0 & 0 & z
	\end{pmatrix},
\end{equation}
and 
\begin{equation}
T_{\downarrow} : f(z) \mapsto f(z) \begin{pmatrix}
1 & 0 & 0 & 0\\
0 & 1 & 0 & 0 \\
0 & 0 & z & 0 \\
0 & 0 & 0 & z^{-1}
\end{pmatrix}.
\end{equation}
\end{comment}
\section{Proof of Theorem \ref{thm r=s=0}}\label{sec thm r=s=0} 
Let us recall the Riemann-Hilbert problem \textbf{RH-$\mathscr{Y}$1} through \textbf{RH-$\mathscr{Y}$4} below, being the specialization $r=s=0$ of the Riemann-Hilbert problem \textbf{RH-$X$1} through \textbf{RH-$X$4}:
\begin{itemize}
	\item \textbf{RH-$\mathscr{Y}$1} \quad  $\mathscr{Y}$ is holomorphic in the complement of $\T \cup \{0\}$.
	\item \textbf{RH-$\mathscr{Y}$2} \quad   For $z \in \T$, $\mathscr{Y}$ satisfies \begin{equation*}\label{Y-jump}
	\mathscr{Y}_+(z;n)=\mathscr{Y}_-(z;n) \begin{pmatrix}
	1 & 0 & z^{-1} \tilde{w}(z) & - z \phi(z) \\
	0 & 1 & z^{-1} \tilde{\phi}(z) & - z w(z) \\
	0 & 0 & 1 & 0 \\
	0 & 0 & 0 & 1
	\end{pmatrix}, 
	\end{equation*} 
	\item \textbf{RH-$\mathscr{Y}$3}  \quad As $z \to \infty$ we have \begin{equation*}\label{Yinfinity}
	\mathscr{Y}(z;n)=\left(\di I+\frac{ \overset{\infty}{  \mathscr{Y}}_1}{z}+\frac{\overset{\infty}{  \mathscr{Y}}_2}{z^2} + O(z^{-3})\right)\begin{pmatrix}
	z^n & 0 & 0 & 0\\
	0 & 1 & 0 & 0 \\
	0 & 0 & z^{-n} & 0 \\
	0 & 0 & 0 & 1
	\end{pmatrix},  
	\end{equation*} 
	\item \textbf{RH-$\mathscr{Y}$4} \quad As $z \to 0$ we have
	\begin{equation*}\label{Yzero}
	\mathscr{Y}(z;n)=\widehat{\mathscr{Y}}\left(I+\overset{\circ}{  \mathscr{Y}}_1z+\overset{\circ}{  \mathscr{Y}}_2z^2+O(z^3)\right)\begin{pmatrix}
	1 & 0 & 0 & 0\\
	0 & z^{-n} & 0 & 0 \\
	0 & 0 & 1 & 0 \\
	0 & 0 & 0 & z^n
	\end{pmatrix}.
	\end{equation*}
\end{itemize}
 Consider the transformation $\mathscr{Y} \mapsto \mathscr{W}$ defined by 
\begin{equation}\label{S Y}
	\mathscr{W}(z;n) := \mathscr{Y}(z;n) \begin{pmatrix}
	1 & 0 & 0 & 0\\
	0 & 1 & 0 & 0 \\
	0 & 0 & z & 0 \\
	0 & 0 & 0 & z^{-1}
	\end{pmatrix}.
\end{equation}
This is the $r=s=0$ specification of the general transformation  $(\ref{W})$, and it  satisfies
\begin{itemize}
	\item \textbf{RH-$\mathscr{W}$1} \quad  $\mathscr{W}$ is holomorphic in the complement of $\T \cup \{0\}$.
	\item \textbf{RH-$\mathscr{W}$2} \quad   For $z \in \T$, $\mathscr{W}$ satisfies \begin{equation*}\label{S-jump}
	\mathscr{W}_+(z;n)=\mathscr{W}_-(z;n) \begin{pmatrix}
	1 & 0 &  \tilde{w}(z) & -  \phi(z) \\
	0 & 1 &  \tilde{\phi}(z) & - w(z) \\
	0 & 0 & 1 & 0 \\
	0 & 0 & 0 & 1
	\end{pmatrix}, 
	\end{equation*} 
	\item \textbf{RH-$\mathscr{W}$3}  \quad As $z \to \infty$ we have \begin{equation*}\label{Sinfinity}
	\mathscr{W}(z;n)=\left(\di I+\frac{ \overset{\infty}{  \mathscr{Y}}_1}{z}+\frac{\overset{\infty}{  \mathscr{Y}}_2}{z^2} + O(z^{-3})\right)\begin{pmatrix}
	z^n & 0 & 0 & 0\\
	0 & 1 & 0 & 0 \\
	0 & 0 & z^{-n} & 0 \\
	0 & 0 & 0 & 1
	\end{pmatrix} \begin{pmatrix}
	1 & 0 & 0 & 0\\
	0 & 1 & 0 & 0 \\
	0 & 0 & z & 0 \\
	0 & 0 & 0 & z^{-1}
	\end{pmatrix},  
	\end{equation*} 
	\item \textbf{RH-$\mathscr{W}$4} \quad As $z \to 0$ we have
	\begin{equation*}\label{Szero}
	\mathscr{W}(z;n)=\widehat{\mathscr{Y}}\left(I+\overset{\circ}{  \mathscr{Y}}_1z+\overset{\circ}{  \mathscr{Y}}_2z^2+O(z^3)\right)\begin{pmatrix}
	1 & 0 & 0 & 0\\
	0 & z^{-n} & 0 & 0 \\
	0 & 0 & 1 & 0 \\
	0 & 0 & 0 & z^n
	\end{pmatrix} \begin{pmatrix}
	1 & 0 & 0 & 0\\
	0 & 1 & 0 & 0 \\
	0 & 0 & z & 0 \\
	0 & 0 & 0 & z^{-1}
	\end{pmatrix}.
	\end{equation*}
\end{itemize}
Since $\mathscr{W}$ and $\mathscr{X}$ have the same jump matrices on the unit circle, their ratio \begin{equation}\label{R S X}
	\mathscr{R}(z;n) = \mathscr{W}(z;n) \mathscr{X}^{-1}(z;n)
\end{equation} must be a rational function with singular behavior only at zero and $\infty$. Let us consider the behavior at $\infty$ using \textbf{RH-$\mathscr{X}$3} and \textbf{RH-$\mathscr{W}$3}
\begin{equation}
\begin{split}
	\mathscr{R}(z;n) & =  \left(\di I+\frac{ \overset{\infty}{  \mathscr{Y}}_1}{z}+\frac{\overset{\infty}{  \mathscr{Y}}_2}{z^2} + O(z^{-3})\right) \begin{pmatrix}
	1 & 0 & 0 & 0\\
	0 & 1 & 0 & 0 \\
	0 & 0 & z & 0 \\
	0 & 0 & 0 & z^{-1}
	\end{pmatrix} \left(\di I - \frac{ \overset{\infty}{  \mathscr{X}}_1}{z}	+ \frac{ \overset{\infty}{  \mathscr{X}}_1^2 - \overset{\infty}{  \mathscr{X}}_2}{z^2} 	+ O(z^{-3})\right) \\ & = \begin{pmatrix}
	1 & 0 & 0 & 0\\
	0 & 1 & 0 & 0 \\
	0 & 0 & z & 0 \\
	0 & 0 & 0 & 0
	\end{pmatrix} + \overset{\infty}{  \mathscr{Y}}_1 \begin{pmatrix}
	0 & 0 & 0 & 0\\
	0 & 0 & 0 & 0 \\
	0 & 0 & 1 & 0 \\
	0 & 0 & 0 & 0
	\end{pmatrix} - \begin{pmatrix}
	0 & 0 & 0 & 0\\
	0 & 0 & 0 & 0 \\
	0 & 0 & 1 & 0 \\
	0 & 0 & 0 & 0
	\end{pmatrix} \overset{\infty}{  \mathscr{X}}_1 + O(z^{-1}).
\end{split}
\end{equation}
Considering the behaviour of $\mathscr{R}$ near zero using \textbf{RH-$\mathscr{X}$4} and \textbf{RH-$\mathscr{W}$4} we find
\begin{equation}
	\begin{split}
\mathscr{R}(z;n) & = \widehat{\mathscr{Y}}\left(I+\overset{\circ}{  \mathscr{Y}}_1z+O(z^2)\right) \begin{pmatrix}
	 1 & 0 & 0 & 0\\
	 0 & 1 & 0 & 0 \\
	 0 & 0 & z & 0 \\
	 0 & 0 & 0 & z^{-1}
	 \end{pmatrix}  \left(I-\overset{\circ}{  \mathscr{X}}_1z+O(z^2)\right) P^{-1}(n) \\ & =  \frac{1}{z} \widehat{\mathscr{Y}} \begin{pmatrix}
	 0 & 0 & 0 & 0\\
	 0 & 0 & 0 & 0 \\
	 0 & 0 & 0 & 0 \\
	 0 & 0 & 0 & 1
	 \end{pmatrix} P^{-1}(n) + O(1).
	\end{split}
\end{equation}
Therefore by the Liouville's theorem we have the following formula for $\mathscr{R}$ in terms of the \textit{unknown} matrices $\widehat{\mathscr{Y}}$ and $\overset{\infty}{  \mathscr{Y}}_1$:
\begin{equation}
	\mathscr{R}(z;n) = \frac{1}{z} \widehat{\mathscr{Y}} \begin{pmatrix}
	0 & 0 & 0 & 0\\
	0 & 0 & 0 & 0 \\
	0 & 0 & 0 & 0 \\
	0 & 0 & 0 & 1
	\end{pmatrix} P^{-1}(n) + \begin{pmatrix}
	1 & 0 & 0 & 0\\
	0 & 1 & 0 & 0 \\
	0 & 0 & z & 0 \\
	0 & 0 & 0 & 0
	\end{pmatrix} + \overset{\infty}{  \mathscr{Y}}_1 \begin{pmatrix}
	0 & 0 & 0 & 0\\
	0 & 0 & 0 & 0 \\
	0 & 0 & 1 & 0 \\
	0 & 0 & 0 & 0
	\end{pmatrix} - \begin{pmatrix}
	0 & 0 & 0 & 0\\
	0 & 0 & 0 & 0 \\
	0 & 0 & 1 & 0 \\
	0 & 0 & 0 & 0
	\end{pmatrix} \overset{\infty}{  \mathscr{X}}_1.
\end{equation}
Since the matrices are sparse in the above formula, to determine $\mathscr{R}$, we only need to determine four entries from each one of $\widehat{\mathscr{Y}}$ and $\overset{\infty}{  \mathscr{Y}}_1$. To this end,  let us introduce  the relevant entries of these matrices by the formulae,
\begin{equation}
	\widehat{\mathscr{Y}} \equiv \begin{pmatrix}
	* & * & * & \widehat{\mathscr{Y}}_{14}\\
	* & * & *  & \widehat{\mathscr{Y}}_{24} \\
	* & * & *  & \de \\
	* & * & *  & \ga
	\end{pmatrix}, \qandq \overset{\infty}{  \mathscr{Y}}_1 \equiv \begin{pmatrix}
	* & * & \overset{\infty}{\mathscr{Y}}_{1,13} & *\\
	* & * & \overset{\infty}{\mathscr{Y}}_{1,23}  & * \\
	* & * & \overset{\infty}{\mathscr{Y}}_{1,33}  & * \\
	* & * & \overset{\infty}{\mathscr{Y}}_{1,43} & *
	\end{pmatrix}.  
\end{equation}
For simplicity of notations, below we suppress the dependence of quantities on $n$. In the notations above, $\mathscr{R}$ can be written as 
\begin{equation}\label{RR}
\mathscr{R}(z) = \frac{1}{z} \begin{pmatrix}
0 & 0 & 0 & \widehat{\mathscr{Y}}_{14}\\
0 & 0 & 0  & \widehat{\mathscr{Y}}_{24} \\
0 & 0 & 0  & \de \\
0 & 0 & 0  & \ga
\end{pmatrix} P^{-1} + \begin{pmatrix}
1 & 0 & \overset{\infty}{\mathscr{Y}}_{1,13} & 0\\
0 & 1 & \overset{\infty}{\mathscr{Y}}_{1,23} & 0 \\
-\overset{\infty}{\mathscr{X}}_{1,31} & -\overset{\infty}{\mathscr{X}}_{1,32} &  \overset{\infty}{\mathscr{Y}}_{1,33} - \overset{\infty}{\mathscr{X}}_{1,33} & - \overset{\infty}{\mathscr{X}}_{1,34} \\
0 & 0 & \overset{\infty}{\mathscr{Y}}_{1,43} & 0
\end{pmatrix} + z \begin{pmatrix}
0 & 0 & 0 & 0\\
0 & 0 & 0  & 0 \\
0 & 0 & 1  & 0 \\
0 & 0 & 0  & 0
\end{pmatrix} .
\end{equation}
We will try to find these unknowns using the above form for $\mathscr{R}$ and trying to satisfy \textbf{RH-$\mathscr{Y}$3} and \textbf{RH-$\mathscr{Y}$4}. 
Using \eqref{S Y} and \eqref{R S X} we have
\begin{equation}\label{Y R X}
	\mathscr{Y}(z)= \mathscr{R}(z)\mathscr{X}(z) \begin{pmatrix}
	1 & 0 & 0 & 0\\
	0 & 1 & 0 & 0 \\
	0 & 0 & z^{-1} & 0 \\
	0 & 0 & 0 & z
	\end{pmatrix},
\end{equation}
so
\begin{equation}\label{Y X}
	\begin{split}
\mathscr{Y}(z)&= \left( \frac{1}{z} \begin{pmatrix}
0 & 0 & 0 & \widehat{\mathscr{Y}}_{14}\\
0 & 0 & 0  & \widehat{\mathscr{Y}}_{24} \\
0 & 0 & 0  & \de \\
0 & 0 & 0  & \ga
\end{pmatrix} P^{-1} + \begin{pmatrix}
1 & 0 & \overset{\infty}{\mathscr{Y}}_{1,13} & 0\\
0 & 1 & \overset{\infty}{\mathscr{Y}}_{1,23} & 0 \\
-\overset{\infty}{\mathscr{X}}_{1,31} & -\overset{\infty}{\mathscr{X}}_{1,32} &  \overset{\infty}{\mathscr{Y}}_{1,33} - \overset{\infty}{\mathscr{X}}_{1,33} & - \overset{\infty}{\mathscr{X}}_{1,34} \\
0 & 0 & \overset{\infty}{\mathscr{Y}}_{1,43} & 0
\end{pmatrix} + z \begin{pmatrix}
0 & 0 & 0 & 0\\
0 & 0 & 0  & 0 \\
0 & 0 & 1  & 0 \\
0 & 0 & 0  & 0
\end{pmatrix} \right) \\ & \times \mathscr{X}(z) \begin{pmatrix}
1 & 0 & 0 & 0\\
0 & 1 & 0 & 0 \\
0 & 0 & z^{-1} & 0 \\
0 & 0 & 0 & z
\end{pmatrix}.
\end{split}
\end{equation}
Using \textbf{RH-$\mathscr{X}$3} we have as $z \to \infty$
\begin{equation}\label{Y X infty}
\begin{split}
\mathscr{Y}(z) & = \left( \frac{1}{z} \begin{pmatrix}
0 & 0 & 0 & \widehat{\mathscr{Y}}_{14}\\
0 & 0 & 0  & \widehat{\mathscr{Y}}_{24} \\
0 & 0 & 0  & \de \\
0 & 0 & 0  & \ga
\end{pmatrix} P^{-1} + \begin{pmatrix}
1 & 0 & \overset{\infty}{\mathscr{Y}}_{1,13} & 0\\
0 & 1 & \overset{\infty}{\mathscr{Y}}_{1,23} & 0 \\
-\overset{\infty}{\mathscr{X}}_{1,31} & -\overset{\infty}{\mathscr{X}}_{1,32} &  \overset{\infty}{\mathscr{Y}}_{1,33} - \overset{\infty}{\mathscr{X}}_{1,33} & - \overset{\infty}{\mathscr{X}}_{1,34} \\
0 & 0 & \overset{\infty}{\mathscr{Y}}_{1,43} & 0
\end{pmatrix} + z \begin{pmatrix}
0 & 0 & 0 & 0\\
0 & 0 & 0  & 0 \\
0 & 0 & 1  & 0 \\
0 & 0 & 0  & 0
\end{pmatrix} \right) \\ & \times \left(\di I+\frac{ \overset{\infty}{  \mathscr{X}}_1}{z}+\frac{\overset{\infty}{  \mathscr{X}}_2}{z^2} + O(z^{-3})\right) \begin{pmatrix}
1 & 0 & 0 & 0\\
0 & 1 & 0 & 0 \\
0 & 0 & z^{-1} & 0 \\
0 & 0 & 0 & z
\end{pmatrix} \begin{pmatrix}
z^n & 0 & 0 & 0\\
0 & 1 & 0 & 0 \\
0 & 0 & z^{-n} & 0 \\
0 & 0 & 0 & 1
\end{pmatrix}. 
\end{split}
\end{equation}
Now, we compare this with \textbf{RH-$\mathscr{Y}$3}, thus we must have
\begin{equation}\label{Y Y infty}
\begin{split}
& \left( \frac{1}{z} \begin{pmatrix}
0 & 0 & 0 & \widehat{\mathscr{Y}}_{14}\\
0 & 0 & 0  & \widehat{\mathscr{Y}}_{24} \\
0 & 0 & 0  & \de \\
0 & 0 & 0  & \ga
\end{pmatrix} P^{-1} + \begin{pmatrix}
1 & 0 & \overset{\infty}{\mathscr{Y}}_{1,13} & 0\\
0 & 1 & \overset{\infty}{\mathscr{Y}}_{1,23} & 0 \\
-\overset{\infty}{\mathscr{X}}_{1,31} & -\overset{\infty}{\mathscr{X}}_{1,32} &  \overset{\infty}{\mathscr{Y}}_{1,33} - \overset{\infty}{\mathscr{X}}_{1,33} & - \overset{\infty}{\mathscr{X}}_{1,34} \\
0 & 0 & \overset{\infty}{\mathscr{Y}}_{1,43} & 0
\end{pmatrix} + z \begin{pmatrix}
0 & 0 & 0 & 0\\
0 & 0 & 0  & 0 \\
0 & 0 & 1  & 0 \\
0 & 0 & 0  & 0
\end{pmatrix} \right) \\ & \times \left(\di I+\frac{ \overset{\infty}{  \mathscr{X}}_1}{z}+\frac{\overset{\infty}{  \mathscr{X}}_2}{z^2} + O(z^{-3})\right) \begin{pmatrix}
1 & 0 & 0 & 0\\
0 & 1 & 0 & 0 \\
0 & 0 & z^{-1} & 0 \\
0 & 0 & 0 & z
\end{pmatrix}  \equiv I+\frac{ \overset{\infty}{  \mathscr{Y}}_1}{z}+\frac{\overset{\infty}{  \mathscr{Y}}_2}{z^2} + O(z^{-3}),
\end{split}
\end{equation}
Therefore we have 
\begin{equation*}
	\begin{pmatrix}
0 & 0 & 0 & \widehat{\mathscr{Y}}_{14}\\
0 & 0 & 0  & \widehat{\mathscr{Y}}_{24} \\
0 & 0 & 0  & \de \\
0 & 0 & 0  & \ga
\end{pmatrix} P^{-1}\begin{pmatrix}
0 & 0 & 0 & 0\\
0 & 0 & 0  & 0 \\
0 & 0 & 0  & 0 \\
0 & 0 & 0  & 1
\end{pmatrix} + \begin{pmatrix}
1 & 0 & \overset{\infty}{\mathscr{Y}}_{1,13} & 0\\
0 & 1 & \overset{\infty}{\mathscr{Y}}_{1,23} & 0 \\
-\overset{\infty}{\mathscr{X}}_{1,31} & -\overset{\infty}{\mathscr{X}}_{1,32} &  \overset{\infty}{\mathscr{Y}}_{1,33} - \overset{\infty}{\mathscr{X}}_{1,33} & - \overset{\infty}{\mathscr{X}}_{1,34} \\
0 & 0 & \overset{\infty}{\mathscr{Y}}_{1,43} & 0
\end{pmatrix}  \begin{pmatrix}
1 & 0 & 0 & 0\\
0 & 1 & 0  & 0 \\
0 & 0 & 0  & 0 \\
0 & 0 & 0  & 0
\end{pmatrix} 
\end{equation*}
\begin{equation*}
 + \begin{pmatrix}
1 & 0 & \overset{\infty}{\mathscr{Y}}_{1,13} & 0\\
0 & 1 & \overset{\infty}{\mathscr{Y}}_{1,23} & 0 \\
-\overset{\infty}{\mathscr{X}}_{1,31} & -\overset{\infty}{\mathscr{X}}_{1,32} &  \overset{\infty}{\mathscr{Y}}_{1,33} - \overset{\infty}{\mathscr{X}}_{1,33} & - \overset{\infty}{\mathscr{X}}_{1,34} \\
0 & 0 & \overset{\infty}{\mathscr{Y}}_{1,43} & 0
\end{pmatrix} \overset{\infty}{\mathscr{X}}_1  \begin{pmatrix}
0 & 0 & 0 & 0\\
0 & 0 & 0  & 0 \\
0 & 0 & 0  & 0 \\
0 & 0 & 0  & 1
\end{pmatrix} 
\end{equation*}
\begin{equation}\label{Eqn}
 +  \begin{pmatrix}
0 & 0 & 0 & 0\\
0 & 0 & 0  & 0 \\
0 & 0 & 1  & 0 \\
0 & 0 & 0  & 0
\end{pmatrix}  + \begin{pmatrix}
0 & 0 & 0 & 0\\
0 & 0 & 0  & 0 \\
0 & 0 & 1  & 0 \\
0 & 0 & 0  & 0
\end{pmatrix} \overset{\infty}{\mathscr{X}}_1 \begin{pmatrix}
1 & 0 & 0 & 0\\
0 & 1 & 0  & 0 \\
0 & 0 & 0  & 0 \\
0 & 0 & 0  & 0
\end{pmatrix}+ \begin{pmatrix}
0 & 0 & 0 & 0\\
0 & 0 & 0  & 0 \\
0 & 0 & 1  & 0 \\
0 & 0 & 0  & 0
\end{pmatrix} \overset{\infty}{\mathscr{X}}_2 \begin{pmatrix}
0 & 0 & 0 & 0\\
0 & 0 & 0  & 0 \\
0 & 0 & 0  & 0 \\
0 & 0 & 0  & 1
\end{pmatrix} \equiv I.
\end{equation}
Simplifying \eqref{Eqn} we get:
\begin{equation}\label{Eqn1}
\begin{split}
&	\begin{pmatrix}
0 & 0 & 0 & \widehat{\mathscr{Y}}_{14} P_{33}\\
0 & 0 & 0  & \widehat{\mathscr{Y}}_{24} P_{33} \\
0 & 0 & 0  & \de P_{33}\\
0 & 0 & 0  & \ga P_{33}
\end{pmatrix}  + \begin{pmatrix}
1 & 0 & 0 & 0\\
0 & 1 & 0 & 0 \\
-\overset{\infty}{\mathscr{X}}_{1,31} & -\overset{\infty}{\mathscr{X}}_{1,32} &  0 & 0 \\
0 & 0 & 0 & 0
\end{pmatrix} +  \begin{pmatrix}
0 & 0 & 0 & 0\\
0 & 0 & 0  & 0 \\
0 & 0 & 1  & 0 \\
0 & 0 & 0  & 0
\end{pmatrix}  + \begin{pmatrix}
0 & 0 & 0 & 0\\
0 & 0 & 0  & 0 \\
\overset{\infty}{\mathscr{X}}_{1,31} & \overset{\infty}{\mathscr{X}}_{1,32} & 0  & 0 \\
0 & 0 & 0  & 0
\end{pmatrix}  \\ &  + \begin{pmatrix}
1 & 0 & \overset{\infty}{\mathscr{Y}}_{1,13} & 0\\
0 & 1 & \overset{\infty}{\mathscr{Y}}_{1,23} & 0 \\
-\overset{\infty}{\mathscr{X}}_{1,31} & -\overset{\infty}{\mathscr{X}}_{1,32} &  \overset{\infty}{\mathscr{Y}}_{1,33} - \overset{\infty}{\mathscr{X}}_{1,33} & - \overset{\infty}{\mathscr{X}}_{1,34} \\
0 & 0 & \overset{\infty}{\mathscr{Y}}_{1,43} & 0
\end{pmatrix} \begin{pmatrix}
0 & 0 & 0 & \overset{\infty}{\mathscr{X}}_{1,14}\\
0 & 0 & 0 & \overset{\infty}{\mathscr{X}}_{1,24} \\
0 & 0 & 0  & \overset{\infty}{\mathscr{X}}_{1,34} \\
0 & 0 & 0 & \overset{\infty}{\mathscr{X}}_{1,44} \\
\end{pmatrix} + \begin{pmatrix}
0 & 0 & 0 & 0\\
0 & 0 & 0  & 0 \\
0 & 0 & 0 & \overset{\infty}{\mathscr{X}}_{2,34} \\
0 & 0 & 0  & 0
\end{pmatrix} \equiv I,
\end{split}
\end{equation}
where we have used $\left(P^{-1}\right)_{44} = P_{33}$, which is due to the representation of $P^{-1}$ as $W P W$ with \[ W =  \begin{pmatrix}
0 & 1 & 0 & 0\\
1 & 0 & 0  & 0 \\
0 & 0 & 0  & 1 \\
0 & 0 & 1  & 0
\end{pmatrix}, \] given by part (b) of Theorem \ref{Thm WW}, also see \cite{GI}. Combining terms and simplifying \eqref{Eqn1} we find
\begin{equation}\label{Eqn1111}
\begin{split}
&	\begin{pmatrix}
1 & 0 & 0 &  P_{33} \widehat{\mathscr{Y}}_{14}  + \overset{\infty}{\mathscr{X}}_{1,34} \overset{\infty}{\mathscr{Y}}_{1,13} + \overset{\infty}{\mathscr{X}}_{1,14} \\
0 & 1 & 0  &  P_{33} \widehat{\mathscr{Y}}_{24} + \overset{\infty}{\mathscr{X}}_{1,34} \overset{\infty}{\mathscr{Y}}_{1,23} + \overset{\infty}{\mathscr{X}}_{1,24} \\
0 & 0 & 1  &  P_{33} \de + \overset{\infty}{\mathscr{X}}_{1,34} \overset{\infty}{\mathscr{Y}}_{1,33} + \overset{\infty}{\mathscr{X}}_{2,34} - \overset{\infty}{\mathscr{X}}_{1,31} \overset{\infty}{\mathscr{X}}_{1,14} - \overset{\infty}{\mathscr{X}}_{1,32} \overset{\infty}{\mathscr{X}}_{1,24}  - \overset{\infty}{\mathscr{X}}_{1,34}\overset{\infty}{\mathscr{X}}_{1,33} - \overset{\infty}{\mathscr{X}}_{1,34}\overset{\infty}{\mathscr{X}}_{1,44} \\
0 & 0 & 0  & P_{33} \ga  + \overset{\infty}{\mathscr{X}}_{1,34} \overset{\infty}{\mathscr{Y}}_{1,43} 
\end{pmatrix}     \equiv I.
\end{split}
\end{equation}
This gives four linear equations in the eight unknowns $\{ \widehat{\mathscr{Y}}_{j4} , \overset{\infty}{\mathscr{Y}}_{1,j3} \}^4_{j=1}$:
\begin{align}
	& P_{33} \widehat{\mathscr{Y}}_{14}  + \overset{\infty}{\mathscr{X}}_{1,34} \overset{\infty}{\mathscr{Y}}_{1,13} + \overset{\infty}{\mathscr{X}}_{1,14}  = 0, \label{1} \\
	& P_{33} \widehat{\mathscr{Y}}_{24} + \overset{\infty}{\mathscr{X}}_{1,34} \overset{\infty}{\mathscr{Y}}_{1,23} + \overset{\infty}{\mathscr{X}}_{1,24}  = 0, \label{2} \\
	& P_{33} \de + \overset{\infty}{\mathscr{X}}_{1,34} \overset{\infty}{\mathscr{Y}}_{1,33} + \overset{\infty}{\mathscr{X}}_{2,34} - \overset{\infty}{\mathscr{X}}_{1,31} \overset{\infty}{\mathscr{X}}_{1,14} - \overset{\infty}{\mathscr{X}}_{1,32} \overset{\infty}{\mathscr{X}}_{1,24}  - \overset{\infty}{\mathscr{X}}_{1,34}\overset{\infty}{\mathscr{X}}_{1,33} - \overset{\infty}{\mathscr{X}}_{1,34}\overset{\infty}{\mathscr{X}}_{1,44}  = 0, \label{3} \\
	& P_{33} \ga  + \overset{\infty}{\mathscr{X}}_{1,34} \overset{\infty}{\mathscr{Y}}_{1,43}  = 1. \label{4}
\end{align}
To find the complementary equations we consider the behavior of \eqref{Y X} near zero. Using \textbf{RH-$\mathscr{X}$4} we have as $z \to 0$
\begin{equation}\label{Y X zero}
\begin{split}
\mathscr{Y}(z&;n)  = \left( \frac{1}{z} \begin{pmatrix}
0 & 0 & 0 & \widehat{\mathscr{Y}}_{14}\\
0 & 0 & 0  & \widehat{\mathscr{Y}}_{24} \\
0 & 0 & 0  & \de \\
0 & 0 & 0  & \ga
\end{pmatrix} P^{-1}(n) +  \begin{pmatrix}
1 & 0 & \overset{\infty}{\mathscr{Y}}_{1,13} & 0\\
0 & 1 & \overset{\infty}{\mathscr{Y}}_{1,23} & 0 \\
-\overset{\infty}{\mathscr{X}}_{1,31} & -\overset{\infty}{\mathscr{X}}_{1,32} &  \overset{\infty}{\mathscr{Y}}_{1,33} - \overset{\infty}{\mathscr{X}}_{1,33} & - \overset{\infty}{\mathscr{X}}_{1,34} \\
0 & 0 & \overset{\infty}{\mathscr{Y}}_{1,43} & 0
\end{pmatrix} + z \begin{pmatrix}
0 & 0 & 0 & 0\\
0 & 0 & 0  & 0 \\
0 & 0 & 1  & 0 \\
0 & 0 & 0  & 0
\end{pmatrix} \right) \\ & \times P(n)\left(I+\overset{\circ}{  \mathscr{X}}_1z+\overset{\circ}{  \mathscr{X}}_2z^2+O(z^3)\right) \begin{pmatrix}
1 & 0 & 0 & 0\\
0 & 1 & 0 & 0 \\
0 & 0 & z^{-1} & 0 \\
0 & 0 & 0 & z
\end{pmatrix}  \begin{pmatrix}
1 & 0 & 0 & 0\\
0 & z^{-n} & 0 & 0 \\
0 & 0 & 1 & 0 \\
0 & 0 & 0 & z^n
\end{pmatrix} \\ & 
= \left( \frac{1}{z} \begin{pmatrix}
0 & 0 & 0 & \widehat{\mathscr{Y}}_{14}\\
0 & 0 & 0  & \widehat{\mathscr{Y}}_{24} \\
0 & 0 & 0  & \de \\
0 & 0 & 0  & \ga
\end{pmatrix} +  \begin{pmatrix}
1 & 0 & \overset{\infty}{\mathscr{Y}}_{1,13} & 0\\
0 & 1 & \overset{\infty}{\mathscr{Y}}_{1,23} & 0 \\
-\overset{\infty}{\mathscr{X}}_{1,31} & -\overset{\infty}{\mathscr{X}}_{1,32} &  \overset{\infty}{\mathscr{Y}}_{1,33} - \overset{\infty}{\mathscr{X}}_{1,33} & - \overset{\infty}{\mathscr{X}}_{1,34} \\
0 & 0 & \overset{\infty}{\mathscr{Y}}_{1,43} & 0
\end{pmatrix} P(n) + z \begin{pmatrix}
0 & 0 & 0 & 0\\
0 & 0 & 0  & 0 \\
0 & 0 & 1  & 0 \\
0 & 0 & 0  & 0
\end{pmatrix}P(n) \right) \\ & \times \left(I+\overset{\circ}{  \mathscr{X}}_1z+\overset{\circ}{  \mathscr{X}}_2z^2+O(z^3)\right) \begin{pmatrix}
1 & 0 & 0 & 0\\
0 & 1 & 0 & 0 \\
0 & 0 & z^{-1} & 0 \\
0 & 0 & 0 & z
\end{pmatrix}  \begin{pmatrix}
1 & 0 & 0 & 0\\
0 & z^{-n} & 0 & 0 \\
0 & 0 & 1 & 0 \\
0 & 0 & 0 & z^n
\end{pmatrix}.
\end{split}
\end{equation}
The coefficient of $z^{-2}$ in the above expression is \[ \begin{pmatrix}
0 & 0 & 0 & \widehat{\mathscr{Y}}_{14}\\
0 & 0 & 0  & \widehat{\mathscr{Y}}_{24} \\
0 & 0 & 0  & \de \\
0 & 0 & 0  & \ga
\end{pmatrix} \begin{pmatrix}
0 & 0 & 0 & 0\\
0 & 0 & 0  & 0 \\
0 & 0 & 1  & 0 \\
0 & 0 & 0  & 0
\end{pmatrix} = 0.  \]
The coefficient of $z^{-1}$ in the \eqref{Y X zero} is
\[ \begin{pmatrix}
0 & 0 &  \overset{\circ}{  \mathscr{X}}_{1,43} \widehat{\mathscr{Y}}_{14} + P_{33}\overset{\infty}{\mathscr{Y}}_{1,13} +  P_{13} & 0\\
0 & 0 & \overset{\circ}{  \mathscr{X}}_{1,43} \widehat{\mathscr{Y}}_{24}   + P_{33}\overset{\infty}{\mathscr{Y}}_{1,23} + P_{23}  & 0 \\
0 & 0 &   \overset{\circ}{  \mathscr{X}}_{1,43} \de + P_{33}\overset{\infty}{\mathscr{Y}}_{1,33}   -\overset{\infty}{\mathscr{X}}_{1,31}P_{13} -\overset{\infty}{\mathscr{X}}_{1,32}P_{23}-\overset{\infty}{\mathscr{X}}_{1,33}P_{33}-\overset{\infty}{\mathscr{X}}_{1,34}P_{43}  & 0 \\
0 & 0 & \overset{\circ}{  \mathscr{X}}_{1,43} \widehat{\mathscr{Y}}_{44} + P_{33}\overset{\infty}{\mathscr{Y}}_{1,43}    & 0
\end{pmatrix}, \]
which we need to set equal to zero according to \textbf{RH-$\mathscr{Y}$4}:
\begin{align}
& \overset{\circ}{  \mathscr{X}}_{1,43} \widehat{\mathscr{Y}}_{14} + P_{33}\overset{\infty}{\mathscr{Y}}_{1,13} +  P_{13}  = 0, \label{5} \\
& \overset{\circ}{  \mathscr{X}}_{1,43} \widehat{\mathscr{Y}}_{24}   + P_{33}\overset{\infty}{\mathscr{Y}}_{1,23} + P_{23}  = 0, \label{6} \\
& \overset{\circ}{  \mathscr{X}}_{1,43} \de + P_{33}\overset{\infty}{\mathscr{Y}}_{1,33}   -\overset{\infty}{\mathscr{X}}_{1,31}P_{13} -\overset{\infty}{\mathscr{X}}_{1,32}P_{23}-\overset{\infty}{\mathscr{X}}_{1,33}P_{33}-\overset{\infty}{\mathscr{X}}_{1,34}P_{43}  = 0, \label{7} \\
& \overset{\circ}{  \mathscr{X}}_{1,43} \widehat{\mathscr{Y}}_{44} + P_{33}\overset{\infty}{\mathscr{Y}}_{1,43}  = 0, \label{8}
\end{align}
Notice that the 8 equations \eqref{1}-\eqref{4} and \eqref{5}-\eqref{8}  \textit{decouples} into four sets of two equations in two unknowns. Solving these equations we will find formulae for $\widehat{\mathscr{Y}}_{jk}$ as given in Theorem  \ref{thm r=s=0}.

Finally, using \eqref{RR} and $P^{-1}=W P W$ with \[ W =  \begin{pmatrix}
0 & 1 & 0 & 0\\
1 & 0 & 0  & 0 \\
0 & 0 & 0  & 1 \\
0 & 0 & 1  & 0
\end{pmatrix}, \]
we arrived at the desired explicit representation of $\mathscr{R}$ in terms of the data from the solution of the Riemann-Hilbert problem \textbf{RH-$\mathscr{X}$1} through \textbf{RH-$\mathscr{X}$4}:

\begin{equation}
	\begin{split}
		\mathscr{R}(z) & =  \frac{1}{z} \begin{pmatrix}
			\widehat{\mathscr{Y}}_{14}  P_{32} & \widehat{\mathscr{Y}}_{14}  P_{31} & \widehat{\mathscr{Y}}_{14}  P_{34} & \widehat{\mathscr{Y}}_{14}  P_{33} \\
			\widehat{\mathscr{Y}}_{24}  P_{32} & \widehat{\mathscr{Y}}_{24}  P_{31} & \widehat{\mathscr{Y}}_{24}  P_{34} & \widehat{\mathscr{Y}}_{24}  P_{33} \\
			\widehat{\mathscr{Y}}_{34}  P_{32} & \widehat{\mathscr{Y}}_{34}  P_{31} & \widehat{\mathscr{Y}}_{34}  P_{34} & \widehat{\mathscr{Y}}_{34}  P_{33} \\
			\widehat{\mathscr{Y}}_{44}  P_{32} & \widehat{\mathscr{Y}}_{44}  P_{31} & \widehat{\mathscr{Y}}_{44}  P_{34} & \widehat{\mathscr{Y}}_{44}  P_{33} \\
		\end{pmatrix} + \begin{pmatrix}
			1 & 0 & \overset{\infty}{\mathscr{Y}}_{1,13} & 0\\
			0 & 1 & \overset{\infty}{\mathscr{Y}}_{1,23} & 0 \\
			-\overset{\infty}{\mathscr{X}}_{1,31} & -\overset{\infty}{\mathscr{X}}_{1,32} &  \overset{\infty}{\mathscr{Y}}_{1,33} - \overset{\infty}{\mathscr{X}}_{1,33} & - \overset{\infty}{\mathscr{X}}_{1,34} \\
			0 & 0 & \overset{\infty}{\mathscr{Y}}_{1,43} & 0
		\end{pmatrix} \\ & + z \begin{pmatrix}
			0 & 0 & 0 & 0\\
			0 & 0 & 0  & 0 \\
			0 & 0 & 1  & 0 \\
			0 & 0 & 0  & 0
		\end{pmatrix}.
	\end{split}
\end{equation}

\section{Proof of Theorem \ref{thm r=0 s=2}}\label{sec thm r=0 s=2} 

For $r=0$ and $s=2$, recall that we denote $X(z;n,0,2)$ by $\mathscr T(z;n)$, which satisfies
\begin{itemize}
	\item \textbf{RH-$\mathscr{T}$1} \quad  $\mathscr{T}(\cdot;n)$ is holomorphic in the complement of $\T \cup \{0\}$.
	
	\item \textbf{RH-$\mathscr{T}$2} \quad   For $z \in \T$, $\mathscr{T}$ satisfies \begin{equation*}\label{hatW-jumpa}
	\mathscr{T}_+(z;n)=\mathscr{T}_-(z;n) \begin{pmatrix}
	1 & 0 & z \tilde{w}(z) & - z \phi(z) \\
	0 & 1 & z^{-1} \tilde{\phi}(z) & - z^{-1} w(z) \\
	0 & 0 & 1 & 0 \\
	0 & 0 & 0 & 1
	\end{pmatrix}, 
	\end{equation*} 
	
	\item \textbf{RH-$\mathscr{T}$3}  \quad As $z \to \infty$ we have \begin{equation*}\label{hatXinfinityaa}
	\mathscr{T}(z;n)=\left(\di I+\frac{\overset{\infty}{\mathscr{T}}_{1}}{z} +\frac{\overset{\infty}{\mathscr{T}}_{2}}{z^2} + O(z^{-3})\right)\begin{pmatrix}
	z^n & 0 & 0 & 0\\
	0 & 1 & 0 & 0 \\
	0 & 0 & z^{-n} & 0 \\
	0 & 0 & 0 & 1
	\end{pmatrix},  
	\end{equation*} 
	\item \textbf{RH-$\mathscr{T}$4} \quad As $z \to 0$ we have
	\begin{equation*}\label{hatXzeroaa}
	\mathscr{T}(z;n)=\widehat{\mathscr{T}}\left(I+\overset{\circ}{\mathscr{T}}_{1} z +\overset{\circ}{\mathscr{T}}_{2} z^2 + O(z^{3})\right)\begin{pmatrix}
	1 & 0 & 0 & 0\\
	0 & z^{-n} & 0 & 0 \\
	0 & 0 & 1 & 0 \\
	0 & 0 & 0 & z^n
	\end{pmatrix}.
	\end{equation*}
\end{itemize}
Let us recall \eqref{X R X} in this case
\begin{equation}\label{T R X}
\mathscr{T}(z)= R(z)\mathscr{X}(z) \begin{pmatrix}
z^{-2} & 0 & 0 & 0\\
0 & 1 & 0 & 0 \\
0 & 0 & z^{-1} & 0 \\
0 & 0 & 0 & z^{-1}
\end{pmatrix}.
\end{equation}
Let us also recall \eqref{V} in this case
\begin{equation}\label{VV}
\mathscr{V}(z) = \mathscr{T}(z)  \begin{pmatrix}
z^{2} & 0 & 0 & 0\\
0 & 1 & 0 & 0 \\
0 & 0 & z & 0 \\
0 & 0 & 0 & z
\end{pmatrix},
\end{equation}
Behavior of $\mathscr V$ as $z \to \infty$
\begin{equation}\label{Tinfinitya}
\mathscr{V}(z)=\left(\di I+\frac{\overset{\infty}{\mathscr{T}}_{1}}{z} +\frac{\overset{\infty}{\mathscr{T}}_{2}}{z^2} + O(z^{-3}) \right)\begin{pmatrix}
z^n & 0 & 0 & 0\\
0 & 1 & 0 & 0 \\
0 & 0 & z^{-n} & 0 \\
0 & 0 & 0 & 1
\end{pmatrix} \begin{pmatrix}
z^{2} & 0 & 0 & 0\\
0 & 1 & 0 & 0 \\
0 & 0 & z & 0 \\
0 & 0 & 0 & z
\end{pmatrix},  
\end{equation} 
Behavior of $\mathscr V$ as $z \to 0$
\begin{equation}\label{Vzeroa}
\mathscr{V}(z)=\widehat{\mathscr T}\left( I+\overset{\circ}{\mathscr{T}}_{1} z +\overset{\circ}{\mathscr{T}}_{2} z^2 + O(z^{3}) \right)\begin{pmatrix}
1 & 0 & 0 & 0\\
0 & z^{-n} & 0 & 0 \\
0 & 0 & 1 & 0 \\
0 & 0 & 0 & z^n
\end{pmatrix} \begin{pmatrix}
z^{2} & 0 & 0 & 0\\
0 & 1 & 0 & 0 \\
0 & 0 & z & 0 \\
0 & 0 & 0 & z
\end{pmatrix}.
\end{equation}
Notice that $R=\mathscr{V} \mathscr{X}^{-1}$ is an entire function, due to \eqref{Vzeroa} and \textbf{RH-$\mathscr{X}$4}. Behavior of $R = \mathscr{V} \mathscr{X}^{-1}$ as $z \to \infty$ is given by
\begin{equation}
\begin{split}
	R(z) & = z^2 \begin{pmatrix}
	1 & 0 & 0 & 0\\
	0 & 0 & 0 & 0 \\
	0 & 0 & 0 & 0 \\
	0 & 0 & 0 & 0
	\end{pmatrix}   + z \left\{ \begin{pmatrix}
	0 & 0 & 0 & 0\\
	0 & 0 & 0 & 0 \\
	0 & 0 & 1 & 0 \\
	0 & 0 & 0 & 1
	\end{pmatrix} + \overset{\infty}{\mathscr{T}}_{1} \begin{pmatrix}
	1 & 0 & 0 & 0\\
	0 & 0 & 0 & 0 \\
	0 & 0 & 0 & 0 \\
	0 & 0 & 0 & 0
	\end{pmatrix} - \begin{pmatrix}
	1 & 0 & 0 & 0\\
	0 & 0 & 0 & 0 \\
	0 & 0 & 0 & 0 \\
	0 & 0 & 0 & 0
	\end{pmatrix} \overset{\infty}{\mathscr{X}}_{1} \right\} \\ & + \begin{pmatrix}
	0 & 0 & 0 & 0\\
	0 & 1 & 0 & 0 \\
	0 & 0 & 0 & 0 \\
	0 & 0 & 0 & 0
	\end{pmatrix}  
	- \begin{pmatrix}
	0 & 0 & 0 & 0\\
	0 & 0 & 0 & 0 \\
	0 & 0 & 1 & 0 \\
	0 & 0 & 0 & 1
	\end{pmatrix} \overset{\infty}{\mathscr{X}}_{1}  + \overset{\infty}{\mathscr{T}}_{1} \begin{pmatrix}
	0 & 0 & 0 & 0\\
	0 & 0 & 0 & 0 \\
	0 & 0 & 1 & 0 \\
	0 & 0 & 0 & 1
	\end{pmatrix} - \overset{\infty}{\mathscr{T}}_{1} \begin{pmatrix}
	1 & 0 & 0 & 0\\
	0 & 0 & 0 & 0 \\
	0 & 0 & 0 & 0 \\
	0 & 0 & 0 & 0
	\end{pmatrix}  \overset{\infty}{\mathscr{X}}_{1} \\ & +  \overset{\infty}{\mathscr{T}}_{2} \begin{pmatrix}
	1 & 0 & 0 & 0\\
	0 & 0 & 0 & 0 \\
	0 & 0 & 0 & 0 \\
	0 & 0 & 0 & 0
	\end{pmatrix} + \begin{pmatrix}
	1 & 0 & 0 & 0\\
	0 & 0 & 0 & 0 \\
	0 & 0 & 0 & 0 \\
	0 & 0 & 0 & 0
	\end{pmatrix} \left( \overset{\infty}{\mathscr{X}}_1^2 - \overset{\infty}{  \mathscr{X}}_2 \right).
	\end{split}
\end{equation}
Further simplification gives
\begin{equation}\label{R E B}
\begin{split}
R(z) & = z^2 \begin{pmatrix}
1 & 0 & 0 & 0\\
0 & 0 & 0 & 0 \\
0 & 0 & 0 & 0 \\
0 & 0 & 0 & 0
\end{pmatrix}   + z  \begin{pmatrix}
\overset{\infty}{\mathscr{T}}_{1,11}- \overset{\infty}{\mathscr{X}}_{1,11} & -\overset{\infty}{\mathscr{X}}_{1,12} & -\overset{\infty}{\mathscr{X}}_{1,13} & -\overset{\infty}{\mathscr{X}}_{1,14}\\
\overset{\infty}{\mathscr{T}}_{1,21} & 0 & 0 & 0 \\
\overset{\infty}{\mathscr{T}}_{1,31} & 0 & 1 & 0 \\
\overset{\infty}{\mathscr{T}}_{1,41} & 0 & 0 & 1
\end{pmatrix}    \\ & + \begin{pmatrix}
\overset{\infty}{\mathscr{T}}_{2,11} + \left[ \overset{\infty}{\mathscr{X}}_1^2 \right]_{11} - \overset{\infty}{  \mathscr{X}}_{2,11} &  \left[ \overset{\infty}{\mathscr{X}}_1^2 \right]_{12} - \overset{\infty}{  \mathscr{X}}_{2,12} &  \overset{\infty}{\mathscr{T}}_{1,13} + \left[ \overset{\infty}{\mathscr{X}}_1^2 \right]_{13} - \overset{\infty}{  \mathscr{X}}_{2,13} & \overset{\infty}{\mathscr{T}}_{1,14} + \left[ \overset{\infty}{\mathscr{X}}_1^2 \right]_{14} - \overset{\infty}{  \mathscr{X}}_{2,14} \\
\overset{\infty}{\mathscr{T}}_{2,21} & 1 & \overset{\infty}{\mathscr{T}}_{1,23} & \overset{\infty}{\mathscr{T}}_{1,24} \\
\overset{\infty}{\mathscr{T}}_{2,31}- \overset{\infty}{\mathscr{X}}_{1,31} & - \overset{\infty}{\mathscr{X}}_{1,32} & \overset{\infty}{\mathscr{T}}_{1,33}- \overset{\infty}{\mathscr{X}}_{1,33} & \overset{\infty}{\mathscr{T}}_{1,34} - \overset{\infty}{\mathscr{X}}_{1,34} \\
\overset{\infty}{\mathscr{T}}_{2,41}- \overset{\infty}{\mathscr{X}}_{1,41} & - \overset{\infty}{\mathscr{X}}_{1,42} & \overset{\infty}{\mathscr{T}}_{1,43} - \overset{\infty}{\mathscr{X}}_{1,43} & \overset{\infty}{\mathscr{T}}_{1,44} - \overset{\infty}{\mathscr{X}}_{1,44}
\end{pmatrix}   \\ &  - \begin{pmatrix}
\overset{\infty}{\mathscr{T}}_{1,11}\overset{\infty}{\mathscr{X}}_{1,11} &  \overset{\infty}{\mathscr{T}}_{1,11}\overset{\infty}{\mathscr{X}}_{1,12} &  \overset{\infty}{\mathscr{T}}_{1,11}\overset{\infty}{\mathscr{X}}_{1,13} &  \overset{\infty}{\mathscr{T}}_{1,11}\overset{\infty}{\mathscr{X}}_{1,14}\\
\overset{\infty}{\mathscr{T}}_{1,21}\overset{\infty}{\mathscr{X}}_{1,11} &  \overset{\infty}{\mathscr{T}}_{1,21}\overset{\infty}{\mathscr{X}}_{1,12} &  \overset{\infty}{\mathscr{T}}_{1,21}\overset{\infty}{\mathscr{X}}_{1,13} &  \overset{\infty}{\mathscr{T}}_{1,21}\overset{\infty}{\mathscr{X}}_{1,14}\\
\overset{\infty}{\mathscr{T}}_{1,31}\overset{\infty}{\mathscr{X}}_{1,11} &  \overset{\infty}{\mathscr{T}}_{1,31}\overset{\infty}{\mathscr{X}}_{1,12} &  \overset{\infty}{\mathscr{T}}_{1,31}\overset{\infty}{\mathscr{X}}_{1,13} &  \overset{\infty}{\mathscr{T}}_{1,31}\overset{\infty}{\mathscr{X}}_{1,14}\\
\overset{\infty}{\mathscr{T}}_{1,41}\overset{\infty}{\mathscr{X}}_{1,11} &  \overset{\infty}{\mathscr{T}}_{1,41}\overset{\infty}{\mathscr{X}}_{1,12} &  \overset{\infty}{\mathscr{T}}_{1,41}\overset{\infty}{\mathscr{X}}_{1,13} &  \overset{\infty}{\mathscr{T}}_{1,41}\overset{\infty}{\mathscr{X}}_{1,14}
\end{pmatrix}  \equiv  z^2 \begin{pmatrix}
1 & 0 & 0 & 0\\
0 & 0 & 0 & 0 \\
0 & 0 & 0 & 0 \\
0 & 0 & 0 & 0
\end{pmatrix} + z E + B.
\end{split}
\end{equation}
To completely determine $R(z)$, in this case we need to  find the sixteen unknowns $\left\{ \overset{\infty}{\mathscr{T}}_{1,j1}, \overset{\infty}{\mathscr{T}}_{1,j3}, \overset{\infty}{\mathscr{T}}_{1,j4}, \overset{\infty}{\mathscr{T}}_{2,j1} \right\}^{4}_{j=1}$  in terms of the data from the solution of the Riemann-Hilbert problem \textbf{RH-$\mathscr{X}$1} through \textbf{RH-$\mathscr{X}$4}. To this end, we substitute the above expression for $R$ into \eqref{T R X} and compare the resulting asymptotics of $\mathscr T$, as $z \to \infty$ and as $z \to 0$, respectively with \textbf{RH-$\mathscr{T}$3}  and \textbf{RH-$\mathscr{T}$4}.  Similar to the proof of  Theorem \ref{thm r=0 s=1} we conclude that \textbf{RH-$\mathscr{T}$3} is satisfied automatically
because of the structure of  $R(z)$, and we only have to take care of  \textbf{RH-$\mathscr{T}$4}.  
It turns out, as shown below, that all sixteen equations to determine the sixteen unknowns come from the facts that there are no terms in $$\mathscr T(z;n) \begin{pmatrix}
1 & 0 & 0 & 0\\
0 & z^{n} & 0 & 0 \\
0 & 0 & 1 & 0 \\
0 & 0 & 0 & z^{-n}
\end{pmatrix} $$ with $z^{-1}$ and $z^{-2}$ as required by \textbf{RH-$\mathscr{T}$4}. Indeed,

\begin{equation}
\begin{split}
&	\mathscr{T}(z) \begin{pmatrix}
1 & 0 & 0 & 0\\
0 & z^{n} & 0 & 0 \\
0 & 0 & 1 & 0 \\
0 & 0 & 0 & z^{-n}
\end{pmatrix} = R(z)\mathscr{X}(z) \begin{pmatrix}
z^{-2} & 0 & 0 & 0\\
0 & 1 & 0 & 0 \\
0 & 0 & z^{-1} & 0 \\
0 & 0 & 0 & z^{-1}
\end{pmatrix} \\ & = \left(z^2 \begin{pmatrix}
1 & 0 & 0 & 0\\
0 & 0 & 0 & 0 \\
0 & 0 & 0 & 0 \\
0 & 0 & 0 & 0
\end{pmatrix} + z E + B \right) P\left(I+\overset{\circ}{  \mathscr{X}}_1z+\overset{\circ}{  \mathscr{X}}_2z^2+O(z^3)\right) \begin{pmatrix}
z^{-2} & 0 & 0 & 0\\
0 & 1 & 0 & 0 \\
0 & 0 & z^{-1} & 0 \\
0 & 0 & 0 & z^{-1}
\end{pmatrix} \\ & = \left(z^2 \begin{pmatrix}
1 & 0 & 0 & 0\\
0 & 0 & 0 & 0 \\
0 & 0 & 0 & 0 \\
0 & 0 & 0 & 0
\end{pmatrix} + z E + B \right) \left(P+P\overset{\circ}{  \mathscr{X}}_1z+P\overset{\circ}{  \mathscr{X}}_2z^2+O(z^3)\right) \\ & \times \left( z^{-2}\begin{pmatrix}
1 & 0 & 0 & 0\\
0 & 0 & 0 & 0 \\
0 & 0 & 0 & 0 \\
0 & 0 & 0 & 0
\end{pmatrix} + z^{-1}\begin{pmatrix}
0 & 0 & 0 & 0\\
0 & 0 & 0 & 0 \\
0 & 0 & 1 & 0 \\
0 & 0 & 0 & 1
\end{pmatrix} + \begin{pmatrix}
0 & 0 & 0 & 0\\
0 & 1 & 0 & 0 \\
0 & 0 & 0 & 0 \\
0 & 0 & 0 & 0
\end{pmatrix} \right). 
\end{split}
\end{equation}
The equations corresponding to the coefficients of of $z^{-2}$ and $z^{-1}$, in the above expression in view of \textbf{RH-$\mathscr{T}$4} are 
\begin{equation}
	BP\begin{pmatrix}
	1 & 0 & 0 & 0\\
	0 & 0 & 0 & 0 \\
	0 & 0 & 0 & 0 \\
	0 & 0 & 0 & 0
	\end{pmatrix} = 0,
\end{equation} and
\begin{equation}\label{410}
BP\begin{pmatrix}
0 & 0 & 0 & 0\\
0 & 0 & 0 & 0 \\
0 & 0 & 1 & 0 \\
0 & 0 & 0 & 1
\end{pmatrix} + \left(  BP\overset{\circ}{\mathscr{X}}_1 + EP \right) \begin{pmatrix}
1 & 0 & 0 & 0\\
0 & 0 & 0 & 0 \\
0 & 0 & 0 & 0 \\
0 & 0 & 0 & 0
\end{pmatrix} = 0,
\end{equation}
respectively. These equations imply that the first, third and fourth columns of $BP$ are zero. So we have the following twelve equations:

\begin{equation}\label{411}
\begin{split}
(BP)_{11} & = \left(	\overset{\infty}{\mathscr{T}}_{2,11} + \left[ \overset{\infty}{\mathscr{X}}_1^2 \right]_{11} - \overset{\infty}{  \mathscr{X}}_{2,11} -  \overset{\infty}{\mathscr{T}}_{1,11}\overset{\infty}{\mathscr{X}}_{1,11} \right) P_{11} + \left(  \left[ \overset{\infty}{\mathscr{X}}_1^2 \right]_{12} - \overset{\infty}{  \mathscr{X}}_{2,12} - \overset{\infty}{\mathscr{T}}_{1,11}\overset{\infty}{\mathscr{X}}_{1,12} \right) P_{21} \\ & + \left(  \overset{\infty}{\mathscr{T}}_{1,13} + \left[ \overset{\infty}{\mathscr{X}}_1^2 \right]_{13} - \overset{\infty}{  \mathscr{X}}_{2,13} - \overset{\infty}{\mathscr{T}}_{1,11}\overset{\infty}{\mathscr{X}}_{1,13} \right) P_{31} + \left( \overset{\infty}{\mathscr{T}}_{1,14} + \left[ \overset{\infty}{\mathscr{X}}_1^2 \right]_{14} - \overset{\infty}{  \mathscr{X}}_{2,14} - \overset{\infty}{\mathscr{T}}_{1,11}\overset{\infty}{\mathscr{X}}_{1,14} \right) P_{41} = 0,
\end{split}
\end{equation}

\begin{equation}\label{412}
\begin{split}
(BP)_{13} & = \left(	\overset{\infty}{\mathscr{T}}_{2,11} + \left[ \overset{\infty}{\mathscr{X}}_1^2 \right]_{11} - \overset{\infty}{  \mathscr{X}}_{2,11} -  \overset{\infty}{\mathscr{T}}_{1,11}\overset{\infty}{\mathscr{X}}_{1,11} \right) P_{13} + \left(  \left[ \overset{\infty}{\mathscr{X}}_1^2 \right]_{12} - \overset{\infty}{  \mathscr{X}}_{2,12} - \overset{\infty}{\mathscr{T}}_{1,11}\overset{\infty}{\mathscr{X}}_{1,12} \right) P_{23} \\ & + \left(  \overset{\infty}{\mathscr{T}}_{1,13} + \left[ \overset{\infty}{\mathscr{X}}_1^2 \right]_{13} - \overset{\infty}{  \mathscr{X}}_{2,13} - \overset{\infty}{\mathscr{T}}_{1,11}\overset{\infty}{\mathscr{X}}_{1,13} \right) P_{33} + \left( \overset{\infty}{\mathscr{T}}_{1,14} + \left[ \overset{\infty}{\mathscr{X}}_1^2 \right]_{14} - \overset{\infty}{  \mathscr{X}}_{2,14} - \overset{\infty}{\mathscr{T}}_{1,11}\overset{\infty}{\mathscr{X}}_{1,14} \right) P_{43} = 0,
\end{split}
\end{equation}

\begin{equation}\label{413}
\begin{split}
(BP)_{14} & = \left(	\overset{\infty}{\mathscr{T}}_{2,11} + \left[ \overset{\infty}{\mathscr{X}}_1^2 \right]_{11} - \overset{\infty}{  \mathscr{X}}_{2,11} -  \overset{\infty}{\mathscr{T}}_{1,11}\overset{\infty}{\mathscr{X}}_{1,11} \right) P_{14} + \left(  \left[ \overset{\infty}{\mathscr{X}}_1^2 \right]_{12} - \overset{\infty}{  \mathscr{X}}_{2,12} - \overset{\infty}{\mathscr{T}}_{1,11}\overset{\infty}{\mathscr{X}}_{1,12} \right) P_{24} \\ & + \left(  \overset{\infty}{\mathscr{T}}_{1,13} + \left[ \overset{\infty}{\mathscr{X}}_1^2 \right]_{13} - \overset{\infty}{  \mathscr{X}}_{2,13} - \overset{\infty}{\mathscr{T}}_{1,11}\overset{\infty}{\mathscr{X}}_{1,13} \right) P_{34} + \left( \overset{\infty}{\mathscr{T}}_{1,14} + \left[ \overset{\infty}{\mathscr{X}}_1^2 \right]_{14} - \overset{\infty}{  \mathscr{X}}_{2,14} - \overset{\infty}{\mathscr{T}}_{1,11}\overset{\infty}{\mathscr{X}}_{1,14} \right) P_{44} = 0,
\end{split}
\end{equation}

\begin{comment}

\red{The above equations are \textit{3 equations in 4 unknowns}:  \[ \overset{\infty}{\mathscr{T}}_{2,11}, \overset{\infty}{\mathscr{T}}_{1,11}, \overset{\infty}{\mathscr{T}}_{1,13}, \overset{\infty}{\mathscr{T}}_{1,14} \]}	content...
\end{comment}

\begin{equation}\label{414}
\begin{split}
(BP)_{21} & = \left(	\overset{\infty}{\mathscr{T}}_{2,21} -  \overset{\infty}{\mathscr{T}}_{1,21}\overset{\infty}{\mathscr{X}}_{1,11} \right) P_{11} + \left(  1 - \overset{\infty}{\mathscr{T}}_{1,21}\overset{\infty}{\mathscr{X}}_{1,12} \right) P_{21} + \left(  \overset{\infty}{\mathscr{T}}_{1,23}  - \overset{\infty}{\mathscr{T}}_{1,21}\overset{\infty}{\mathscr{X}}_{1,13} \right) P_{31} \\ & + \left( \overset{\infty}{\mathscr{T}}_{1,24}  - \overset{\infty}{\mathscr{T}}_{1,21}\overset{\infty}{\mathscr{X}}_{1,14} \right) P_{41} = 0,
\end{split}
\end{equation}

\begin{equation}\label{415}
\begin{split}
(BP)_{23} & = \left(	\overset{\infty}{\mathscr{T}}_{2,21} -  \overset{\infty}{\mathscr{T}}_{1,21}\overset{\infty}{\mathscr{X}}_{1,11} \right) P_{13} + \left(  1 - \overset{\infty}{\mathscr{T}}_{1,21}\overset{\infty}{\mathscr{X}}_{1,12} \right) P_{23} + \left(  \overset{\infty}{\mathscr{T}}_{1,23}  - \overset{\infty}{\mathscr{T}}_{1,21}\overset{\infty}{\mathscr{X}}_{1,13} \right) P_{33} \\&+ \left( \overset{\infty}{\mathscr{T}}_{1,24}  - \overset{\infty}{\mathscr{T}}_{1,21}\overset{\infty}{\mathscr{X}}_{1,14} \right) P_{43} = 0,
\end{split}
\end{equation}

\begin{equation}\label{416}
\begin{split}
(BP)_{24} & = \left(	\overset{\infty}{\mathscr{T}}_{2,21} -  \overset{\infty}{\mathscr{T}}_{1,21}\overset{\infty}{\mathscr{X}}_{1,11} \right) P_{14} + \left(  1 - \overset{\infty}{\mathscr{T}}_{1,21}\overset{\infty}{\mathscr{X}}_{1,12} \right) P_{24} + \left(  \overset{\infty}{\mathscr{T}}_{1,23}  - \overset{\infty}{\mathscr{T}}_{1,21}\overset{\infty}{\mathscr{X}}_{1,13} \right) P_{34}  \\&+ \left( \overset{\infty}{\mathscr{T}}_{1,24}  - \overset{\infty}{\mathscr{T}}_{1,21}\overset{\infty}{\mathscr{X}}_{1,14} \right) P_{44} = 0,
\end{split}
\end{equation}

\begin{comment}

\red{The above equations are \textit{3 equations in 4 unknowns}:  \[ \overset{\infty}{\mathscr{T}}_{2,21}, \overset{\infty}{\mathscr{T}}_{1,21}, \overset{\infty}{\mathscr{T}}_{1,23}, \overset{\infty}{\mathscr{T}}_{1,24} \]}	content...
\end{comment}

\begin{equation}\label{417}
\begin{split}
(BP)_{31} & = \left(	\overset{\infty}{\mathscr{T}}_{2,31} - \overset{\infty}{  \mathscr{X}}_{1,31} -  \overset{\infty}{\mathscr{T}}_{1,31}\overset{\infty}{\mathscr{X}}_{1,11} \right) P_{11} + \left( - \overset{\infty}{  \mathscr{X}}_{1,32} - \overset{\infty}{\mathscr{T}}_{1,31}\overset{\infty}{\mathscr{X}}_{1,12} \right) P_{21} \\ & + \left(  \overset{\infty}{\mathscr{T}}_{1,33}  - \overset{\infty}{  \mathscr{X}}_{1,33} - \overset{\infty}{\mathscr{T}}_{1,31}\overset{\infty}{\mathscr{X}}_{1,13} \right) P_{31} + \left( \overset{\infty}{\mathscr{T}}_{1,34} - \overset{\infty}{  \mathscr{X}}_{1,34} - \overset{\infty}{\mathscr{T}}_{1,31}\overset{\infty}{\mathscr{X}}_{1,14} \right) P_{41} = 0,
\end{split}
\end{equation}

\begin{equation}\label{418}
\begin{split}
(BP)_{33} & = \left(	\overset{\infty}{\mathscr{T}}_{2,31} - \overset{\infty}{  \mathscr{X}}_{1,31} -  \overset{\infty}{\mathscr{T}}_{1,31}\overset{\infty}{\mathscr{X}}_{1,11} \right) P_{13} + \left( - \overset{\infty}{  \mathscr{X}}_{1,32} - \overset{\infty}{\mathscr{T}}_{1,31}\overset{\infty}{\mathscr{X}}_{1,12} \right) P_{23} \\ & + \left(  \overset{\infty}{\mathscr{T}}_{1,33}  - \overset{\infty}{  \mathscr{X}}_{1,33} - \overset{\infty}{\mathscr{T}}_{1,31}\overset{\infty}{\mathscr{X}}_{1,13} \right) P_{33} + \left( \overset{\infty}{\mathscr{T}}_{1,34} - \overset{\infty}{  \mathscr{X}}_{1,34} - \overset{\infty}{\mathscr{T}}_{1,31}\overset{\infty}{\mathscr{X}}_{1,14} \right) P_{43} = 0,
\end{split}
\end{equation}

\begin{equation}\label{419}
\begin{split}
(BP)_{34} & = \left(	\overset{\infty}{\mathscr{T}}_{2,31} - \overset{\infty}{  \mathscr{X}}_{1,31} -  \overset{\infty}{\mathscr{T}}_{1,31}\overset{\infty}{\mathscr{X}}_{1,11} \right) P_{14} + \left( - \overset{\infty}{  \mathscr{X}}_{1,32} - \overset{\infty}{\mathscr{T}}_{1,31}\overset{\infty}{\mathscr{X}}_{1,12} \right) P_{24} \\ & + \left(  \overset{\infty}{\mathscr{T}}_{1,33}  - \overset{\infty}{  \mathscr{X}}_{1,33} - \overset{\infty}{\mathscr{T}}_{1,31}\overset{\infty}{\mathscr{X}}_{1,13} \right) P_{34} + \left( \overset{\infty}{\mathscr{T}}_{1,34} - \overset{\infty}{  \mathscr{X}}_{1,34} - \overset{\infty}{\mathscr{T}}_{1,31}\overset{\infty}{\mathscr{X}}_{1,14} \right) P_{44} = 0,
\end{split}
\end{equation}

\begin{comment}

\red{The above equations are \textit{3 equations in 4 unknowns}:  \[ \overset{\infty}{\mathscr{T}}_{2,31}, \overset{\infty}{\mathscr{T}}_{1,31}, \overset{\infty}{\mathscr{T}}_{1,33}, \overset{\infty}{\mathscr{T}}_{1,34} \]}	content...
\end{comment}

\begin{equation}\label{420}
\begin{split}
(BP)_{41} & = \left(	\overset{\infty}{\mathscr{T}}_{2,41} - \overset{\infty}{  \mathscr{X}}_{1,41} -  \overset{\infty}{\mathscr{T}}_{1,41}\overset{\infty}{\mathscr{X}}_{1,11} \right) P_{11} + \left( - \overset{\infty}{  \mathscr{X}}_{1,42} - \overset{\infty}{\mathscr{T}}_{1,41}\overset{\infty}{\mathscr{X}}_{1,12} \right) P_{21} \\ & + \left(  \overset{\infty}{\mathscr{T}}_{1,43}  - \overset{\infty}{  \mathscr{X}}_{1,43} - \overset{\infty}{\mathscr{T}}_{1,41}\overset{\infty}{\mathscr{X}}_{1,13} \right) P_{31} + \left( \overset{\infty}{\mathscr{T}}_{1,44} - \overset{\infty}{  \mathscr{X}}_{1,44} - \overset{\infty}{\mathscr{T}}_{1,41}\overset{\infty}{\mathscr{X}}_{1,14} \right) P_{41} = 0,
\end{split}
\end{equation}

\begin{equation}\label{421}
\begin{split}
(BP)_{43} & = \left(	\overset{\infty}{\mathscr{T}}_{2,41} - \overset{\infty}{  \mathscr{X}}_{1,41} -  \overset{\infty}{\mathscr{T}}_{1,41}\overset{\infty}{\mathscr{X}}_{1,11} \right) P_{13} + \left( - \overset{\infty}{  \mathscr{X}}_{1,42} - \overset{\infty}{\mathscr{T}}_{1,41}\overset{\infty}{\mathscr{X}}_{1,12} \right) P_{23} \\ & + \left(  \overset{\infty}{\mathscr{T}}_{1,43}  - \overset{\infty}{  \mathscr{X}}_{1,43} - \overset{\infty}{\mathscr{T}}_{1,41}\overset{\infty}{\mathscr{X}}_{1,13} \right) P_{33} + \left( \overset{\infty}{\mathscr{T}}_{1,44} - \overset{\infty}{  \mathscr{X}}_{1,44} - \overset{\infty}{\mathscr{T}}_{1,41}\overset{\infty}{\mathscr{X}}_{1,14} \right) P_{43} = 0,
\end{split}
\end{equation}

\begin{equation}\label{422}
\begin{split}
(BP)_{44} & = \left(	\overset{\infty}{\mathscr{T}}_{2,41} - \overset{\infty}{  \mathscr{X}}_{1,41} -  \overset{\infty}{\mathscr{T}}_{1,41}\overset{\infty}{\mathscr{X}}_{1,11} \right) P_{14} + \left( - \overset{\infty}{  \mathscr{X}}_{1,42} - \overset{\infty}{\mathscr{T}}_{1,41}\overset{\infty}{\mathscr{X}}_{1,12} \right) P_{24} \\ & + \left(  \overset{\infty}{\mathscr{T}}_{1,43}  - \overset{\infty}{  \mathscr{X}}_{1,43} - \overset{\infty}{\mathscr{T}}_{1,41}\overset{\infty}{\mathscr{X}}_{1,13} \right) P_{34} + \left( \overset{\infty}{\mathscr{T}}_{1,44} - \overset{\infty}{  \mathscr{X}}_{1,44} - \overset{\infty}{\mathscr{T}}_{1,41}\overset{\infty}{\mathscr{X}}_{1,14} \right) P_{44} = 0.
\end{split}
\end{equation}
\begin{comment}
\red{The above equations are \textit{3 equations in 4 unknowns}:  \[ \overset{\infty}{\mathscr{T}}_{2,41}, \overset{\infty}{\mathscr{T}}_{1,41}, \overset{\infty}{\mathscr{T}}_{1,43}, \overset{\infty}{\mathscr{T}}_{1,44} \]}
	content...
\end{comment}
The complementary equations come from setting the first column of \eqref{410} equal to zero, which are

\begin{equation}
	 \mathscr{A}_{11} B_{11} + \mathscr{A}_{21} B_{12} +
	\mathscr{A}_{31} B_{13} + \mathscr{A}_{41} B_{14} +
	P_{11} \left(\overset{\infty}{\mathscr{T}}_{1,11}- \overset{\infty}{\mathscr{X}}_{1,11}\right) - \overset{\infty}{\mathscr{X}}_{1,12} P_{21}  - \overset{\infty}{\mathscr{X}}_{1,13} 
	P_{31}  - \overset{\infty}{\mathscr{X}}_{1,14} P_{41} = 0,
\end{equation}

\begin{equation}
	\mathscr{A}_{11} B_{21} + \mathscr{A}_{21} B_{22} +
	\mathscr{A}_{31} B_{23} + \mathscr{A}_{41} B_{24} +
	P_{11} \overset{\infty}{\mathscr{T}}_{1,21} = 0,
\end{equation}

\begin{equation}
	\mathscr{A}_{11} B_{31} + \mathscr{A}_{21} B_{32} +
	\mathscr{A}_{31} B_{33} + \mathscr{A}_{41} B_{34} +
	P_{11} \overset{\infty}{\mathscr{T}}_{1,31}  + 
	P_{31} = 0,
\end{equation}
and
\begin{equation}
	\mathscr{A}_{11} B_{41} + \mathscr{A}_{21} B_{42} +
	\mathscr{A}_{31} B_{43} + \mathscr{A}_{41} B_{44} +
	P_{11} \overset{\infty}{\mathscr{T}}_{1,41} +  P_{41} = 0,
\end{equation}
where $\mathscr A \equiv P\overset{\circ}{\mathscr{X}}_1$. Using the explicit expression for $B_{jk}$ in  \eqref{R E B} we can write the last four equations as
\begin{equation}\label{427}
	\begin{split}
	& \mathscr{A}_{11} \left(\overset{\infty}{\mathscr{T}}_{2,11} + \left[ \overset{\infty}{\mathscr{X}}_1^2 \right]_{11} - \overset{\infty}{  \mathscr{X}}_{2,11} - \overset{\infty}{\mathscr{T}}_{1,11}\overset{\infty}{\mathscr{X}}_{1,11} \right) + \mathscr{A}_{21} \left(\left[ \overset{\infty}{\mathscr{X}}_1^2 \right]_{12} - \overset{\infty}{  \mathscr{X}}_{2,12} - \overset{\infty}{\mathscr{T}}_{1,11}\overset{\infty}{\mathscr{X}}_{1,12} \right) + \\ &
	\mathscr{A}_{31} \left(\overset{\infty}{\mathscr{T}}_{1,13} + \left[ \overset{\infty}{\mathscr{X}}_1^2 \right]_{13} - \overset{\infty}{  \mathscr{X}}_{2,13} - \overset{\infty}{\mathscr{T}}_{1,11}\overset{\infty}{\mathscr{X}}_{1,13} \right) + \mathscr{A}_{41} \left(\overset{\infty}{\mathscr{T}}_{1,14} + \left[ \overset{\infty}{\mathscr{X}}_1^2 \right]_{14} - \overset{\infty}{  \mathscr{X}}_{2,14} - \overset{\infty}{\mathscr{T}}_{1,11}\overset{\infty}{\mathscr{X}}_{1,14} \right) \\ & +
	P_{11} \left(\overset{\infty}{\mathscr{T}}_{1,11}- \overset{\infty}{\mathscr{X}}_{1,11}\right) - \overset{\infty}{\mathscr{X}}_{1,12} P_{21}  - \overset{\infty}{\mathscr{X}}_{1,13} 
	P_{31}  - \overset{\infty}{\mathscr{X}}_{1,14} P_{41} = 0,
	\end{split}
\end{equation}

\begin{comment}

\red{The above equation is in 4 unknowns:  \[ \overset{\infty}{\mathscr{T}}_{2,11}, \overset{\infty}{\mathscr{T}}_{1,11}, \overset{\infty}{\mathscr{T}}_{1,13}, \overset{\infty}{\mathscr{T}}_{1,14} \]
Thus, it makes a complete four by four system to solve for the above unknowns together with equations \eqref{411}-\eqref{413}.
}

	content...
\end{comment}

\begin{comment}
	content...

\begin{equation}
\mathscr{A}_{11} B_{21} + \mathscr{A}_{21} B_{22} +
\mathscr{A}_{31} B_{23} + \mathscr{A}_{41} B_{24} +
P_{11} \overset{\infty}{\mathscr{T}}_{1,21} = 0,
\end{equation}
\end{comment}

\begin{equation}\label{428}
\begin{split}
& \mathscr{A}_{11} \left(\overset{\infty}{\mathscr{T}}_{2,21} - \overset{\infty}{\mathscr{T}}_{1,21}\overset{\infty}{\mathscr{X}}_{1,11} \right) + \mathscr{A}_{21} \left(1-\overset{\infty}{\mathscr{T}}_{1,21}\overset{\infty}{\mathscr{X}}_{1,12}\right) +
\mathscr{A}_{31} \left(\overset{\infty}{\mathscr{T}}_{1,23}-\overset{\infty}{\mathscr{T}}_{1,21}\overset{\infty}{\mathscr{X}}_{1,13}\right) + \\ & \mathscr{A}_{41} \left(\overset{\infty}{\mathscr{T}}_{1,24} - \overset{\infty}{\mathscr{T}}_{1,21}\overset{\infty}{\mathscr{X}}_{1,14} \right) +
P_{11} \overset{\infty}{\mathscr{T}}_{1,21} = 0,
\end{split}
\end{equation}

\begin{comment}

\red{The above equation is in 4 unknowns:  \[ \overset{\infty}{\mathscr{T}}_{2,21}, \overset{\infty}{\mathscr{T}}_{1,21}, \overset{\infty}{\mathscr{T}}_{1,23}, \overset{\infty}{\mathscr{T}}_{1,24} \]
	Thus, it makes a complete four by four system to solve for the above unknowns together with equations \eqref{414}-\eqref{416}.
}
	content...
\end{comment}

\begin{comment}
	content...

\begin{equation}
\mathscr{A}_{11} B_{31} + \mathscr{A}_{21} B_{32} +
\mathscr{A}_{31} B_{33} + \mathscr{A}_{41} B_{34} +
P_{11} \overset{\infty}{\mathscr{T}}_{1,31}  + 
P_{31} = 0,
\end{equation}
\end{comment}

\begin{equation}\label{429}
\begin{split}
& \mathscr{A}_{11} \left(\overset{\infty}{\mathscr{T}}_{2,31}- \overset{\infty}{\mathscr{X}}_{1,31}-\overset{\infty}{\mathscr{T}}_{1,31}\overset{\infty}{\mathscr{X}}_{1,11}\right) + \mathscr{A}_{21} \left(- \overset{\infty}{\mathscr{X}}_{1,32} - \overset{\infty}{\mathscr{T}}_{1,31}\overset{\infty}{\mathscr{X}}_{1,12}\right) \\ & +
\mathscr{A}_{31} \left(\overset{\infty}{\mathscr{T}}_{1,33}- \overset{\infty}{\mathscr{X}}_{1,33}-\overset{\infty}{\mathscr{T}}_{1,31}\overset{\infty}{\mathscr{X}}_{1,13}\right) + \mathscr{A}_{41} \left(\overset{\infty}{\mathscr{T}}_{1,34} - \overset{\infty}{\mathscr{X}}_{1,34} - \overset{\infty}{\mathscr{T}}_{1,31}\overset{\infty}{\mathscr{X}}_{1,14} \right) +
P_{11} \overset{\infty}{\mathscr{T}}_{1,31}  + 
P_{31} = 0,
\end{split}
\end{equation}

\begin{comment}

\red{The above equation is in 4 unknowns:  \[ \overset{\infty}{\mathscr{T}}_{2,31}, \overset{\infty}{\mathscr{T}}_{1,31}, \overset{\infty}{\mathscr{T}}_{1,33}, \overset{\infty}{\mathscr{T}}_{1,34} \]
	Thus, it makes a complete four by four system to solve for the above unknowns together with equations \eqref{417}-\eqref{419}.
}
	content...
\end{comment}

\begin{comment}
	content...

\begin{equation}
\mathscr{A}_{11} B_{41} + \mathscr{A}_{21} B_{42} +
\mathscr{A}_{31} B_{43} + \mathscr{A}_{41} B_{44} +
P_{11} \overset{\infty}{\mathscr{T}}_{1,41} +  P_{41} = 0,
\end{equation}
\end{comment}

\begin{equation}\label{430}
\begin{split}
& \mathscr{A}_{11} \left(\overset{\infty}{\mathscr{T}}_{2,41}- \overset{\infty}{\mathscr{X}}_{1,41} - \overset{\infty}{\mathscr{T}}_{1,41}\overset{\infty}{\mathscr{X}}_{1,11}\right) + \mathscr{A}_{21} \left(- \overset{\infty}{\mathscr{X}}_{1,42}-\overset{\infty}{\mathscr{T}}_{1,41}\overset{\infty}{\mathscr{X}}_{1,12}\right) + \\ &
\mathscr{A}_{31} \left(\overset{\infty}{\mathscr{T}}_{1,43} - \overset{\infty}{\mathscr{X}}_{1,43}-\overset{\infty}{\mathscr{T}}_{1,41}\overset{\infty}{\mathscr{X}}_{1,13}\right) + \mathscr{A}_{41} \left(\overset{\infty}{\mathscr{T}}_{1,44} - \overset{\infty}{\mathscr{X}}_{1,44} - \overset{\infty}{\mathscr{T}}_{1,41}\overset{\infty}{\mathscr{X}}_{1,14} \right) +
P_{11} \overset{\infty}{\mathscr{T}}_{1,41} +  P_{41} = 0.
\end{split}
\end{equation}

\begin{comment}
	content...

\red{The above equation is in 4 unknowns:  \[ \overset{\infty}{\mathscr{T}}_{2,41}, \overset{\infty}{\mathscr{T}}_{1,41}, \overset{\infty}{\mathscr{T}}_{1,43}, \overset{\infty}{\mathscr{T}}_{1,44} \]
	Thus, it makes a complete four by four system to solve for the above unknowns together with equations \eqref{420}-\eqref{422}.
}
\end{comment}

Now we observe that the sixteen equations \eqref{411} - \eqref{422} and \eqref{427} - \eqref{430} can be split into four sets of mutually decoupled equations, where solving each set of equations gives four unknowns in $$\left\{ \overset{\infty}{\mathscr{T}}_{1,j1}, \overset{\infty}{\mathscr{T}}_{1,j3}, \overset{\infty}{\mathscr{T}}_{1,j4}, \overset{\infty}{\mathscr{T}}_{2,j1} \right\}^{4}_{j=1}.$$ To this end,  we categorize these equations below. \begin{itemize}
	\item The four equations \eqref{411}, \eqref{412}, \eqref{413}, and \eqref{427} are in four unknowns $ \overset{\infty}{\mathscr{T}}_{2,11}, \overset{\infty}{\mathscr{T}}_{1,11}, \overset{\infty}{\mathscr{T}}_{1,13}, \overset{\infty}{\mathscr{T}}_{1,14}, $ which can be written as \begin{equation}\label{sys1}
	\begin{pmatrix}
	P_{11} & -\mathscr{B}_{11} & P_{31} & P_{41}\\
	P_{13} & -\mathscr{B}_{13} & P_{33} & P_{43} \\
	P_{14} & -\mathscr{B}_{14} & P_{34} & P_{44} \\
	\mathscr{A}_{11} & \mathscr{D}_{11} & \mathscr{A}_{31} & \mathscr{A}_{41}
	\end{pmatrix} \begin{pmatrix}
	\overset{\infty}{\mathscr{T}}_{2,11}\\ \overset{\infty}{\mathscr{T}}_{1,11}\\ \overset{\infty}{\mathscr{T}}_{1,13}\\ \overset{\infty}{\mathscr{T}}_{1,14}
	\end{pmatrix} = \begin{pmatrix}
	\left[\mathscr{C}P\right]_{11} \\
	\left[\mathscr{C}P\right]_{13} \\
	\left[\mathscr{C}P\right]_{14} \\
	\left[\mathscr{C}\mathscr{A} + \mathscr{B} \right]_{11}
	\end{pmatrix} ,
	\end{equation}
	where for simplicity of notations we have introduced
	
	\begin{equation}
	\mathscr A \equiv P\overset{\circ}{\mathscr{X}}_1, \qquad \mathscr B \equiv \overset{\infty}{\mathscr{X}}_1 P \qquad \mathscr C \equiv  \overset{\infty}{  \mathscr{X}}_2 - \overset{\infty}{  \mathscr{X}}_1^2 , \qandq \mathscr{D} \equiv  P -\overset{\infty}{\mathscr{X}}_{1}\mathscr A.
	\end{equation}
	\item The four equations \eqref{414}, \eqref{415}, \eqref{416}, and \eqref{428} are in four unknowns $\overset{\infty}{\mathscr{T}}_{2,21}, \overset{\infty}{\mathscr{T}}_{1,21}, \overset{\infty}{\mathscr{T}}_{1,23}, \overset{\infty}{\mathscr{T}}_{1,24},$ which can be written as
	\begin{equation}\label{sys2}
	\begin{pmatrix}
	P_{11} & -\mathscr{B}_{11} & P_{31} & P_{41}\\
	P_{13} & -\mathscr{B}_{13} & P_{33} & P_{43} \\
	P_{14} & -\mathscr{B}_{14} & P_{34} & P_{44} \\
	\mathscr{A}_{11} & \mathscr{D}_{11} & \mathscr{A}_{31} & \mathscr{A}_{41}
	\end{pmatrix} \begin{pmatrix}
	\overset{\infty}{\mathscr{T}}_{2,21}\\ \overset{\infty}{\mathscr{T}}_{1,21}\\ \overset{\infty}{\mathscr{T}}_{1,23}\\ \overset{\infty}{\mathscr{T}}_{1,24}
	\end{pmatrix} = \begin{pmatrix}
	-P_{21} \\
	-P_{23} \\
	-P_{24} \\
	-\mathscr{A}_{21}
	\end{pmatrix}. 
	\end{equation}
	\item The four equations \eqref{417}, \eqref{418}, \eqref{419}, and \eqref{429} are in four unknowns $\overset{\infty}{\mathscr{T}}_{2,31}, \overset{\infty}{\mathscr{T}}_{1,31}, \overset{\infty}{\mathscr{T}}_{1,33}, \overset{\infty}{\mathscr{T}}_{1,34},$ which can be written as
	\begin{equation}\label{sys3}
	\begin{pmatrix}
	P_{11} & -\mathscr{B}_{11} & P_{31} & P_{41}\\
	P_{13} & -\mathscr{B}_{13} & P_{33} & P_{43} \\
	P_{14} & -\mathscr{B}_{14} & P_{34} & P_{44} \\
	\mathscr{A}_{11} & \mathscr{D}_{11} & \mathscr{A}_{31} & \mathscr{A}_{41}
	\end{pmatrix} \begin{pmatrix}
	\overset{\infty}{\mathscr{T}}_{2,31}\\ \overset{\infty}{\mathscr{T}}_{1,31}\\ \overset{\infty}{\mathscr{T}}_{1,33}\\ \overset{\infty}{\mathscr{T}}_{1,34}
	\end{pmatrix} = \begin{pmatrix}
	\mathscr{B}_{31} \\
	\mathscr{B}_{33} \\
	\mathscr{B}_{34} \\
	-\mathscr{D}_{31}
	\end{pmatrix}. 
	\end{equation}
	\item The four equations \eqref{420}, \eqref{421}, \eqref{422}, and \eqref{430} are in four unknowns $\overset{\infty}{\mathscr{T}}_{2,41}, \overset{\infty}{\mathscr{T}}_{1,41}, \overset{\infty}{\mathscr{T}}_{1,43}, \overset{\infty}{\mathscr{T}}_{1,44}$, which can be written as \begin{equation}\label{sys4}
	\begin{pmatrix}
	P_{11} & -\mathscr{B}_{11} & P_{31} & P_{41}\\
	P_{13} & -\mathscr{B}_{13} & P_{33} & P_{43} \\
	P_{14} & -\mathscr{B}_{14} & P_{34} & P_{44} \\
	\mathscr{A}_{11} & \mathscr{D}_{11} & \mathscr{A}_{31} & \mathscr{A}_{41}
	\end{pmatrix} \begin{pmatrix}
	\overset{\infty}{\mathscr{T}}_{2,41}\\ \overset{\infty}{\mathscr{T}}_{1,41}\\ \overset{\infty}{\mathscr{T}}_{1,43}\\ \overset{\infty}{\mathscr{T}}_{1,44}
	\end{pmatrix} = \begin{pmatrix}
	\mathscr{B}_{41} \\
	\mathscr{B}_{43} \\
	\mathscr{B}_{44} \\
	-\mathscr{D}_{41}
	\end{pmatrix}.
	\end{equation}
\end{itemize}
We see that all four linear systems above have the same matrix coefficient. So we try to invert the following linear system

 \begin{equation}\label{4by4 system}
\begin{pmatrix}
P_{11} & -\mathscr{B}_{11} & P_{31} & P_{41}\\
P_{13} & -\mathscr{B}_{13} & P_{33} & P_{43} \\
P_{14} & -\mathscr{B}_{14} & P_{34} & P_{44} \\
\mathscr{A}_{11} & \mathscr{D}_{11} & \mathscr{A}_{31} & \mathscr{A}_{41}
\end{pmatrix} \begin{pmatrix}
\mathscr{F}_1\\ 
\mathscr{F}_2\\ 
\mathscr{F}_3\\ 
\mathscr{F}_4
\end{pmatrix} = \begin{pmatrix}
x \\
y \\
w \\
z
\end{pmatrix},
\end{equation}
and then find the desired unknowns by replacing $\begin{pmatrix}
x & y & w & z
\end{pmatrix}^T$ by the appropriate right hand side in \eqref{sys1}, \eqref{sys2}, \eqref{sys3}, or \eqref{sys4}.
Now, we introduce  the objects (cf. the formulation of Theorem \ref{thm r=0 s=2})

\noindent\begin{minipage}{.3\linewidth}
	\begin{alignat*}{2}
		& \al  && := \left(\frac{\mathscr{A}_{11} \mathscr{B}_{11}}{P_{11}} + \mathscr{D}_{11}\right)^{-1},  \\
		& \om_{jk} &&:=\frac{P_{1j} P_{k1}}{P_{11}}, 
	\end{alignat*}	
\end{minipage}
\noindent\begin{minipage}{.3\linewidth}
	\begin{alignat*}{2}
		&	\theta &&:=\frac{\mathscr{A}_{11}}{P_{11}}, \\
		&\eta_j && :=\frac{P_{1j}}{P_{11}},
	\end{alignat*}	
\end{minipage}
\noindent\begin{minipage}{.3\linewidth}
	\begin{alignat*}{2}
		& \rho_j &&:= \mathscr{A}_{j1} - \frac{\mathscr{A}_{11} P_{j1}}{P_{11}},   \\
		&\nu_j && := \frac{\mathscr{B}_{11} P_{1j}}{P_{11}} - \mathscr{B}_{1j} ,
	\end{alignat*}	
\end{minipage}

\noindent assuming that they are well defined. Using these objects define

\noindent\begin{minipage}{.5\linewidth}
	\begin{alignat*}{2}
	& \mathscr{M}_{jk}  && = -\al\rho_k \nu_j-\om_{jk}+P_{kj},   
	\end{alignat*}	
\end{minipage}
\begin{minipage}{.5\linewidth}
	\begin{alignat*}{2}
	&f_j(x,y,z)  &&= \al\nu_j
	\left(z-\theta x\right)+\eta_j x-y, 
	\end{alignat*}	
\end{minipage}
and
\begin{equation}
\Delta := \mathscr{M}_{34}\mathscr{M}_{43}  -\mathscr{M}_{33}\mathscr{M}_{44},
\end{equation}
and it is assumed that $\Delta\neq 0$. Then, inverting \eqref{4by4 system} yields
\begin{align}
\mathscr{F}_1(x,y,w,z) & =  \frac{x}{P_{11}} + \frac{\al \mathscr{B}_{11}}{P_{11}}  \left(z-\theta x\right)  + \left( \left( \frac{P_{41}+\al \mathscr{B}_{11}\rho_4}{P_{11}\Delta} \right) \mathscr{M}_{43}   -\left(   \frac{P_{31}+\al \mathscr{B}_{11}\rho_3}{P_{11}\Delta}   \right)  \mathscr{M}_{44}   \right) f_3(x,y,z) \\ &  + \left(    \left(   \frac{P_{31}+\al \mathscr{B}_{11}\rho_3}{P_{11}\Delta}   \right)   \mathscr{M}_{34}   - \left( \frac{P_{41}+\al \mathscr{B}_{11}\rho_4}{P_{11}\Delta} \right) \mathscr{M}_{33}  \right) f_4(x,w,z), \nonumber \\
\mathscr{F}_2(x,y,w,z) & = \al \left( z - \theta x + \left(\frac{\rho_4 \mathscr{M}_{43} - \rho_3\mathscr{M}_{44}}{\Delta} \right) f_3(x,y,z) + \left(  \frac{\rho_3\mathscr{M}_{34}   -\rho_4 \mathscr{M}_{33}}{\Delta}   \right) f_4(x,w,z)   \right), \\
\mathscr{F}_3(x,y,w,z) &=\frac{ 1}{\Delta}  \left( \frac{1}{\mathscr{M}_{43}} f_3(x,y,z)   - \mathscr{M}_{34} f_4(x,w,z)    \right),  \\
\mathscr{F}_4(x,y,w,z) & = \frac{\mathscr{M}_{33} f_4(x,w,z) - f_3(x,y,z)}{\Delta}.
\end{align}
These are exactly the four functions that appear in Theorem \ref{thm r=0 s=2}. This finishes the proof of  this theorem.

\section{Proof of Theorem \ref{Thm WW}}\label{proof Thm{WW}}
 From \eqref{hatW-jumpab} and \eqref{WWW} one can directly check that 
\begin{equation}\label{WJWJ}
	W J_X^{-1}(z^{-1};r,s) W = J_X(z;r,s),
\end{equation} which is part (a) of Theorem \ref{Thm WW}. To prove the other two parts, let us consider the function 
\begin{equation}\label{BB}
	B(z;n,r,s):=W P^{-1}(n,r,s) X(z^{-1};n,r,s) W.
\end{equation}
We have
\begin{equation}
	B^{-1}_-(z;n,r,s)B_+(z;n,r,s)=W \left( X(z^{-1};n,r,s) \right)^{-1}_- \left(X(z^{-1};n,r,s)\right)_+ W.
\end{equation}
So
\begin{equation}
	\begin{split}
		B^{-1}_-(z;n,r,s)B_+(z;n,r,s) & =W \left( \underset{\ze \underset{-}{\to} z}{\lim} X^{-1}(\ze^{-1};n,r,s) \right) \left( \underset{\ze \underset{+}{\to} z}{\lim} X(\ze^{-1};n,r,s) \right) W \\ & =W \left( \underset{\ze^{-1} \underset{+}{\to} z^{-1}}{\lim} X^{-1}(\ze^{-1};n,r,s) \right) \left( \underset{\ze^{-1} \underset{-}{\to} z^{-1}}{\lim} X(\ze^{-1};n,r,s) \right) W \\ & =W \left( \underset{\tau \underset{+}{\to} z^{-1}}{\lim} X^{-1}(\tau;n,r,s) \right) \left( \underset{\tau \underset{-}{\to} z^{-1}}{\lim} X(\tau;n,r,s) \right) W \\ & =W X_+^{-1}(z^{-1};n,r,s)  X_-(z^{-1};n,r,s) W \\ & = W J_X^{-1}(z^{-1};r,s) W.
	\end{split}
\end{equation}
Therefore, from \eqref{WJWJ} we have
\begin{equation}\label{B jump}
	B_+(z;n,r,s)= B_-(z;n,r,s) J_X(z;r,s).
\end{equation}
Let us now remind the asymptotic behavior of $X(z;n,r,s)$ near zero and infinity with the following notations for the subleading terms:	\begin{itemize}
	\item \textbf{RH-X3}  \quad As $z \to \infty$  \begin{equation*}\label{hatXinfinityab}
		X(z;n,r,s)=\left(\di I+\frac{ \overset{\infty}{  X}_1(n,r,s)}{z}+\frac{\overset{\infty}{X}_2(n,r,s)}{z^2} + O(z^{-3})\right)\begin{pmatrix}
			z^n & 0 & 0 & 0\\
			0 & 1 & 0 & 0 \\
			0 & 0 & z^{-n} & 0 \\
			0 & 0 & 0 & 1
		\end{pmatrix},  
	\end{equation*} 
	
	\item \textbf{RH-X4} \quad As $z \to 0$ 
	\begin{equation*}\label{hatXzeroab}
		X(z;n,r,s)=P(n,r,s)\left(I+\overset{\circ}{X}_1(n,r,s)z+\overset{\circ}{ X}_2(n,r,s)z^2+O(z^3)\right)\begin{pmatrix}
			1 & 0 & 0 & 0\\
			0 & z^{-n} & 0 & 0 \\
			0 & 0 & 1 & 0 \\
			0 & 0 & 0 & z^n
		\end{pmatrix}.
	\end{equation*}
\end{itemize}
Let us consider the asymptotic behavior of $B$ as $z \to \infty$:
\begin{equation}\label{B infty}
	\begin{split}
		B(z;n,r,s)& =W P^{-1}(n,r,s) X(z^{-1};n,r,s) W \\ &=W  \left(I+\overset{\circ}{X}_1(n,r,s)z^{-1}+O(z^{-2})\right)\begin{pmatrix}
			1 & 0 & 0 & 0\\
			0 & z^{n} & 0 & 0 \\
			0 & 0 & 1 & 0 \\
			0 & 0 & 0 & z^{-n}
		\end{pmatrix} W \\ & =W \left(I+\overset{\circ}{X}_1(n,r,s)z^{-1}+O(z^{-2})\right)WW\begin{pmatrix}
			1 & 0 & 0 & 0\\
			0 & z^{n} & 0 & 0 \\
			0 & 0 & 1 & 0 \\
			0 & 0 & 0 & z^{-n}
		\end{pmatrix} W \\ & = \left(I+W\overset{\circ}{X}_1(n,r,s)Wz^{-1}+O(z^{-2})\right)\begin{pmatrix}
			z^{n} & 0 & 0 & 0\\
			0 & 1 & 0 & 0 \\
			0 & 0 & z^{-n} & 0 \\
			0 & 0 & 0 & 1
		\end{pmatrix},k
	\end{split}
\end{equation}
where we have used $W^2=I$. Next, consider the asymptotic behavior of $B$ as $z \to 0$:
\begin{equation}\label{B zero}
	\begin{split}
		B(z;n,r,s)& =W P^{-1}(n,r,s) X(z^{-1};n,r,s) W \\ &=W P^{-1}(n,r,s) \left(I+\overset{\infty}{X}_1(n,r,s)z+O(z)\right)\begin{pmatrix}
			z^{-n} & 0 & 0 & 0\\
			0 & 1 & 0 & 0 \\
			0 & 0 & z^{n} & 0 \\
			0 & 0 & 0 & 1
		\end{pmatrix} W \\ & =W P^{-1}(n,r,s)WW \left(I+\overset{\infty}{X}_1(n,r,s)z+O(z)\right)WW\begin{pmatrix}
			z^{-n} & 0 & 0 & 0\\
			0 & 1 & 0 & 0 \\
			0 & 0 & z^{n} & 0 \\
			0 & 0 & 0 & 1
		\end{pmatrix} W \\ & =W P^{-1}(n,r,s)W \left(I+W\overset{\infty}{X}_1(n,r,s)Wz+O(z)\right)\begin{pmatrix}
			1 & 0 & 0 & 0\\
			0 & z^{-n} & 0 & 0 \\
			0 & 0 & 1 & 0 \\
			0 & 0 & 0 & z^{n}
		\end{pmatrix} . 
	\end{split}
\end{equation}
By definition \eqref{BB}, $B$ is analytic in $\C\setminus \left(\T \cup\{0\}\right)$. This, together with \eqref{B jump}, \eqref{B infty}, and \eqref{B zero} and Lemma \ref{Lemma X unique} gives \begin{equation}
	B(z;n,r,s) \equiv X(z;n,r,s),
\end{equation}
and as a result we immediately confirm parts (b) and (c) of Theorem \ref{Thm WW}:
\begin{equation}
	P(n,r,s) = W P^{-1}(n,r,s)W,
\end{equation}
and
\begin{equation}
	\overset{\circ}{X}_1(n,r,s) = W\overset{\infty}{X}_1(n,r,s)W.
\end{equation}

\section{Asymptotics of the Norms of the Orthogonal Polynomials $\mathcal{P}_n(z;0,1)$}\label{Sec asym norms r=0 s=1}
In this section we illustrate how one can use Theorem \ref{Thm WW} and the expression of $\mathscr{U}$ in terms of $\mathscr{X}$, as given in Theorem \ref{thm r=0 s=1}, to find the large-$n$ asymptotics of the norm $h^{(0,1)}_{n-1}$. 
This is a case study as we focus on $(r,s)=(0,1)$, but we would like to highlight that the same procedure can be followed for other choices of $(r,s)$ once we have expressions for $X(z;n,r,s)$ in terms of $\mathscr{X}(z;n)$, such as those in Theorems \ref{thm r=s=0} and \ref{thm r=0 s=2}. What is presented below partially follows section 4.2 of \cite{GI}. Let us start with 
formula (\ref{T+H h_n}) for the norm $h^{(0,1)}_{n-1}$ which can be rewritten as

\begin{equation}\label{h to y 01}
	\frac{-1}{h^{(0,1)}_{n-1}} = \lim_{z \to 0} z^{n-1} \mathcal{Y}_{21}(z^{-1};n,0,1).
\end{equation}
In view of  \textbf{RH-$\mathscr{U}$4} let us define
\begin{equation}\label{H}
	\mathscr{H}(z;n) := \widehat{\mathscr{U}}(n)^{-1} \mathscr{U}(z;n) \begin{pmatrix}
		1 & 0 & 0 & 0\\
		0 & z^n & 0 & 0 \\
		0 & 0 & 1 & 0 \\
		0 & 0 & 0 & z^{-n}
	\end{pmatrix},
\end{equation}
which implies
\begin{equation}\label{H0}
	\mathscr{H}(z;n) = I+\overset{\circ}{  \mathscr{U}}_1(n)z+\overset{\circ}{  \mathscr{U}}_2(n)z^2+O(z^3), \qasq z \to 0.
\end{equation}
From \eqref{H}, \eqref{24-to-44}, \eqref{mathfrakR}, and \eqref{C's-to-P}  and remembering that $\mathscr{U}(z;n) = X(z;n,0,1)$ we have
\begin{equation}
	\widehat{X}(z;n,0,1) = \begin{pmatrix}
		C_1(n,0,1) & 1 & C_3(n,0,1) & 0  \\
		C_2(n,0,1)  & 0 & C_4(n,0,1) & 1 
	\end{pmatrix}\mathscr{H}(z;n)\begin{pmatrix}
	1 & 0 & 0 & 0\\
	0 & z^{-n} & 0 & 0 \\
	0 & 0 & 1 & 0 \\
	0 & 0 & 0 & z^{n}
\end{pmatrix},
\end{equation}
recalling that, by definition, $\widehat{\mathscr{U}}(n) = P(n,0,1)$. From \eqref{X naught to Y} it follows that \begin{equation}
	\mathcal{Y}_{21}(z^{-1};n,0,1) = 	\widehat{X}_{22}(z;n,0,1),
\end{equation}
hence,
\begin{equation}
	z^{n-1}\mathcal{Y}_{21}(z^{-1};n,0,1) = C_2(n,0,1)  z^{-1}\mathscr{H}_{12}(z;n) + C_4(n,0,1)  z^{-1}\mathscr{H}_{32}(z;n) +   z^{-1}\mathscr{H}_{42}(z;n).
\end{equation}
From \eqref{H0} we conclude that 
\begin{equation}
	z^{-1}\mathscr{H}_{j2}(z;n) = \overset{\circ}{\mathscr{U}}_{1,j2}(n)+O(z), \qquad j=1,3,4.
\end{equation}
 Therefore, \eqref{h to y 01} yields the formula,
\begin{equation}\label{h C U}
		\frac{-1}{h^{(0,1)}_{n-1}} =  C_2(n,0,1)  \overset{\circ}{\mathscr{U}}_{1,12}(n) + C_4(n,0,1)  \overset{\circ}{\mathscr{U}}_{1,32}(n) +   \overset{\circ}{\mathscr{U}}_{1,42}(n).
\end{equation}
\subsection{$h^{(0,1)}_{n-1}$ and the $\mathscr{X}$-RHP Data}
In order to find the asymptotics of $h^{(0,1)}_{n-1}$, we need to find the expressions for the objects on the right hand side of \eqref{h C U} in terms of the  $\mathscr{X}$-RHP data. From part (c) of Theorem \ref{Thm WW} we have $\overset{\circ}{\mathscr{U}}_1(n)=W  \overset{\infty}{\mathscr{U}}_1(n) W$, which can be written as
\begin{equation}
	\left(
	\begin{array}{cccc}
		\overset{\circ}{\mathscr{U}}_{1,11} & \overset{\circ}{\mathscr{U}}_{1,12} & \overset{\circ}{\mathscr{U}}_{1,13} & \overset{\circ}{\mathscr{U}}_{1,14} \\
		\overset{\circ}{\mathscr{U}}_{1,21} & \overset{\circ}{\mathscr{U}}_{1,22} & \overset{\circ}{\mathscr{U}}_{1,23} & \overset{\circ}{\mathscr{U}}_{1,24} \\
		\overset{\circ}{\mathscr{U}}_{1,31} & \overset{\circ}{\mathscr{U}}_{1,32} & \overset{\circ}{\mathscr{U}}_{1,33} & \overset{\circ}{\mathscr{U}}_{1,34} \\
		\overset{\circ}{\mathscr{U}}_{1,41} & \overset{\circ}{\mathscr{U}}_{1,42} & \overset{\circ}{\mathscr{U}}_{1,43} & \overset{\circ}{\mathscr{U}}_{1,44} \\
	\end{array}
	\right) = \left(
	\begin{array}{cccc}
		\overset{\infty}{\mathscr{U}}_{1,22} & \overset{\infty}{\mathscr{U}}_{1,21} & \overset{\infty}{\mathscr{U}}_{1,24} & \overset{\infty}{\mathscr{U}}_{1,23} \\
		\overset{\infty}{\mathscr{U}}_{1,12} & \overset{\infty}{\mathscr{U}}_{1,11} & \overset{\infty}{\mathscr{U}}_{1,14} & \overset{\infty}{\mathscr{U}}_{1,13} \\
		\overset{\infty}{\mathscr{U}}_{1,42} & \overset{\infty}{\mathscr{U}}_{1,41} & \overset{\infty}{\mathscr{U}}_{1,44} & \overset{\infty}{\mathscr{U}}_{1,43} \\
		\overset{\infty}{\mathscr{U}}_{1,32} & \overset{\infty}{\mathscr{U}}_{1,31} & \overset{\infty}{\mathscr{U}}_{1,34} & \overset{\infty}{\mathscr{U}}_{1,33} \\
	\end{array}
	\right).
\end{equation}
So we have
\begin{align}
	\overset{\circ}{\mathscr{U}}_{1,12} & = \overset{\infty}{\mathscr{U}}_{1,21} = \frac{P_{31}P_{23}  -P_{33}P_{21}}{P_{11}P_{33}  -P_{13}P_{31}}, \label{U1012 to Xdata} \\
		\overset{\circ}{\mathscr{U}}_{1,32} & = \overset{\infty}{\mathscr{U}}_{1,41} = \frac{P_{43}P_{31}  -P_{33}P_{41}}{P_{11}P_{33}  -P_{13}P_{31}}, \label{U1032 to Xdata}\\
			\overset{\circ}{\mathscr{U}}_{1,42} & = \overset{\infty}{\mathscr{U}}_{1,31} = \overset{\infty}{\mathscr{X}}_{1,31} + \frac{P_{33}\underset{\footnotesize{j \in\{2,4\}}}{\sum} \overset{\infty}{\mathscr{X}}_{1,3j} P_{j1} -P_{31}\underset{\footnotesize{j \in\{2,4\}}}{\sum} \overset{\infty}{\mathscr{X}}_{1,3j} P_{j3}}{P_{11}P_{33}  -P_{13}P_{31}}, \label{U1042 to Xdata}
\end{align}
where we have used Theorem \ref{thm r=0 s=1}.

Now move on to find $C_2$ and $C_4$ in terms of the $\mathscr{X}$-RHP data. We are not going to study each and every one of conditions \eqref{Csolvcond01} through \eqref{gencond2}  of Lemma \ref{12}, rather like what is presented in \cite{GI} we consider\footnote{In Section~\ref{sec asymp h_n 01} we impose conditions on the symbols to ensure this; see~\eqref{E} and~\eqref{generic symbols 2}--\eqref{E asymp 2}, as well as the assumptions in Theorem~\ref{main thm}.
} the condition \eqref{gencond}:
\begin{equation}
		(1-\widehat{\mathscr{U}}_{21}(n))\widehat{\mathscr{U}}_{42}(n) + \widehat{\mathscr{U}}_{22}(n)\widehat{\mathscr{U}}_{41}(n) \neq 0.
\end{equation}
Then, from \eqref{C's-to-P} we can find:
	\begin{equation}\label{C2 C4}
	C_2(n,0,1) = \frac{\widehat{\mathscr{U}}_{42}(n)\widehat{\mathscr{U}}_{31}(n)-\widehat{\mathscr{U}}_{41}(n)\widehat{\mathscr{U}}_{32}(n)}{(1-\widehat{\mathscr{U}}_{21}(n))\widehat{\mathscr{U}}_{42}(n)+\widehat{\mathscr{U}}_{41}(n)\widehat{\mathscr{U}}_{22}(n)}, \qquad C_4(n,0,1) =- \frac{\widehat{\mathscr{U}}_{22}(n)\widehat{\mathscr{U}}_{31}(n)+[1-\widehat{\mathscr{U}}_{21}(n)]\widehat{\mathscr{U}}_{32}(n)}{(1-\widehat{\mathscr{U}}_{21}(n))\widehat{\mathscr{U}}_{42}(n)+\widehat{\mathscr{U}}_{41}(n)\widehat{\mathscr{U}}_{22}(n)}.
\end{equation}

 Let us recall from Theorem \ref{thm r=0 s=1} that

	\begin{equation}\label{URXrecall}
	\mathscr{U}(z)= R(z)\mathscr{X}(z) \begin{pmatrix}
		z^{-1} & 0 & 0 & 0\\
		0 & 1 & 0 & 0 \\
		0 & 0 & z^{-1} & 0 \\
		0 & 0 & 0 & 1
	\end{pmatrix},
\end{equation}
where $R(z)=A+z \begin{pmatrix}
	1 & 0 & 0 & 0\\
	0 & 0 & 0  & 0 \\
	0 & 0 & 1  & 0 \\
	0 & 0 & 0  & 0
\end{pmatrix}$ with
\begin{equation}\label{A}
	A =  \begin{pmatrix}
		\overset{\infty}{\mathscr{U}}_{1,11} - \overset{\infty}{\mathscr{X}}_{1,11} & -\overset{\infty}{\mathscr{X}}_{1,12} & 	\overset{\infty}{\mathscr{U}}_{1,13} - \overset{\infty}{\mathscr{X}}_{1,13} & - \overset{\infty}{\mathscr{X}}_{1,14}\\
		\overset{\infty}{\mathscr{U}}_{1,21} & 1 & \overset{\infty}{\mathscr{U}}_{1,23} & 0 \\
		\overset{\infty}{\mathscr{U}}_{1,31} - \overset{\infty}{\mathscr{X}}_{1,31} & -\overset{\infty}{\mathscr{X}}_{1,32} & 	\overset{\infty}{\mathscr{U}}_{1,33} - \overset{\infty}{\mathscr{X}}_{1,33} & - \overset{\infty}{\mathscr{X}}_{1,34}\\
		\overset{\infty}{\mathscr{U}}_{1,41} & 0 & \overset{\infty}{\mathscr{U}}_{1,43} & 1
	\end{pmatrix}.
\end{equation}
From \eqref{URXrecall} and \textbf{RH-$\mathscr{X}$4} we have
	\begin{equation}\label{u al be}
				\mathscr{U}(z) \begin{pmatrix}
					1 & 0 & 0 & 0\\
					0 & z^{n} & 0 & 0 \\
					0 & 0 & 1 & 0 \\
					0 & 0 & 0 & z^{-n}
				\end{pmatrix} = \al_n z^{-1} + \be_n + O(z), \qasq z \to 0,
\end{equation}
where
\begin{equation}
	\al_n = A P(n) \begin{pmatrix}
		1 & 0 & 0 & 0\\
		0 & 0 & 0 & 0 \\
		0 & 0 & 1 & 0 \\
		0 & 0 & 0 & 0
	\end{pmatrix},
\end{equation} and
\begin{equation}\label{betan}
	\be_n = A P(n) \begin{pmatrix}
		0 & 0 & 0 & 0\\
		0 & 1 & 0 & 0 \\
		0 & 0 & 0 & 0 \\
		0 & 0 & 0 & 1
	\end{pmatrix} + A P(n) \overset{\circ}{  \mathscr{X}}_1 \begin{pmatrix}
	1 & 0 & 0 & 0\\
	0 & 0 & 0 & 0 \\
	0 & 0 & 1 & 0 \\
	0 & 0 & 0 & 0
\end{pmatrix} +  \begin{pmatrix}
1 & 0 & 0 & 0\\
0 & 0 & 0 & 0 \\
0 & 0 & 1 & 0 \\
0 & 0 & 0 & 0
\end{pmatrix}P(n)\begin{pmatrix}
1 & 0 & 0 & 0\\
0 & 0 & 0 & 0 \\
0 & 0 & 1 & 0 \\
0 & 0 & 0 & 0
\end{pmatrix}.
\end{equation}
\begin{comment}
	content...
and
\begin{equation}
	\begin{split}
	\gamma_n & =  \begin{pmatrix}
	1 & 0 & 0 & 0\\
	0 & 0 & 0 & 0 \\
	0 & 0 & 1 & 0 \\
	0 & 0 & 0 & 0
\end{pmatrix} P(n) \overset{\circ}{  \mathscr{X}}_1 \begin{pmatrix}
		1 & 0 & 0 & 0\\
		0 & 0 & 0 & 0 \\
		0 & 0 & 1 & 0 \\
		0 & 0 & 0 & 0
	\end{pmatrix} +  \begin{pmatrix}
		1 & 0 & 0 & 0\\
		0 & 0 & 0 & 0 \\
		0 & 0 & 1 & 0 \\
		0 & 0 & 0 & 0
	\end{pmatrix}P(n)\begin{pmatrix}
		0 & 0 & 0 & 0\\
		0 & 1 & 0 & 0 \\
		0 & 0 & 0 & 0 \\
		0 & 0 & 0 & 1
	\end{pmatrix} \\ & + A P(n) \overset{\circ}{  \mathscr{X}}_1 \begin{pmatrix}
	0 & 0 & 0 & 0\\
	0 & 1 & 0 & 0 \\
	0 & 0 & 0 & 0 \\
	0 & 0 & 0 & 1
\end{pmatrix}+ A P(n) \overset{\circ}{  \mathscr{X}}_2 \begin{pmatrix}
1 & 0 & 0 & 0\\
0 & 0 & 0 & 0 \\
0 & 0 & 1 & 0 \\
0 & 0 & 0 & 0
\end{pmatrix}.
	\end{split}
\end{equation} 
\end{comment}
\noindent Notice that by \eqref{conditions in u problem} we have 
\begin{equation}\label{alphan=0}
	\al_n=0.
\end{equation}
 Comparing \eqref{u al be} with \textbf{RH-$\mathscr{U}$4} gives
\begin{equation}\label{betaUhat}
	\be_n  \equiv \widehat{\mathscr{U}}.
\end{equation}
Using \eqref{alphan=0} we can simplify  the second term on the right hand side of \eqref{betan}, indeed
\begin{equation}
	\begin{split}
		A P(n) \overset{\circ}{  \mathscr{X}}_1 \begin{pmatrix}
			1 & 0 & 0 & 0\\
			0 & 0 & 0 & 0 \\
			0 & 0 & 1 & 0 \\
			0 & 0 & 0 & 0
		\end{pmatrix} & = 	A P(n) \left[ \begin{pmatrix}
		1 & 0 & 0 & 0\\
		0 & 0 & 0 & 0 \\
		0 & 0 & 1 & 0 \\
		0 & 0 & 0 & 0
	\end{pmatrix}+ \begin{pmatrix}
	0 & 0 & 0 & 0\\
	0 & 1 & 0 & 0 \\
	0 & 0 & 0 & 0 \\
	0 & 0 & 0 & 1
\end{pmatrix} \right] \overset{\circ}{  \mathscr{X}}_1 \begin{pmatrix}
		1 & 0 & 0 & 0\\
		0 & 0 & 0 & 0 \\
		0 & 0 & 1 & 0 \\
		0 & 0 & 0 & 0
	\end{pmatrix} \\ & = A P(n) \begin{pmatrix}
0 & 0 & 0 & 0\\
0 & 1 & 0 & 0 \\
0 & 0 & 0 & 0 \\
0 & 0 & 0 & 1
\end{pmatrix} \overset{\circ}{  \mathscr{X}}_1 \begin{pmatrix}
1 & 0 & 0 & 0\\
0 & 0 & 0 & 0 \\
0 & 0 & 1 & 0 \\
0 & 0 & 0 & 0
\end{pmatrix}  = AP(n) \left(
\begin{array}{cccc}
0 & 0 & 0 & 0 \\
\overset{\circ}{  \mathscr{X}}_{1,21} & 0 & \overset{\circ}{  \mathscr{X}}_{1,23} & 0 \\
0 & 0 & 0 & 0 \\
\overset{\circ}{  \mathscr{X}}_{1,41} & 0 & \overset{\circ}{  \mathscr{X}}_{1,43} & 0 \\
\end{array}
\right). 
	\end{split}
\end{equation}
Combining this, with \eqref{betan} and \eqref{betaUhat} and after some simplifications we find
\begin{equation}\label{Uhat}
	\widehat{\mathscr{U}} = A P(n) \left(
	\begin{array}{cccc}
		0 & 0 & 0 & 0 \\
		\overset{\circ}{  \mathscr{X}}_{1,21} & 1 & \overset{\circ}{  \mathscr{X}}_{1,23} & 0 \\
		0 & 0 & 0 & 0 \\
		\overset{\circ}{  \mathscr{X}}_{1,41} & 0 & \overset{\circ}{  \mathscr{X}}_{1,43} & 1 \\
	\end{array}
	\right)  +  \left(
	\begin{array}{cccc}
		P_{11} & 0 & P_{13} & 0 \\
		0 & 0 & 0 & 0 \\
		P_{31} & 0 & P_{33} & 0 \\
		0 & 0 & 0 & 0 \\
	\end{array}
	\right).
\end{equation}
We therefore find
\begin{eqnarray}
	\widehat{\mathscr{U}}_{21} &=&		\overset{\infty}{  \mathscr{X}}_{1,12} \left(A_{21} P_{12}+A_{23}
	P_{32}+P_{22}\right)+	\overset{\infty}{  \mathscr{X}}_{1,32} \left(A_{21}
	P_{14}+A_{23} P_{34}+P_{24}\right), \label{634} \\
	\widehat{\mathscr{U}}_{42} &=&	A_{41} P_{12}+A_{43} P_{32}+P_{42}, \\
	\widehat{\mathscr{U}}_{41} &=&		\overset{\infty}{  \mathscr{X}}_{1,12} \left(A_{41} P_{12}+A_{43}
	P_{32}+P_{42}\right)+	\overset{\infty}{  \mathscr{X}}_{1,32} \left(A_{41}
	P_{14}+A_{43} P_{34}+P_{44}\right), \\
	\widehat{\mathscr{U}}_{22} &=&	A_{21} P_{12}+A_{23} P_{32}+P_{22}, \\
	\widehat{\mathscr{U}}_{32}  &=&	A_{31} P_{12}+A_{32} P_{22}+A_{33} P_{32}+A_{34}
	P_{42}, \label{639}\\
	\widehat{\mathscr{U}}_{31} &=&		\overset{\infty}{  \mathscr{X}}_{1,12} \left(A_{31} P_{12}+A_{32} P_{22}+A_{33}
	P_{32}+A_{34} P_{42}\right) \nonumber \\ &+&	\overset{\infty}{  \mathscr{X}}_{1,32} \left(A_{31}
	P_{14}+A_{32} P_{24}+A_{33} P_{34}+A_{34}
	P_{44}\right)+P_{31}, \label{638}
\end{eqnarray}
where we have used that $\overset{\circ}{  \mathscr{X}}_{1,21}=\overset{\infty}{  \mathscr{X}}_{1,12}$ and $\overset{\circ}{  \mathscr{X}}_{1,41}=\overset{\infty}{  \mathscr{X}}_{1,32}$ using part (c) of Theorem \ref{Thm WW}. Below, we recall the expressions for the relevant entries of the matrix \( A \) in terms of the \(\mathscr{X}\)-RHP data, using notations that simplify the resulting formulas.

To streamline the expressions, we define the determinant of \(2 \times 2\) minors of \(P(n)\) as follows:
\begin{equation}
	D^{rs}_{jk} := P_{jk}P_{rs} - P_{js}P_{rk}, \qquad 1 \leq j < r \leq 4, \quad 1 \leq k < s \leq 4.
\end{equation}
Assuming the generic condition, $D^{33}_{11} \neq 0$ (cf. note that this is  the condition we assumed in Theorem \ref {thm r=0 s=1}),  we have

\noindent\begin{minipage}{.23\linewidth}
\begin{alignat*}{2}
	&	A_{21} &&= -\frac{D^{33}_{21}}{D^{33}_{11}},  \\
	& A_{23}&&= -\frac{D^{23}_{11}}{D^{33}_{11}}, 
\end{alignat*}	
\end{minipage}
\noindent\begin{minipage}{.23\linewidth}
	\begin{alignat*}{2}
		&	A_{41} &&= \frac{D^{43}_{31}}{D^{33}_{11}},  \\
		& A_{43}&&= -\frac{D^{43}_{11}}{D^{33}_{11}}, 
	\end{alignat*}	
\end{minipage}
\noindent\begin{minipage}{.23\linewidth}
	\begin{alignat*}{2}
		&	A_{32} &&= -\overset{\infty}{\mathscr{X}}_{1,32} ,  \\
		& A_{34}&&= -\overset{\infty}{\mathscr{X}}_{1,34} , 
	\end{alignat*}	
\end{minipage}
\noindent\begin{minipage}{.23\linewidth}
	\begin{alignat*}{2}
		&	A_{31} &&= \frac{\overset{\infty}{\mathscr{X}}_{1,32}D^{33}_{21} -\overset{\infty}{\mathscr{X}}_{1,34}D^{43}_{31} }{D^{33}_{11}},  \\
		& A_{33}&&= \frac{\overset{\infty}{\mathscr{X}}_{1,32}D^{23}_{11} +\overset{\infty}{\mathscr{X}}_{1,34}D^{43}_{11} }{D^{33}_{11}}, 
	\end{alignat*}	
\end{minipage}
\begin{comment}
	\begin{equation}\label{U121inf}
		A_{21}  = \frac{P_{31}P_{23}  -P_{33}P_{21}}{P_{11}P_{33}  -P_{13}P_{31}} ,
	\end{equation}
	
	\begin{equation}
		A_{23} = \frac{P_{13}P_{21}  -P_{11}P_{23}}{P_{11}P_{33}  -P_{13}P_{31}} 
	\end{equation}
	
	\begin{equation}\label{U141inf}
		A_{41}  = \frac{P_{43}P_{31}  -P_{33}P_{41}}{P_{11}P_{33}  -P_{13}P_{31}} 
	\end{equation}
	
	\begin{equation}
		A_{43} = \frac{P_{13}P_{41}  -P_{11}P_{43}}{P_{11}P_{33}  -P_{13}P_{31}} 
	\end{equation}
	
	\begin{equation}\label{X1inf32}
		A_{32} = -\overset{\infty}{\mathscr{X}}_{1,32} 
	\end{equation}
	
	\begin{equation}\label{X1inf34}
		A_{34}  = -\overset{\infty}{\mathscr{X}}_{1,34} 
	\end{equation}
	
	\begin{equation}
		A_{31}  = \frac{\overset{\infty}{\mathscr{X}}_{1,32}[P_{33}P_{21} -P_{31}P_{23}]+\overset{\infty}{\mathscr{X}}_{1,34}[P_{33}  P_{41} -P_{31} P_{43}]}{P_{11}P_{33}  -P_{13}P_{31}} 
	\end{equation}
	
	\begin{equation}
		A_{33}  = \frac{\overset{\infty}{\mathscr{X}}_{1,32}[P_{11}P_{23} -P_{13}P_{21}]+\overset{\infty}{\mathscr{X}}_{1,34}[P_{11}  P_{43} -P_{13} P_{41}]}{P_{11}P_{33}  -P_{13}P_{31}} 
	\end{equation}
\end{comment}

\noindent Using \eqref{A} and Theorem \ref{thm r=0 s=1} we find the following expression dor the numerator of $C_2(n)$:

\begin{equation}\label{num C2}
	\begin{split}
&	\widehat{\mathscr{U}}_{42}(n)\widehat{\mathscr{U}}_{31}(n)-\widehat{\mathscr{U}}_{41}(n)\widehat{\mathscr{U}}_{32}(n) = \\ & \frac{-1}{D^{33}_{11}} \left[ P_{31}   \left( P_{41} D_{12}^{33}-P_{42} D_{11}^{33}+P_{43} D_{11}^{32} \right)  \right]   \\ &   -\frac{\left(\overset{\infty}{\mathscr{X}}_{1,32}\right)^2}{D^{33}_{11}} \left[ D_{11}^{32} D_{23}^{44} -  D_{23}^{34} D_{11}^{42} - D_{11}^{22} D_{33}^{44} - D_{21}^{32} D_{13}^{44} + P_{14}P_{41}D_{22}^{33} - P_{14}P_{42}D_{21}^{33} - P_{13}P_{41}D_{22}^{34} + P_{13}P_{42}D_{21}^{34}  \right].
	\end{split}
\end{equation}
Similarly, for the numerator of $C_4(n)$ we have
\begin{equation}\label{num C4}
	\begin{split}
	& -\widehat{\mathscr{U}}_{22}(n)\widehat{\mathscr{U}}_{31}(n)-[1-\widehat{\mathscr{U}}_{21}(n)]\widehat{\mathscr{U}}_{32}(n) = \\ & \frac{-1}{D^{33}_{11}} \left[ (\overset{\infty}{\mathscr{X}}_{1,32} - P_{31}) \left( P_{32} D_{11}^{23}-P_{31} D_{12}^{23}-P_{33} D_{11}^{22} \right) + \overset{\infty}{\mathscr{X}}_{1,34}  \left( P_{41} D_{12}^{33}-P_{42} D_{11}^{33}+P_{43} D_{11}^{32} \right)  \right]   \\ &   -\frac{\overset{\infty}{\mathscr{X}}_{1,32}\overset{\infty}{\mathscr{X}}_{1,34}}{D^{33}_{11}} \left[ D_{11}^{32} D_{23}^{44} -  D_{23}^{34} D_{11}^{42} - D_{11}^{22} D_{33}^{44} - D_{21}^{32} D_{13}^{44} + P_{14}P_{41}D_{22}^{33} - P_{14}P_{42}D_{21}^{33} - P_{13}P_{41}D_{22}^{34} + P_{13}P_{42}D_{21}^{34}  \right],
	\end{split}
\end{equation}
while for the shared denominator of $C_2(n)$ and $C_4(n)$ we have
\begin{equation}\label{den C}
	\begin{split}
	& 	(1-\widehat{\mathscr{U}}_{21}(n))\widehat{\mathscr{U}}_{42}(n)+\widehat{\mathscr{U}}_{41}(n)\widehat{\mathscr{U}}_{22}(n)  = \frac{-1}{D^{33}_{11}} \left( P_{41} D_{12}^{33}-P_{42} D_{11}^{33}+P_{43} D_{11}^{32} \right) \\ & -\frac{\overset{\infty}{\mathscr{X}}_{1,32}}{D^{33}_{11}} \left( D_{11}^{32} D_{23}^{44} -  D_{23}^{34} D_{11}^{42} - D_{11}^{22} D_{33}^{44} - D_{21}^{32} D_{13}^{44} + P_{14}P_{41}D_{22}^{33} - P_{14}P_{42}D_{21}^{33} - P_{13}P_{41}D_{22}^{34} + P_{13}P_{42}D_{21}^{34}    \right).
	\end{split}
\end{equation}
Let the quantity
\begin{equation}\label{E}
	E(n) := - D^{33}_{11} \left( (1-\widehat{\mathscr{U}}_{21}(n))\widehat{\mathscr{U}}_{42}(n)+\widehat{\mathscr{U}}_{41}(n)\widehat{\mathscr{U}}_{22}(n) \right)
\end{equation}
denote the remaining expression after clearing out the denominator of \eqref{den C}.  Noticing the common terms among the expressions \eqref{num C2}, \eqref{num C4}, and \eqref{den C}, we obtain (generically,  $E(n) \neq 0$)
\begin{equation}
	C_2(n) = \overset{\infty}{\mathscr{X}}_{1,32} + \frac{\left(P_{31} -\overset{\infty}{\mathscr{X}}_{1,32} \right)  \left( P_{41} D_{12}^{33}-P_{42} D_{11}^{33}+P_{43} D_{11}^{32} \right)}{E(n)},
\end{equation}
and
\begin{equation}
	C_4(n) = \overset{\infty}{\mathscr{X}}_{1,34} + \frac{(P_{31}-\overset{\infty}{\mathscr{X}}_{1,32}) \left(P_{33} D_{11}^{22} +P_{31} D_{12}^{23} -P_{32} D_{11}^{23} \right)}{E(n)}.
\end{equation}
Combining these with equations \eqref{h C U},  \eqref{U1012 to Xdata}, \eqref{U1032 to Xdata}, and \eqref{U1042 to Xdata} and straightforward simplifications we obtain the following exact formula for $h^{(0,1)}_{n-1}$:
\begin{equation}\label{h exact}
	\frac{-1}{h^{(0,1)}_{n-1}} =  \overset{\infty}{\mathscr{X}}_{1,31} + \frac{P_{31} -\overset{\infty}{\mathscr{X}}_{1,32}}{E(n)D^{33}_{11}} \left[D^{43}_{31}\left(P_{33} D_{11}^{22} +P_{31} D_{12}^{23} -P_{32} D_{11}^{23} \right) - D^{33}_{21}\left( P_{41} D_{12}^{33}-P_{42} D_{11}^{33}+P_{43} D_{11}^{32} \right)\right].
\end{equation}
Now we focus on finding the large $n$ asymptotics of the right hand side of the above equation by recalling equation (4.17) of \cite{GI} which gives the asymptotic formula for $P(n)$,
\begin{equation}\label{T+H P(n) Expansion 1}
	P(n)=\begin{pmatrix}
		-C_{\rho}(0)\al(0)\mathcal{R}_{1,14}(0;n)-\mathcal{R}_{1,12}(0;n) & 0 & \mathcal{R}_{1,14}(0;n) & -\al(0)\\[7pt]
		-1 & -\di \frac{\mathcal{R}_{1,23}(0;n)}{\al(0)} & 0 & -\al(0)\mathcal{R}_{1,21}(0;n) \\[7pt]
		-C_{\rho}(0)\al(0)\mathcal{R}_{1,34}(0;n)-\mathcal{R}_{1,32}(0;n) & -\di \frac{1}{\al(0)} & \mathcal{R}_{1,34}(0;n) & 0 \\[7pt]
		-C_{\rho}(0)\al(0) & -\di \frac{\mathcal{R}_{1,43}(0;n)}{\al(0)} & 1 & -\al(0)\mathcal{R}_{1,41}(0;n) 
	\end{pmatrix} + O{(e^{-2cn})},
\end{equation}
as $n \to \infty$.

Notice that 
\begin{align}
D^{33}_{11} = P_{11}P_{33}  -P_{13}P_{31} & =  \mathcal{R}_{1,32}(0;n)\mathcal{R}_{1,14}(0;n) - \mathcal{R}_{1,12}(0;n)\mathcal{R}_{1,34}(0;n)+ O(e^{-3cn}), \label{11 33} \\
D^{43}_{31} =	P_{31} P_{43} -	P_{33}  P_{41}  & =-\mathcal{R}_{1,32}(0;n)+ O(e^{-2cn}), \\
D^{22}_{11} =	P_{11}P_{22} -P_{12}P_{21} & = \left(\frac{\mathcal{R}_{1,23}(0;n)}{\al(0)} \right)\left(C_{\rho}(0)\al(0)\mathcal{R}_{1,14}(0;n)+\mathcal{R}_{1,12}(0;n)\right)+ O(e^{-3cn}) ,\\
D^{23}_{12} =	P_{12}P_{23} -P_{13}P_{22} & = \frac{\mathcal{R}_{1,14}(0;n)\mathcal{R}_{1,23}(0;n)}{\al(0)}+ O(e^{-3cn}),\\
 D^{23}_{11} =	P_{11}P_{23} -P_{13}P_{21} & = \mathcal{R}_{1,14}(0;n)+ O(e^{-2cn}),\\
 D^{33}_{21} =	P_{21}P_{33}  -P_{31}P_{23} & =  -\mathcal{R}_{1,34}(0;n)+ O(e^{-2cn})    , \\
  D^{33}_{12} =	P_{12}P_{33}  -P_{13}P_{32} & =   \frac{\mathcal{R}_{1,14}(0;n)}{\al(0)}+ O(e^{-2cn})    , \\
  D^{32}_{11} =	P_{11}P_{32} -P_{12}P_{31} & =  C_{\rho}(0)\mathcal{R}_{1,14}(0;n)+\frac{\mathcal{R}_{1,12}(0;n)}{\al(0)}+ O(e^{-2cn}) ,\label{11 32}
\end{align}
Notice also that
\begin{align}
P_{33}	D^{43}_{31}D^{22}_{11} & = -\frac{\mathcal{R}_{1,34}(0;n)\mathcal{R}_{1,23}(0;n)\mathcal{R}_{1,32}(0;n)}{\al(0)} \left(C_{\rho}(0)\al(0)\mathcal{R}_{1,14}(0;n)+\mathcal{R}_{1,12}(0;n)\right)+ O(e^{-5cn}) , \\
P_{31}	D^{43}_{31}D^{23}_{12}  & = (C_{\rho}(0)\al(0)\mathcal{R}_{1,34}(0;n)+\mathcal{R}_{1,32}(0;n)) \frac{\mathcal{R}_{1,32}(0;n)\mathcal{R}_{1,14}(0;n)\mathcal{R}_{1,23}(0;n)}{\al(0)}+ O(e^{-5cn}),\\
-P_{32}	D^{43}_{31}D^{23}_{11}  & = -\frac{\mathcal{R}_{1,32}(0;n)\mathcal{R}_{1,14}(0;n)}{\al(0)}+ O(e^{-3cn}),\\
-P_{41} D^{33}_{21}	D^{33}_{12}  & =   -C_{\rho}(0) \mathcal{R}_{1,34}(0;n)\mathcal{R}_{1,14}(0;n)+ O(e^{-3cn})    , \\
P_{42}D^{33}_{21}	D^{33}_{11}  & = \frac{\mathcal{R}_{1,43}(0;n)}{\al(0)}\left( \mathcal{R}_{1,34}(0;n) \mathcal{R}_{1,32}(0;n)\mathcal{R}_{1,14}(0;n) -  \left(\mathcal{R}_{1,34}(0;n)\right)^2 \mathcal{R}_{1,12}(0;n)\right)+ O(e^{-5cn}),  \\
-P_{43} D^{33}_{21}	D^{32}_{11} & =  C_{\rho}(0)\mathcal{R}_{1,34}(0;n)\mathcal{R}_{1,14}(0;n)+\frac{\mathcal{R}_{1,34}(0;n)\mathcal{R}_{1,12}(0;n)}{\al(0)}+ O(e^{-3cn}).
\end{align}
Therefore we observe that 
\begin{equation}\label{remaining part} \boxed{
	D^{43}_{31}\left(P_{33} D_{11}^{22} +P_{31} D_{12}^{23} -P_{32} D_{11}^{23} \right) - D^{33}_{21}\left( P_{41} D_{12}^{33}-P_{42} D_{11}^{33}+P_{43} D_{11}^{32} \right) = -\frac{\mathcal{R}_{1,32}(0;n)\mathcal{R}_{1,14}(0;n)}{\al(0)}+ O(e^{-3cn}),}
\end{equation}
where the left hand side appears in \eqref{h exact}.
\subsection{Asymptotics of Relevant Entries in $\overset{\infty}{ \mathscr{X}}_1(n)$}\label{sec Xinfty}
In view of equations \eqref{den C}, \eqref{E}, and \eqref{h exact} we need large $n$  asymptotics for $\overset{\infty}{\mathscr{X}}_{1,31}$ and $\overset{\infty}{\mathscr{X}}_{1,32}$. To this end, let us recall some relevant information from \cite{GI}. First of all let us start with \textit{the model Riemann-Hilbert problem for the pair $(\phi,w)$}:

\begin{itemize}
	\item \textbf{RH-$\La$1} \quad  $\La$  is holomorphic in $\C \setminus  \T$.
	
	\item \textbf{RH-$\La$2} \quad   $\La_+(z)=\La_-(z)J_{\La}(z)$, for $z \in \T$, where 
	\begin{equation*}\label{T+H model Jump}
		J_{\La}(z) = \begin{pmatrix}
			0 & 0 & 0 & -\phi(z) \\
			\di	-\frac{w(z)}{\phi(z)} & 0 & \di \tilde{\phi}(z) - \frac{w(z)\tilde{w}(z)}{\phi(z)} & 0 \\
			0 & \di -\frac{1}{\tilde{\phi}(z)} & 0 & 0 \\
			\di	\frac{1}{\phi(z)} & 0 &  \di  \frac{\tilde{w}(z)}{\phi(z)} & 0
		\end{pmatrix}.
	\end{equation*}  
	
	\item \textbf{RH-$\La$3} \quad As $z \to \infty$, we have $\La(z)=\di I+\frac{ \overset{\infty}{\La}_1}{z}+\frac{\overset{\infty}{\La}_2}{z^2} + O(z^{-3})$.
\end{itemize}
In \cite{GI} it was shown that if we consider $D_n(\phi,d\phi;1,1)$, where $d$ is of Szeg{\H o}-type and further satisfies the condition $d(z)\tilde{d}(z)=1$ on the unit circle\footnote{which makes $J_{\La, 23} \equiv 0$.}, then this model problem is explicitly solvable and its solution can be written as 
\begin{equation}\label{Lambda}
	\La(z)= \La^{-1}_{\infty}\!\begin{pmatrix}
		1 & 0 & 0 & 0 \\
		C_{\rho}(z) & 1 & 0 & 0 \\
		0 & 0 & 1 & 0 \\
		0 & 0 & 0 & 1
	\end{pmatrix} \di \!\times \!\begin{cases}
		\begin{pmatrix}
			\di -\be(z) & 0 & 0 & 0 \\
			0 & 0 & \di \frac{1}{\tilde{\al}(z)\be(z) \al(z)} & 0 \\
			0 & \di -\tilde{\al}(z) & 0 & 0 \\
			0 & 0 & 0 & \di -\al(z)
		\end{pmatrix}\!, & |z|<1, \\
		\begin{pmatrix}
			0 & \be(z) & 0 & 0 \\
			0 & 0 & 0 & \di \frac{1}{\be(z)\tilde{\al}(z)\al(z)} \\
			0 & 0 & \tilde{\al}(z) & 0 \\
			\al(z) & 0 & 0 & 0
		\end{pmatrix}\!, & |z|>1,
	\end{cases}\hspace{-10mm}
\end{equation}
where 
\begin{equation}\label{Lambda infty}
	\La^{-1}_{\infty} = \begin{pmatrix}
		0 & 0 & 0 & 1 \\
		1 & 0 & 0 & 0 \\
		0 & 0 & \di \frac{1}{\al(0)} & 0 \\
		0 & \al(0) & 0 & 0
	\end{pmatrix},
\end{equation}
and the functions $\al$, $\be$, and $C_{\rho}$ are defined in \eqref{al be} and  \eqref{C rho}.

In what follows, we connect the desired quantities 
$\overset{\infty}{\mathscr{X}}_{1,31}$ and $\overset{\infty}{\mathscr{X}}_{1,32}$ 
to data from the $\Lambda$-RHP as well as $\mathcal{R}$, the solution of the small-norm Riemann–Hilbert problem:

\begin{itemize}
	\item \textbf{RH-$\mathcal{R}$1} \quad $\mathcal{R}$ is holomorphic in $\C \setminus \Gamma_{\mathcal{R}}$.
	\item \textbf{RH-$\mathcal{R}$2} \quad $\mathcal{R}_+(z;n)=\mathcal{R}_-(z;n)J_{\mathcal{R}}(z;n)$, for $z \in \Gamma_{\mathcal{R}}$.
	\item \textbf{RH-$\mathcal{R}$3} \quad As $z \to \infty$,
	\[
	\mathcal{R}(z;n) = I + \frac{\overset{\infty}{\mathcal{R}}_1(n)}{z} + \frac{\overset{\infty}{\mathcal{R}}_2(n)}{z^2} + O(z^{-3}).
	\]
\end{itemize}

The contour $\Gamma_{\mathcal{R}} := \Gamma_i' \cup \Gamma_o'$ consists of two counter-clockwise oriented circles:  
$\Gamma'_i$ has radius $r_* \in (r_0,1)$ and $\Gamma'_o$ has radius $1/r_*$.  
Here $r_*$ is any number satisfying $r_0 < r_* < 1$ (see \eqref{annulus} and \eqref{r0} for the definition and meaning of $r_0$).  
The jump matrix $J_{\mathcal{R}}$ is given by
\begin{equation}\label{JR000}
	J_{\mathcal{R}}(z;n)-I =
	\begin{cases}
		z^n \!\cdot\!
		\begin{pmatrix}
			0 & g_{12}(z) & 0 & g_{14}(z) \\
			0 & 0 & g_{23}(z) & 0 \\
			0 & 0 & 0 & 0 \\
			0 & 0 & g_{43}(z) & 0
		\end{pmatrix}, & z \in \Gamma_i', \\[1em]
		z^{-n} \!\cdot\!
		\begin{pmatrix}
			0 & 0 & 0 & 0 \\
			g_{21}(z) & 0 & 0 & 0 \\
			0 & g_{32}(z) & 0 & g_{34}(z) \\
			g_{41}(z) & 0 & 0 & 0
		\end{pmatrix}, & z \in \Gamma_o',
	\end{cases}
\end{equation}
with
\begin{align*}
	g_{12}(z) &= - \frac{\alpha(z)}{\phi(z)\beta(z)} - \frac{\tilde{w}(z) C_{\rho}(z)}{\phi(z)\beta(z)\tilde{\alpha}(z)}, &
	g_{14}(z) &= \frac{\tilde{w}(z)}{\phi(z)\beta(z)\tilde{\alpha}(z) \alpha(0)}, \\
	g_{23}(z) &= - \frac{\alpha(0) \tilde{w}(z) \beta(z)}{\tilde{\phi}(z)\tilde{\alpha}(z)}, &
	g_{43}(z) &= - \alpha^2(0) \left( \frac{\alpha(z)\beta(z)}{\tilde{\phi}(z)} + \frac{\beta(z)\tilde{w}(z) C_{\rho}(z)}{\tilde{\alpha}(z)\tilde{\phi}(z)} \right), \\
	g_{21}(z) &= \frac{w(z)\beta(z)}{\phi(z)\alpha(z)}, &
	g_{32}(z) &= - \frac{1}{\alpha(0)\tilde{\phi}(z)} \left( \frac{\tilde{\alpha}(z)}{\beta(z)} - w(z)\tilde{\alpha}^2(z)\beta(z)\alpha(z)C_{\rho}(z) \right), \\
	g_{34}(z) &= \frac{w(z)\tilde{\alpha}^2(z)\beta(z)\alpha(z)}{\tilde{\phi}(z)\alpha^2(0)}, &
	g_{41}(z) &= -\frac{\alpha(0)}{\phi(z)} \left( \frac{1}{\tilde{\alpha}(z)\beta(z)\alpha^2(z)} - \frac{w(z)\beta(z)C_{\rho}(z)}{\alpha(z)} \right).
\end{align*}
From (\ref{JR000}) it follows  that  the jump matrix $J_{\mathcal{R}}$  satisfies on $\Gamma_{\mathcal{R}}$ 
the small-norm estimate,
\begin{equation}\label{smallnorn000}
	||J_{\mathcal{R}} - I||_{L_{2} \cap L_{\infty}} \leq C e^{-cn},
\end{equation}
for some positive $C$ and $c = -\log r_*$. Therefore, by standard theory of small-norm Riemann-Hilbert problems \cite{Deiftetal,Deiftetal2}, there exists $n_*$  such that  for all $n > n_*$  the $\mathcal{R}$ - RH problem is solvable and 
\begin{equation}\label{Appendix Small norm R asymptotics}
	\mathcal{R}(z;n) = I + \mathcal{R}_1(z;n) + \mathcal{R}_2(z;n) + \mathcal{R}_3(z;n) + \cdots, \qquad \hspace{0.8cm} z \in \C \setminus \Gamma_{\mathcal{R}}, \qquad  n \geq n_*,
\end{equation}
where each $\mathcal{R}_k$  is of order $O(e^{-kcn})$ and they can be found recursively from	
\begin{equation}\label{Rk}
	\mathcal{R}_k(z;n) = \frac{1}{2\pi i}\int_{\Gamma_{\mathcal{R}}} \frac{\left[ \mathcal{R}_{k-1}(\mu;n)\right]_-  \left( J_{\mathcal{R}}(\mu;n)-I\right) }{\mu-z}d\mu,  \qquad  z \in \C \setminus \Gamma_{\mathcal{R}}, \qquad k\geq1.
\end{equation}
Note that this recurrence also means that
\begin{equation*}\label{RkRk+1}
	\mathcal{R}_{k+1}(z;n) = o(\mathcal{R}_k(z;n)),\quad n \to \infty, \qquad  z \in \C \setminus \Gamma_{\mathcal{R}}, \qquad k\geq1.
\end{equation*}
More precisely we have 
\begin{equation}\label{Rk est}
	\mathcal{R}_{k,ij}(z;n) = \frac{O{(e^{-kcn})}}{|z|+1}, \qquad n \to \infty, \quad k\geq 1,
\end{equation}
uniformly for $z \in \C \setminus \Gamma_R$, and the positive constant $c$ is the same as in (\ref{smallnorn000}).  
From \eqref{Rk} we have 
\begin{equation}\label{R1000}
	\mathcal{R}_1(z;n)=\frac{1}{2\pi \ic}\int_{\Gamma_{\mathcal{R}}} \frac{J_{\mathcal{R}}(\mu;n)-I}{\mu-z} d\mu = \begin{pmatrix}
		0 & \mathcal{R}_{1,12}(z;n) & 0 & \mathcal{R}_{1,14}(z;n) \\
		\mathcal{R}_{1,21}(z;n) & 0 & \mathcal{R}_{1,23}(z;n) & 0 \\
		0 & \mathcal{R}_{1,32}(z;n) & 0 & \mathcal{R}_{1,34}(z;n) \\
		\mathcal{R}_{1,41}(z;n) & 0 & \mathcal{R}_{1,43}(z;n) & 0
	\end{pmatrix}, \end{equation}
where	\begin{equation}\label{T+H entries of R_1}
	\begin{split}
		& \mathcal{R}_{1,jk}(z;n) = \frac{1}{2\pi i}\int_{\Gamma'_i} \frac{\mu^ng_{jk}(\mu)}{\mu-z}d\mu, \qquad \ \  jk=12,14,23,43, \\
		&  \mathcal{R}_{1,jk}(z;n) = \frac{1}{2\pi i}\int_{\Gamma'_o} \frac{\mu^{-n}g_{jk}(\mu)}{\mu-z}d\mu, \qquad jk=21,32,34,41.
	\end{split}
\end{equation}
So, in view of \eqref{smallnorn000} and \eqref{Rk}, there exists a positive constant $C_*$ such that 
	\begin{equation}\label{upper bound on R_1 jk 0;n}
		|\mathcal{R}_{1,jk}(0;n)| \leq C_*r^{n}_*, \qquad jk=12,14,23,43,21,32,34,41, \qquad n \geq n_*.
	\end{equation}
	Recalling \textbf{RH-$\mathscr{X}$3}, let us consider
\begin{equation}\label{GG}
	\mathscr{G}(z;n) := \mathscr{X}(z;n) \begin{pmatrix}
		z^{-n} & 0 & 0 & 0\\
		0 & 1 & 0 & 0 \\
		0 & 0 & z^{n} & 0 \\
		0 & 0 & 0 & 1
	\end{pmatrix},
\end{equation}
whose asymptotic behavior as $z \to \infty$ reads
\begin{equation}
	\mathscr{G}(z;n) = I+\frac{ \overset{\infty}{  \mathscr{X}}_1}{z}+\frac{\overset{\infty}{  \mathscr{X}}_2}{z^2} + O(z^{-3}).
\end{equation}
As it is shown in \cite{GI}, for $z \in \Omega_{\infty}$, we can write
\begin{equation}\label{GG}
	\mathscr{G}(z;n) \equiv \mathcal{R}(z;n)\La(z), \qquad z \in \Omega_{\infty},
\end{equation}
and thus from  \textbf{RH-$\mathcal{R}$3} and  \textbf{RH-$\Lambda$3} which it is clear that
\begin{equation}\label{X1inf to R and La}
	\overset{\infty}{  \mathscr{X}}_1(n) = \overset{\infty}{  \mathcal{R}}_1(n) + \overset{\infty}{\La}_1.
\end{equation}
From \eqref{Appendix Small norm R asymptotics}, \eqref{Rk}, and \eqref{Rk est} we obtain that 
\begin{equation}\label{R1infty}
	\overset{\infty}{  \mathcal{R}}_1(n)  = \begin{pmatrix}
		0 & \overset{\infty}{  \mathcal{R}}_{1,12}(n) & 0 & \overset{\infty}{  \mathcal{R}}_{1,14}(n) \\
		\overset{\infty}{  \mathcal{R}}_{1,21}(n) & 0 & \overset{\infty}{  \mathcal{R}}_{1,23}(n) & 0 \\
		0 & \overset{\infty}{  \mathcal{R}}_{1,32}(n) & 0 & \overset{\infty}{  \mathcal{R}}_{1,34}(n) \\
		\overset{\infty}{  \mathcal{R}}_{1,41}(n) & 0 & \overset{\infty}{  \mathcal{R}}_{1,43}(n) & 0
	\end{pmatrix} = -\frac{1}{2\pi \ic} \int_{\Gamma_{\mathcal{R}}} (J_{\mathcal{R}}(\mu;n)-I) \dd \mu + O(e^{-2cn}),
\end{equation}
so
\begin{equation}\label{T+H Entries of the subleading term of R in expansion near infty}
	\begin{split}
		& \overset{\infty}{  \mathcal{R}}_{1,jk}(n) = -\frac{1}{2\pi {\ic}}\int_{\Gamma'_i} \mu^{n}g_{jk}(\mu)\,{\rm d}\mu + O(e^{-2cn}) , \qquad \ \ jk=12,14,23,43, \\
		& \overset{\infty}{  \mathcal{R}}_{1,jk}(n) = -\frac{1}{2\pi {\ic}}\int_{\Gamma'_o} \mu^{-n}g_{jk}(\mu)\,{\rm d}\mu + O(e^{-2cn}), \qquad jk=21,32,34,41.
	\end{split}
\end{equation}
Recalling \eqref{T+H entries of R_1}, the equations \eqref{T+H Entries of the subleading term of R in expansion near infty} can be written as
\begin{equation}\label{T+H Entries of the subleading term of R in expansion near infty1}
	\begin{split}
		& \overset{\infty}{  \mathcal{R}}_{1,jk}(n) = - \mathcal{R}_{1,jk}(0;n+1) + O(e^{-2cn}) , \qquad \ \ jk=12,14,23,43, \\
		& \overset{\infty}{  \mathcal{R}}_{1,jk}(n) = - \mathcal{R}_{1,jk}(0;n-1) + O(e^{-2cn}), \qquad jk=21,32,34,41.
	\end{split}
\end{equation}

Now let us turn our attention to $\overset{\infty}{\La}_1$. From \eqref{T+H integrals in the norm}, as $z \to \infty$ we have

\begin{equation}\label{11111a}
	\al(z) = 1 -\frac{1}{2\pi \ic z} \int_{\T} \ln (\phi(\tau)) \dd \tau + O(z^{-2}), \qquad \be(z) = 1 -\frac{1}{2\pi \ic z} \int_{\T} \ln (d(\tau)) \dd \tau + O(z^{-2}),
\end{equation}
\begin{equation}\label{2222a}
	\tilde{\al}(z) = \al(0) \left( 1 +\frac{1}{2\pi \ic z} \int_{\T} \ln (\phi(\tau)) \frac{\dd \tau}{\tau^2} + O(z^{-2}) \right), \qquad C_{\rho}(z) = -\frac{1}{2\pi \ic z} \int_{\T} \rho(\tau) \dd \tau + O(z^{-2}),
\end{equation}
where
\begin{equation}
	\rho(\tau) = \di -\frac{1}{\be_-(\tau) \be_+(\tau) \tilde{\al}_-(\tau) \al_+(\tau)}, \qquad \tau \in \T.
\end{equation}
For $|z|>1$, we can write \eqref{Lambda} as
\begin{equation}\label{La z large}
	\La(z)=\left(
	\begin{array}{cccc}
		\alpha(z)  & 0 & 0 & 0 \\
		0 & \beta(z)  & 0 & 0 \\
		0 & 0 & \frac{\tilde{\al}(z)}{\alpha(0)} & 0 \\
		0 & \alpha(0) \beta(z)  C_{\rho}(z) & 0 & \frac{\alpha(0)}{\tilde{\al}(z) \alpha(z)  \beta(z) } \\
	\end{array}
	\right).
\end{equation}
Therefore in view of \eqref{X1inf to R and La} and \eqref{La z large} we find
\begin{equation}\label{X1inftyjk}
	\overset{\infty}{  \mathscr{X}}_{1,jk}(n) = \overset{\infty}{  \mathcal{R}}_{1,jk}(n), \qquad jk \in \{ 12, 13, 14, 21, 23, 24, 31, 32, 34, 41, 43\}. 
\end{equation}
Therefore, in particular, we obtain
\begin{equation}\label{relevant Xinfs} \boxed{
	\begin{split}
	\overset{\infty}{  \mathscr{X}}_{1,31}(n) & =  O(e^{-2cn}), \\
\overset{\infty}{  \mathscr{X}}_{1,32}(n) &  =  - \mathcal{R}_{1,32}(0;n-1) + O(e^{-2cn}),		
	\end{split}}
\end{equation}
in view of \eqref{R1infty} and \eqref{T+H Entries of the subleading term of R in expansion near infty1}. Recalling \eqref{T+H P(n) Expansion 1} we have
\begin{equation}\label{P31-X1,32}
	\boxed{P_{31} -\overset{\infty}{\mathscr{X}}_{1,32} = -C_{\rho}(0)\al(0)\mathcal{R}_{1,34}(0;n)-\mathcal{R}_{1,32}(0;n) + \mathcal{R}_{1,32}(0;n-1) + O(e^{-2cn}). }
\end{equation}

\subsection{Asymptotics of $E(n)$}
 Recalling \eqref{den C}, \eqref{E} we have
\begin{equation}\label{E1}
	\begin{split}
		& 	E(n)  =   P_{41} D_{12}^{33}-P_{42} D_{11}^{33}+P_{43} D_{11}^{32}  \\ & +\overset{\infty}{\mathscr{X}}_{1,32} \left( D_{11}^{32} D_{23}^{44} -  D_{23}^{34} D_{11}^{42} - D_{11}^{22} D_{33}^{44} - D_{21}^{32} D_{13}^{44} + P_{14}P_{41}D_{22}^{33} - P_{14}P_{42}D_{21}^{33} - P_{13}P_{41}D_{22}^{34} + P_{13}P_{42}D_{21}^{34}    \right).
	\end{split}
\end{equation}
The asymptotics of the $D^{rs}_{jk}$ terms in the above expression that are not among those already considered in  equations \eqref{11 33}-\eqref{11 32} are given below:
\begin{align}
	D^{44}_{23} = P_{23}P_{44}  -P_{24}P_{43} & =  \al(0)\mathcal{R}_{1,21}(0;n)+ O(e^{-2cn}), \\
	D^{34}_{23} = P_{23}P_{34}  -P_{24}P_{33}  & =\al(0)\mathcal{R}_{1,21}(0;n)\mathcal{R}_{1,34}(0;n)+ O(e^{-3cn}), \\
	D^{44}_{33} =	P_{33}P_{44} -P_{34}P_{43} & = -\al(0)\mathcal{R}_{1,41}(0;n)\mathcal{R}_{1,34}(0;n)+ O(e^{-3cn}) ,\\
	D^{32}_{21} =	P_{21}P_{32} -P_{22}P_{31} & = \frac{1}{\al_0} + O(e^{-2cn}),\\
	D^{44}_{13} =	P_{13}P_{44} -P_{14}P_{43} & = \al(0) + O(e^{-2cn}),\\
	D^{33}_{22} =	P_{22}P_{33}  -P_{23}P_{32} & =  -\frac{1}{\al(0)}\mathcal{R}_{1,23}(0;n)\mathcal{R}_{1,34}(0;n)+ O(e^{-3cn})    , \\
	D^{34}_{22} =	P_{22}P_{34}  -P_{24}P_{32} & =  -\mathcal{R}_{1,21}(0;n)+ O(e^{-2cn})    , \\
	D^{34}_{21} =	P_{21}P_{34} -P_{24}P_{31} & = -\al(0)\mathcal{R}_{1,21}(0;n)\left(
		C_{\rho}(0)\al(0)\mathcal{R}_{1,34}(0;n)+\mathcal{R}_{1,32}(0;n)\right)+ O(e^{-3cn}) .\\
\end{align}
Using these we observe that the leading order asymptotics of \eqref{E1} is given by the following three terms:
\begin{align}
  P_{41}D^{33}_{12}  & =  - C_{\rho}(0)\mathcal{R}_{1,14}(0;n)+ O(e^{-2cn}), \\
P_{43} D_{11}^{32} & = C_{\rho}(0)\mathcal{R}_{1,14}(0;n)+\frac{\mathcal{R}_{1,12}(0;n)}{\al(0)} + O(e^{-2cn}) ,\\
\overset{\infty}{\mathscr{X}}_{1,32}	D^{32}_{21} 	D^{44}_{13} & = - \mathcal{R}_{1,32}(0;n-1) (1+ O(e^{-2cn})),
\end{align}
where in the last equation we have used \eqref{relevant Xinfs}. We thus have
\begin{equation}\label{E asymp}\boxed{
	E(n) = \frac{\mathcal{R}_{1,12}(0;n)}{\al(0)} - \mathcal{R}_{1,32}(0;n-1) + O(e^{-2cn}).}
\end{equation}

\subsection{Asymptotics of  $h^{(0,1)}_{n-1}$}\label{sec asymp h_n 01}

In this subsection we combine the equations \eqref{h exact}, \eqref{11 33}, \eqref{remaining part}, \eqref{relevant Xinfs}, \eqref{P31-X1,32}, and   \eqref{E asymp} to obtain the asymptotics of $h^{(0,1)}_{n-1}$. To that end, we consider generic symbols $\phi$ and $d$ for which the following four properties hold
\begin{enumerate}
	\item  There exists $n_1 \geq n_*$, $r_1 \in   (r^3_*,r^2_*)$\footnote{Recall that the choice of $r_*$ was fixed right below equation \eqref{smallnorn000}.} and a constant $C_1>0$ such that 
	\begin{equation}\label{generic symbols 1}
		|\mathcal{R}_{1,32}(0;n)\mathcal{R}_{1,14}(0;n)| \geq C_1r^{n}_1.	
	\end{equation}
	Using $r_1>r^3_*$, we can rewrite the r.h.s. of \eqref{remaining part} as \begin{equation}
		-\frac{1}{\al(0)}\mathcal{R}_{1,32}(0;n)\mathcal{R}_{1,14}(0;n) \left(1+O(e^{-c_1n})\right), \qquad \mbox{with} \qquad c_1 = - \log\left(\frac{r^{3}_*}{r_1}\right)>0.
	\end{equation} Notice that we need $r_1< r^2_*$ in order to make the estimate \eqref{generic symbols 1} compatible with the following upper bound obtained from \eqref{upper bound on R_1 jk 0;n}
	\begin{equation}\label{upper bound denominator}
		|\mathcal{R}_{1,32}(0;n)\mathcal{R}_{1,14}(0;n)| < C_0^2r^{2n}_*.	
	\end{equation} 
	
\item	There exists $n_2 \geq n_1$, $r_2 \in  (r^3_*,r^2_*)$ and a constant $C_2>0$ such that 
	\begin{equation}\label{generic symbols 2}
		|\mathcal{R}_{1,32}(0;n)\mathcal{R}_{1,14}(0;n) - \mathcal{R}_{1,12}(0;n)\mathcal{R}_{1,34}(0;n)| \geq C_2r^{n}_2, \qquad \mbox{for all} \qquad n>n_2.	
	\end{equation}
	Using $r_2>r^3_*$, we can now rewrite \eqref{11 33} as 
	\begin{equation}\label{P det den}
		D^{33}_{11}  = \left( \mathcal{R}_{1,32}(0;n)\mathcal{R}_{1,14}(0;n) - \mathcal{R}_{1,12}(0;n)\mathcal{R}_{1,34}(0;n) \right) \left( 1 + O(e^{-c_2n}) \right),
	\end{equation}
	with $c_2 = - \log\left(\frac{r^{3}_*}{r_2}\right)>0$.
\item	There exists $n_3 \geq n_2$, $r_3 \in  (r^2_*,r_*)$ and a constant $C_3>0$ such that 
\begin{equation}\label{generic symbols 3}
	\left|\frac{\mathcal{R}_{1,12}(0;n)}{\al(0)} - \mathcal{R}_{1,32}(0;n-1)\right| \geq C_3r^{n}_3, \qquad \mbox{for all} \qquad n>n_3.	
\end{equation}	
	Using $r_3>r^2_*$, we can now rewrite \eqref{E asymp} as 
		\begin{equation}\label{E asymp 2}
		E(n)  = \left( \frac{\mathcal{R}_{1,12}(0;n)}{\al(0)} - \mathcal{R}_{1,32}(0;n-1) \right) \left( 1 + O(e^{-c_3n}) \right),
	\end{equation}
	with $c_3 = - \log\left(\frac{r^{2}_*}{r_3}\right)>0.$
	\item	There exists $n_4 \geq n_3$, $r_4 \in  (r^2_*,r_*)$ and a constant $C_4>0$ such that 
	\begin{equation}\label{generic symbols 4}
		\left|-C_{\rho}(0)\al(0)\mathcal{R}_{1,34}(0;n)-\mathcal{R}_{1,32}(0;n) + \mathcal{R}_{1,32}(0;n-1)\right| \geq C_4r^{n}_4, \qquad \mbox{for all} \qquad n>n_4.	
	\end{equation}	
	Using $r_4>r^2_*$, we can now rewrite \eqref{P31-X1,32} as 
	\begin{equation}\label{P31-X132 asymp}
		P_{31} -\overset{\infty}{\mathscr{X}}_{1,32}  = \left( -C_{\rho}(0)\al(0)\mathcal{R}_{1,34}(0;n)-\mathcal{R}_{1,32}(0;n) + \mathcal{R}_{1,32}(0;n-1) \right) \left( 1 + O(e^{-c_4n}) \right),
	\end{equation} with $c_4 = - \log\left(\frac{r^{2}_*}{r_4}\right)>0$.
\end{enumerate}
Let \begin{equation}
	F(n):= \frac{\mathcal{R}_{1,32}(0;n)\mathcal{R}_{1,14}(0;n)\left( C_{\rho}(0)\al(0)\mathcal{R}_{1,34}(0;n)+\mathcal{R}_{1,32}(0;n) - \mathcal{R}_{1,32}(0;n-1) \right)}{\left( \mathcal{R}_{1,12}(0;n) - \al(0) \mathcal{R}_{1,32}(0;n-1) \right)\left( \mathcal{R}_{1,32}(0;n)\mathcal{R}_{1,14}(0;n) - \mathcal{R}_{1,12}(0;n)\mathcal{R}_{1,34}(0;n) \right)},
\end{equation}
and
\begin{equation}
	\mathfrak{c} := \min\{c_1,c_2,c_3,c_4\}.
\end{equation}
Recalling \eqref{h exact}, \eqref{11 33}, \eqref{remaining part}, \eqref{P31-X1,32}, \eqref{E asymp} we have
\begin{equation}\label{h asymp3}
	\frac{-1}{h^{(0,1)}_{n-1}} - \overset{\infty}{\mathscr{X}}_{1,31}  =  F(n)\left( 1 + O(e^{-\mathfrak{c}n}) \right).
\end{equation}
Recalling \eqref{relevant Xinfs} and the above assumptions we observe that there exists $\widehat{C}$ such that
\begin{equation}\label{h exact1}
	\left|\frac{\overset{\infty}{\mathscr{X}}_{1,31}}{F_n}\right| \leq \widehat{C} \left(\frac{r^5_*}{r_4r_1}\right)^n = \widehat{C} e^{-(c_1+c_4)n}. 
\end{equation}
So we can rewrite \eqref{h asymp3} as
\begin{equation}\label{h asymp4}
	\frac{-1}{h^{(0,1)}_{n-1}}  =  F(n)\left( 1 + O(e^{-\mathfrak{c}n}) \right),
\end{equation}
and therefore
\begin{equation}\label{last}
	h^{(0,1)}_{n-1} = - F^{-1}(n)\left( 1 + O(e^{-\mathfrak{c}n}) \right).
\end{equation}
We have just concluded the proof of Theorem \ref{main thm}.

\begin{comment}
	content...

\section{Dictionary translating notation from \cite{Chelkak} to notation from \cite{GI}}\label{section Dictionary}

\begin{center}
	\begin{tabular}{ || c |  c || } 
		\hline
		Notation from \cite{GI} & Notation from \cite{Chelkak} \\
		\hline \hline
		$\phi$ & $w$  \\
		\hline  
		$\dd$ & $-\xi$  \\
		\hline  
		$w$ & $-\xi w$ \\ 
		\hline  
		$\phi_s$ & $\al_s$  \\
		\hline  
		$w_s$ & $-\be_s$ \\ 
		\hline
	\end{tabular}
\end{center}
\end{comment}

\begin{comment}

\newpage
\section*{\textbf{List of Symbols}}\label{Sec list of symbols}

\footnotesize
\begin{tabular}{lll}
	\textbf{Symbol} & \textbf{Description} & \textbf{Definition} \\
	$D_n[\phi,w;r,s]$ & The Toeplitz+Hankel determinant with symbols $\phi$ and $w$ and offsets $r$ and  $s$ & \eqref{Det} \\
	$\mathcal{P}_n(z;r,s)$ & The orthogonal polynomial associated with the Toeplitz+Hankel determinants $D_n[\phi,w;r,s]$. & \eqref{T+H OP}, \eqref{T+H OP Det rep}

\end{tabular}
	content...
\end{comment}
\color{black}

\end{document}